\expandafter \def \csname CHAPLABELintro\endcsname {1}
\expandafter \def \csname EQLABELduality\endcsname {1.1?}
\expandafter \def \csname EQLABELkclasses\endcsname {1.2?}
\expandafter \def \csname CHAPLABELpoly\endcsname {2}
\expandafter \def \csname EQLABELWei\endcsname {2.1?}
\expandafter \def \csname EQLABELdis\endcsname {2.2?}
\expandafter \def \csname TABLABELscaling\endcsname {2.1?}
\expandafter \def \csname CHAPLABELalgo\endcsname {3}
\expandafter \def \csname EQLABELh11\endcsname {3.1?}
\expandafter \def \csname FIGLABELfibrations\endcsname {3.1?}
\expandafter \def \csname EQLABELbetti\endcsname {3.2?}
\expandafter \def \csname EQLABELh11b\endcsname {3.3?}
\expandafter \def \csname EQLABELnt\endcsname {3.4?}
\expandafter \def \csname FIGLABELtorus\endcsname {3.2?}
\expandafter \def \csname EQLABELnik\endcsname {3.5?}
\expandafter \def \csname TABLABELbottoms\endcsname {3.1?}
\expandafter \def \csname CHAPLABELmir\endcsname {4}
\expandafter \def \csname TABLABEL2si\endcsname {4.1?}
\expandafter \def \csname FIGLABELfan\endcsname {4.1?}
\expandafter \def \csname CHAPLABELsubtle\endcsname {5}
\expandafter \def \csname CHAPLABELcon\endcsname {6}
\expandafter \def \csname CHAPLABELappendix\endcsname {-2}

\font\eightrm=cmr8 at 8pt

\font\seventeenrm=cmr17 at 17pt
\font\twentyonerm=cmr17 at 21pt

\font\ss=cmss10

\font\csc=cmcsc10

\font\twelvecal=cmsy10 at 12pt

\font\twelvemath=cmmi12

\font\seventeenbold=cmbx7 at 17pt

\font\fively=lasy5
\font\sevenly=lasy7
\font\tenly=lasy10

\textfont10=\tenly
\scriptfont10=\sevenly
\scriptscriptfont10=\fively
\magnification=1200
\parskip=10pt
\parindent=20pt
\def\today{\ifcase\month\or January\or February\or March\or April\or May\or
June
       \or July\or August\or September\or October\or November\or December\fi
       \space\number\day, \number\year}

\def\title#1{\footline={\ifnum\pageno<2\hfil
       \else\hss\tenrm\folio\hss\fi}\vskip1truein\centerline{{#1}
       \footnote{\raise1ex\hbox{*}}{\eightrm Supported in part
       by the Robert A. Welch Foundation and N.S.F. Grants
       PHY-880637 and\break PHY-8605978.}}}

\def\newpage{\vfill\eject}
\def\abstract#1{\centerline{\bf ABSTRACT}\vskip.2truein{\narrower\noindent#1
       \smallskip}}

\def\runninghead#1#2{\voffset=2\baselineskip\nopagenumbers
       \headline={\ifodd\pageno\rightheadline\else \leftheadline\fi}
       \def\rightheadline{{\sl#1}\hfill{\rm\folio}}
       \def\leftheadline{{\rm\folio}\hfill{\sl#2}}}
\def\SS{\mathhexbox278}

\newcount\footnoteno
\def\Footnote#1{\advance\footnoteno by 1
                \let\SF=\empty
                \ifhmode\edef\SF{\spacefactor=\the\spacefactor}\/\fi
                $^{\the\footnoteno}$\ignorespaces
                \SF\vfootnote{$^{\the\footnoteno}$}{#1}}

\def\figbox#1#2#3{\vbox{\vskip15pt
                   \vbox{\hrule
                    \hbox{\vrule
                     \vbox{\vskip12truept\centerline #1 \vskip6truept
                          {\hskip.4truein\vbox{\hsize=5truein\noindent
                          {\bf Figure\hskip5truept#2:}\hskip5truept#3}}
                     \vskip18truept}
                    \vrule}
                   \hrule}}}
\def\place#1#2#3{\vbox to0pt{\kern-\parskip\kern-7pt
                             \kern-#2truein\hbox{\kern#1truein #3}
                             \vss}\nointerlineskip}
\def\figurecaption#1#2{\kern.75truein\vbox{\hsize=5truein\noindent{\bf Figure
    \figlabel{#1}:} #2}}
\def\tablecaption#1#2{\kern.75truein\lower12truept\hbox{\vbox{\hsize=5truein
    \noindent{\bf Table\hskip5truept\tablabel{#1}:} #2}}}
\def\boxed#1{\lower3pt\hbox{
                       \vbox{\hrule\hbox{\vrule

\vbox{\kern2pt\hbox{\kern3pt#1\kern3pt}\kern3pt}\vrule}
                         \hrule}}}

\def\g{\gamma}
\def\d{\delta}\def\D{\Delta}

\def\l{\lambda}\def\L{\Lambda}
\def\m{\mu}
\def\n{\nu}

\def\P{\Pi}

\def\S{\Sigma}

\def\ca#1{\relax\ifmmode {{\cal #1}}\else $\cal #1$\fi}

\def\calb{{\cal B}}

\def\calm{{\cal M}}

\def\inbar{\vrule height1.5ex width.4pt depth0pt}
\def\IB{\relax{\rm I\kern-.18em B}}
\def\IC{\relax\hbox{\kern.25em$\inbar\kern-.3em{\rm C}$}}
\def\ID{\relax{\rm I\kern-.18em D}}
\def\IE{\relax{\rm I\kern-.18em E}}
\def\IF{\relax{\rm I\kern-.18em F}}
\def\IG{\relax\hbox{\kern.25em$\inbar\kern-.3em{\rm G}$}}
\def\IH{\relax{\rm I\kern-.18em H}}
\def\II{\relax{\rm I\kern-.18em I}}
\def\IK{\relax{\rm I\kern-.18em K}}
\def\IL{\relax{\rm I\kern-.18em L}}
\def\IM{\relax{\rm I\kern-.18em M}}
\def\IN{\relax{\rm I\kern-.18em N}}
\def\IO{\relax\hbox{\kern.25em$\inbar\kern-.3em{\rm O}$}}
\def\IP{\relax{\rm I\kern-.18em P}}
\def\IQ{\relax\hbox{\kern.25em$\inbar\kern-.3em{\rm Q}$}}
\def\IR{\relax{\rm I\kern-.18em R}}
\def\IZ{\relax\ifmmode\hbox{\ss Z\kern-.4em Z}\else{\ss Z\kern-.4em Z}\fi}
\def\IGa{\relax{\rm I}\kern-.18em\Gamma}
\def\IPi{\relax{\rm I}\kern-.18em\Pi}
\def\ITh{\relax\hbox{\kern.25em$\inbar\kern-.3em\Theta$}}
\def\IOm{\relax\thinspace\inbar\kern1.95pt\inbar\kern-5.525pt\Omega}


\def\ie{{\it i.e.,\ \/}}
\def\eg{{\it e.g.,\ \/}}
\def\noblackboxes{\overfullrule=0pt}

\def\cy{Calabi--Yau}
\def\cym{Calabi--Yau manifold}
\def\cys{Calabi--Yau manifolds}
\def\cyt{Calabi--Yau threefold}

\def\K{K\"ahler}

\def\H#1#2{\relax\ifmmode {H^{#1#2}}\else $H^{#1 #2}$\fi}
\def\M{\relax\ifmmode{\calm}\else $\calm$\fi}

\def\Bigcheck{\lower3.8pt\hbox{\smash{\hbox{{\twentyonerm \v{}}}}}}
\def\bigboldcheck{\smash{\hbox{{\seventeenbold\v{}}}}}

\def\Bighat{\lower3.8pt\hbox{\smash{\hbox{{\twentyonerm \^{}}}}}}

\def\Msharp{\relax\ifmmode{\calm^\sharp}\else $\smash{\calm^\sharp}$\fi}
\def\Mflat{\relax\ifmmode{\calm^\flat}\else $\smash{\calm^\flat}$\fi}
\def\preMcheck{\kern2pt\hbox{\Bigcheck\kern-12pt{$\cal M$}}}
\def\Mcheck{\relax\ifmmode\preMcheck\else $\preMcheck$\fi}
\def\preMhat{\kern2pt\hbox{\Bighat\kern-12pt{$\cal M$}}}
\def\Mhat{\relax\ifmmode\preMhat\else $\preMhat$\fi}

\def\Bsharp{\relax\ifmmode{\calb^\sharp}\else $\calb^\sharp$\fi}
\def\Bflat{\relax\ifmmode{\calb^\flat}\else $\calb^\flat$ \fi}
\def\preBcheck{\hbox{\Bigcheck\kern-9pt{$\cal B$}}}
\def\Bcheck{\relax\ifmmode\preBcheck\else $\preBcheck$\fi}
\def\preBhat{\hbox{\Bighat\kern-9pt{$\cal B$}}}
\def\Bhat{\relax\ifmmode\preBhat\else $\preBhat$\fi}

\def\figBcheck{\kern3pt\hbox{\raise1pt\hbox{\bigboldcheck}\kern-11pt
    {\twelvecal B}}}
\def\figBsharp{{\twelvecal B}\raise5pt\hbox{$\twelvemath\sharp$}}
\def\figBflat{{\twelvecal B}\raise5pt\hbox{$\twelvemath\flat$}}

\def\gcheck{\hbox{\lower2.5pt\hbox{\Bigcheck}\kern-8pt$\g$}}
\def\lhat{\hbox{\raise.5pt\hbox{\Bighat}\kern-8pt$\l$}}

\def\Fcheck{\kern2pt\hbox{\raise1pt\hbox{\Bigcheck}\kern-10pt{$\cal F$}}}
\def\Fhat{\kern2pt\hbox{\raise1pt\hbox{\Bighat}\kern-10pt{$\cal F$}}}

\def\cp#1{\relax\ifmmode {\IP\kern-2pt{}_{#1}}\else $\IP\kern-2pt{}_{#1}$\fi}
\def\h#1#2{\relax\ifmmode {b_{#1#2}}\else $b_{#1#2}$\fi}

\def\frac#1#2{{#1\over #2}}

\def\cone{\relax\thinspace\hbox{$<\kern-.8em{)}$}}
\mathchardef\mho"0A30

\def\-{\hphantom{-}}



\def\picture #1 by #2 (#3){\vbox to #2{\hrule width #1 height 0pt depth 0pt
                                       \vfill\special{picture #3}}}
\def\scaledpicture #1 by #2 (#3 scaled #4){{\dimen0=#1 \dimen1=#2
           \divide\dimen0 by 1000 \multiply\dimen0 by #4
            \divide\dimen1 by 1000 \multiply\dimen1 by #4
            \picture \dimen0 by \dimen1 (#3 scaled #4)}}
\def\illustration #1 by #2 (#3){\vbox to #2{\hrule width #1 height 0pt depth
0pt
                                       \vfill\special{illustration #3}}}
\def\scaledillustration #1 by #2 (#3 scaled #4){{\dimen0=#1 \dimen1=#2
           \divide\dimen0 by 1000 \multiply\dimen0 by #4
            \divide\dimen1 by 1000 \multiply\dimen1 by #4
            \illustration \dimen0 by \dimen1 (#3 scaled #4)}}


\def\delaOssa{\nobreak\vskip1truein\hbox to\hsize
       {\hskip 4truein Xenia de la Ossa\hfill}}

\def\hoy{\number\day\space de \ifcase\month\or enero\or febrero\or marzo\or
       abril\or mayo\or junio\or julio\or agosto\or septiembre\or octubre\or
       noviembre\or diciembre\fi\space de \number\year}

\def\cropen#1{\crcr\noalign{\vskip #1}}

\newif\ifproofmode
\proofmodefalse

\newif\ifforwardreference
\forwardreferencefalse

\newif\ifchapternumbers
\chapternumbersfalse

\newif\ifcontinuousnumbering
\continuousnumberingfalse

\newif\iffigurechapternumbers
\figurechapternumbersfalse

\newif\ifcontinuousfigurenumbering
\continuousfigurenumberingfalse

\newif\iftablechapternumbers
\tablechapternumbersfalse

\newif\ifcontinuoustablenumbering
\continuoustablenumberingfalse

\font\eqsixrm=cmr6

\def\marginstyle{\eqsixrm}

\newtoks\chapletter
\newcount\chapno
\newcount\eqlabelno
\newcount\figureno
\newcount\tableno

\chapno=0
\eqlabelno=0
\figureno=0
\tableno=0

\def\chapfolio{\ifnum\chapno>0 \the\chapno\else\the\chapletter\fi}

\def\bumpchapno{\ifnum\chapno>-1 \global\advance\chapno by 1
\else\global\advance\chapno by -1 \setletter\chapno\fi
\ifcontinuousnumbering\else\global\eqlabelno=0 \fi
\ifcontinuousfigurenumbering\else\global\figureno=0 \fi
\ifcontinuoustablenumbering\else\global\tableno=0 \fi}

\def\setletter#1{\ifcase-#1{}\or{}%
\or\global\chapletter={A}%
\or\global\chapletter={B}%
\or\global\chapletter={C}%
\or\global\chapletter={D}%
\or\global\chapletter={E}%
\or\global\chapletter={F}%
\or\global\chapletter={G}%
\or\global\chapletter={H}%
\or\global\chapletter={I}%
\or\global\chapletter={J}%
\or\global\chapletter={K}%
\or\global\chapletter={L}%
\or\global\chapletter={M}%
\or\global\chapletter={N}%
\or\global\chapletter={O}%
\or\global\chapletter={P}%
\or\global\chapletter={Q}%
\or\global\chapletter={R}%
\or\global\chapletter={S}%
\or\global\chapletter={T}%
\or\global\chapletter={U}%
\or\global\chapletter={V}%
\or\global\chapletter={W}%
\or\global\chapletter={X}%
\or\global\chapletter={Y}%
\or\global\chapletter={Z}\fi}

\def\tempsetletter#1{\ifcase-#1{}\or{}%
\or\global\chapletter={A}%
\or\global\chapletter={B}%
\or\global\chapletter={C}%
\or\global\chapletter={D}%
\or\global\chapletter={E}%
\or\global\chapletter={F}%
\or\global\chapletter={G}%
\or\global\chapletter={H}%
\or\global\chapletter={I}%
\or\global\chapletter={J}%
\or\global\chapletter={K}%
\or\global\chapletter={L}%
\or\global\chapletter={M}%
\or\global\chapletter={N}%
\or\global\chapletter={O}%
\or\global\chapletter={P}%
\or\global\chapletter={Q}%
\or\global\chapletter={R}%
\or\global\chapletter={S}%
\or\global\chapletter={T}%
\or\global\chapletter={U}%
\or\global\chapletter={V}%
\or\global\chapletter={W}%
\or\global\chapletter={X}%
\or\global\chapletter={Y}%
\or\global\chapletter={Z}\fi}

\def\chapshow#1{\ifnum#1>0 \relax#1%
\else{\tempsetletter{\number#1}\chapno=#1\chapfolio}\fi}

\def\chaplabel#1{\bumpchapno\ifproofmode\ifforwardreference
\immediate\write1{\noexpand\expandafter\noexpand\def
\noexpand\csname CHAPLABEL#1\endcsname{\the\chapno}}\fi\fi
\global\expandafter\edef\csname CHAPLABEL#1\endcsname
{\the\chapno}\ifproofmode\llap{\hbox{\marginstyle #1\ }}\fi\chapfolio}

\def\eqnum{\global\advance\eqlabelno by 1
\eqno(\ifchapternumbers\chapfolio.\fi\the\eqlabelno)}

\def\eqlabel#1{\global\advance\eqlabelno by 1 \ifproofmode\ifforwardreference
\immediate\write1{\noexpand\expandafter\noexpand\def
\noexpand\csname EQLABEL#1\endcsname{\the\chapno.\the\eqlabelno?}}\fi\fi
\global\expandafter\edef\csname EQLABEL#1\endcsname
{\the\chapno.\the\eqlabelno?}\eqno(\ifchapternumbers\chapfolio.\fi
\the\eqlabelno)\ifproofmode\rlap{\hbox{\marginstyle #1}}\fi}

\def\eqalignnum{\global\advance\eqlabelno by 1
&(\ifchapternumbers\chapfolio.\fi\the\eqlabelno)}

\def\eqalignlabel#1{\global\advance\eqlabelno by 1 \ifproofmode
\ifforwardreference\immediate\write1{\noexpand\expandafter\noexpand\def
\noexpand\csname EQLABEL#1\endcsname{\the\chapno.\the\eqlabelno?}}\fi\fi
\global\expandafter\edef\csname EQLABEL#1\endcsname
{\the\chapno.\the\eqlabelno?}&(\ifchapternumbers\chapfolio.\fi
\the\eqlabelno)\ifproofmode\rlap{\hbox{\marginstyle #1}}\fi}

\def\eqref#1{\hbox{(\ifundefined{EQLABEL#1}***)\ifproofmode\ifforwardreference%
\else\write16{ ***Undefined Equation Reference #1*** }\fi
\else\write16{ ***Undefined Equation Reference #1*** }\fi
\else\edef\LABxx{\getlabel{EQLABEL#1}}%
\def\LAByy{\expandafter\stripchap\LABxx}\ifchapternumbers%
\chapshow{\LAByy}.\expandafter\stripeq\LABxx%
\else\ifnum\number\LAByy=\chapno\relax\expandafter\stripeq\LABxx%
\else\chapshow{\LAByy}.\expandafter\stripeq\LABxx\fi\fi)\fi}%
\ifproofmode\write2{Equation #1}\fi}

\def\fignum{\global\advance\figureno by 1
\relax\iffigurechapternumbers\chapfolio.\fi\the\figureno}

\def\figlabel#1{\global\advance\figureno by 1
\relax\ifproofmode\ifforwardreference
\immediate\write1{\noexpand\expandafter\noexpand\def
\noexpand\csname FIGLABEL#1\endcsname{\the\chapno.\the\figureno?}}\fi\fi
\global\expandafter\edef\csname FIGLABEL#1\endcsname
{\the\chapno.\the\figureno?}\iffigurechapternumbers\chapfolio.\fi
\ifproofmode\llap{\hbox{\marginstyle#1
\kern1.2truein}}\relax\fi\the\figureno}

\def\figref#1{\hbox{\ifundefined{FIGLABEL#1}!!!!\ifproofmode\ifforwardreference%
\else\write16{ ***Undefined Figure Reference #1*** }\fi
\else\write16{ ***Undefined Figure Reference #1*** }\fi
\else\edef\LABxx{\getlabel{FIGLABEL#1}}%
\def\LAByy{\expandafter\stripchap\LABxx}\iffigurechapternumbers%
\chapshow{\LAByy}.\expandafter\stripeq\LABxx%
\else\ifnum \number\LAByy=\chapno\relax\expandafter\stripeq\LABxx%
\else\chapshow{\LAByy}.\expandafter\stripeq\LABxx\fi\fi\fi}%
\ifproofmode\write2{Figure #1}\fi}

\def\tabnum{\global\advance\tableno by 1
\relax\iftablechapternumbers\chapfolio.\fi\the\tableno}

\def\tablabel#1{\global\advance\tableno by 1
\relax\ifproofmode\ifforwardreference
\immediate\write1{\noexpand\expandafter\noexpand\def
\noexpand\csname TABLABEL#1\endcsname{\the\chapno.\the\tableno?}}\fi\fi
\global\expandafter\edef\csname TABLABEL#1\endcsname
{\the\chapno.\the\tableno?}\iftablechapternumbers\chapfolio.\fi
\ifproofmode\llap{\hbox{\marginstyle#1
\kern1.2truein}}\relax\fi\the\tableno}

\def\tabref#1{\hbox{\ifundefined{TABLABEL#1}!!!!\ifproofmode\ifforwardreference%
\else\write16{ ***Undefined Table Reference #1*** }\fi
\else\write16{ ***Undefined Table Reference #1*** }\fi
\else\edef\LABtt{\getlabel{TABLABEL#1}}%
\def\LABTT{\expandafter\stripchap\LABtt}\iftablechapternumbers%
\chapshow{\LABTT}.\expandafter\stripeq\LABtt%
\else\ifnum\number\LABTT=\chapno\relax\expandafter\stripeq\LABtt%
\else\chapshow{\LABTT}.\expandafter\stripeq\LABtt\fi\fi\fi}%
\ifproofmode\write2{Table#1}\fi}

\newdimen\sectionskip     \sectionskip=20truept
\newcount\sectno
\def\section#1#2{\sectno=0 \null\vskip\sectionskip
    \centerline{\chaplabel{#1}.~~{\bf#2}}\nobreak\vskip.2truein
    \noindent\ignorespaces}

\def\advancesectno{\global\advance\sectno by 1}
\def\sectfolio{\number\sectno}
\def\subsection#1{\goodbreak\advancesectno\null\vskip10pt
                  \noindent\chapfolio.~\sectfolio.~{\bf #1}
                  \nobreak\vskip.05truein\noindent\ignorespaces}

\def\uttg#1{\null\vskip.1truein
    \ifproofmode \line{\hfill{\bf Draft}:
    UTTG--{#1}--\number\year}\line{\hfill\today}
    \else \line{\hfill UTTG--{#1}--\number\year}
    \line{\hfill\ifcase\month\or January\or February\or March\or April\or
May\or June
    \or July\or August\or September\or October\or November\or December\fi
    \space\number\year}\fi}

\def\getlabel#1{\csname#1\endcsname}
\def\ifundefined#1{\expandafter\ifx\csname#1\endcsname\relax}
\def\stripchap#1.#2?{#1}
\def\stripeq#1.#2?{#2}

%
\catcode`@=11 
\def\space@ver#1{\let\@sf=\empty\ifmmode#1\else\ifhmode%
\edef\@sf{\spacefactor=\the\spacefactor}\unskip${}#1$\relax\fi\fi}
\newcount\referencecount     \referencecount=0
\newif\ifreferenceopen       \newwrite\referencewrite
\newtoks\rw@toks
\def\refmark#1{\relax[#1]}
\def\refend{\refmark{\number\referencecount}}
\newcount\lastrefsbegincount \lastrefsbegincount=0
\def\refsend{\refmark{\count255=\referencecount%
\advance\count255 by -\lastrefsbegincount%
\ifcase\count255 \number\referencecount%
\or\number\lastrefsbegincount,\number\referencecount%
\else\number\lastrefsbegincount-\number\referencecount\fi}}
\def\refch@ck{\chardef\rw@write=\referencewrite
\ifreferenceopen\else\referenceopentrue
\immediate\openout\referencewrite=referenc.texauxil \fi}
%
{\catcode`\^^M=\active 
  \gdef\obeyendofline{\catcode`\^^M\active \let^^M\ }}%
%
{\catcode`\^^M=\active 
  \gdef\ignoreendofline{\catcode`\^^M=5}}
{\obeyendofline\gdef\rw@start#1{\def\t@st{#1}\ifx\t@st\blankend%
\endgroup\@sf\relax\else\ifx\t@st\bl@nkend\endgroup\@sf\relax%
\else\rw@begin#1
\backtotext
\fi\fi}}
{\obeyendofline\gdef\rw@begin#1
{\def\n@xt{#1}\rw@toks={#1}\relax%
\rw@next}}
\def\blankend{}
{\obeylines\gdef\bl@nkend{
}}
\newif\iffirstrefline  \firstreflinetrue
\def\rwr@teswitch{\ifx\n@xt\blankend\let\n@xt=\rw@begin%
\else\iffirstrefline\global\firstreflinefalse%
\immediate\write\rw@write{\noexpand\obeyendofline\the\rw@toks}%
\let\n@xt=\rw@begin%
\else\ifx\n@xt\rw@@d \def\n@xt{\immediate\write\rw@write{%
\noexpand\ignoreendofline}\endgroup\@sf}%
\else\immediate\write\rw@write{\the\rw@toks}%
\let\n@xt=\rw@begin\fi\fi\fi}
\def\rw@next{\rwr@teswitch\n@xt}
\def\rw@@d{\backtotext} \let\rw@end=\relax
\let\backtotext=\relax

\newdimen\refindent     \refindent=30pt
\def\Textindent#1{\noindent\llap{#1\enspace}\ignorespaces}
\def\refitem#1{\par\hangafter=0 \hangindent=\refindent\Textindent{#1}}
\def\REFNUM#1{\space@ver{}\refch@ck\firstreflinetrue%
\global\advance\referencecount by 1 \xdef#1{\the\referencecount}}
\def\refnum#1{\space@ver{}\refch@ck\firstreflinetrue%
\global\advance\referencecount by 1\xdef#1{\the\referencecount}\refend}

\def\REF#1{\REFNUM#1%
\immediate\write\referencewrite{%
\noexpand\refitem{#1.}}%
\begingroup\obeyendofline\rw@start}
\def\ref{\refnum\?%
\immediate\write\referencewrite{\noexpand\refitem{\?.}}%
\begingroup\obeyendofline\rw@start}
\def\Ref#1{\refnum#1%
\immediate\write\referencewrite{\noexpand\refitem{#1.}}%
\begingroup\obeyendofline\rw@start}
\def\REFS#1{\REFNUM#1\global\lastrefsbegincount=\referencecount%
\immediate\write\referencewrite{\noexpand\refitem{#1.}}%
\begingroup\obeyendofline\rw@start}

\def\REFSCON#1{\REF#1}

\def\cite#1{\refmark#1}
\catcode`@=12 
%

\input epsf.tex
\baselineskip=15pt plus 1pt minus 1pt
\parskip=5pt
\chapternumberstrue
\figurechapternumberstrue
\tablechapternumberstrue
\ifproofmode
\immediate\openout2=allcrossreferfile \fi
\ifforwardreference\input labelfile
\ifproofmode\immediate\openout1=labelfile \fi\fi
\noblackboxes
\hfuzz=1pt
\vfuzz=2pt


\def\hourandminute{\count255=\time\divide\count255 by 60
\xdef\hour{\number\count255}
\multiply\count255 by -60\advance\count255 by\time
\hour:\ifnum\count255<10 0\fi\the\count255}

\def\subsection#1{\goodbreak\advancesectno\null\vskip10pt
                  \noindent{\it \chapfolio.\sectfolio.~#1}
                  \nobreak\vskip.05truein\noindent\ignorespaces}
\def\cite#1{\refmark{#1}}
\def\\{\hfill\break}
\def\cropen#1{\crcr\noalign{\vskip #1}}

\def\point#1{\noindent\setbox0=\hbox{#1}\kern-\wd0\box0}

\def\nab#1{{}^#1\nabla}
\def\etal{{\it et al.\/}}
\nopagenumbers\pageno=0
\rightline{\eightrm UTTG-13-97}\vskip-5pt
\rightline{\eightrm hep-th/9704097}\vskip-5pt
\rightline{\eightrm 15 May, 1997}

\vskip1truein
\centerline{\seventeenrm Toric Geometry and Enhanced Gauge}
\vskip10pt
\centerline{\seventeenrm Symmetry of F-Theory/Heterotic Vacua}
\vskip30pt
\centerline{\csc Philip~Candelas$^1$, Eugene~Perevalov$^2$ and 
Govindan~Rajesh$^3$}
\vfootnote{$^{\eightrm 1}$}{\eightrm candelas@physics.utexas.edu}
\vfootnote{$^{\eightrm 2}$}{\eightrm pereval@physics.utexas.edu}
\vfootnote{$^{\eightrm 3}$}{\eightrm rajesh@physics.utexas.edu}

\vskip.4truein\bigskip
\centerline{\it Theory Group}
\centerline{\it Department of Physics}
\centerline{\it University of Texas}
\centerline{\it Austin, TX 78712, USA}
\vskip1in\bigskip
\nobreak\vbox{
\centerline{\bf ABSTRACT}
\vskip.25truein
\noindent{We study F-theory compactified on elliptic \cyt s that are 
realised as hypersurfaces in toric varieties. The enhanced gauge group
as well as the number of massless tensor multiplets has a very simple
description in terms of toric geometry. We find a large number of
examples where the gauge group is not a subgroup of $E_8\times E_8$, but
rather, is much bigger (with rank as high as 296). The largest of these groups
is the group recently found by Aspinwall and Gross. Our algorithm can also be
applied to elliptic fourfolds, for which the groups can become very large
indeed (with rank as high as~121328). We present the gauge content for two of
the fourfolds recently studied by Klemm~\etal}
} 
\newpage
    {\bf Contents}
\vskip5pt

    1. Introduction 
\vskip3pt

    2. F-Theory Compactified on Elliptic \cy\ Threefolds 
\vskip3pt

    3. Identifying the Groups
\vskip3pt
 
\hskip10pt {\it 3.1 Generalities}
\vskip3pt

\hskip10pt {\it 3.2 The Algorithm}
\vskip3pt   

\hskip10pt {\it 3.3 Subtleties}
\vskip3pt

    4. Examples
\vskip3pt

\hskip10pt {\it 4.1 A simple example in detail}
\vskip3pt
\hskip10pt {\it 4.2 The mirror of the manifold with Hodge numbers (3,243)} 
\vskip3pt
\hskip10pt {\it 4.3 The mirror of the manifold with Hodge numbers (11,491)}
\vskip3pt
\hskip10pt {\it 4.4 The self-mirror manifold with Hodge numbers (251,251)}
\vskip3pt
\hskip10pt {\it 4.5 The elliptic fourfold
$\IP_5^{(1,1,84,516,1204,1806)}[3612]$}
\vskip3pt
\hskip10pt {\it 4.6 The elliptic fourfold
$\IP_5^{(1,1806, 75894, 466206, 1087814, 1631721)}[3263442]$}
\vskip3pt

    5. Subtleties Revisited
\vskip3pt

    6. Discussion
\vskip3pt
  
    Appendix: Tables of Gauge Groups
\newpage
\pageno=1
\headline={\ifproofmode\hfil\eightrm draft:\ \today\
\hourandminute\else\hfil\fi}
\footline={\rm\hfil\folio\hfil}
\section{intro}{Introduction}
The dualities of String Theory have been the subject of extensive study during
the last two years. Of particular interest to us here is the duality~ 
\REFS\rCV{S. Kachru and C. Vafa, Nucl. Phys. {\bf B450} (1995) 69, 
hep-th/9505105.}
\REFSCON\rFSHV{S. Ferrara, J. Harvey, A. Strominger and C. Vafa,\\
 Phys. Lett. {\bf 361B} (1995) 59, hep-th/9505162.}
\refsend\
$$\hbox{Het}[K3\times T^2, G] = \hbox{IIA}[\ca{M}] \eqlabel{duality}$$
between a $(0,4)$ heterotic compactification on $K3\times T^2$ with gauge
group $G$, and a type IIA compactification on a \cym, \ca{M}.
This is a very exciting arena in which to explore nonperturbative phenomena
in String Theory and has been the focus of much recent work~
\REFS\rKLM{A. Klemm, W. Lerche and P. Mayr, Phys .Lett. {\bf 357B} (1995) 
113, hep-th/9506112.}
\REFSCON\rVW{C. Vafa and E. Witten, hep-th/9507050.}
\REFSCON\rGL{G. G\'{o}mez and E. L\'{o}pez, Phys. Lett. {\bf 356B} (1995) 487, 
hep-th/9506024. \\
M. Bill\'{o} et. al. Class. Quant. Grav. {\bf 13} (1996) 831, hep-th/9506075.\\
I. Antoniadis, E. Gava, K. S. Narain and T. R. Taylor,\\ Nucl. Phys. {\bf B455}
(1995) 109, hep-th/9507115.\\
G. Lopes Cardoso, D. Lust and T. Mohaupt,\\ Nucl. Phys. {\bf B455} (1995) 131,
hep-th/9507113.\\
G. Curio, Phys. Lett. {\bf 368B} (1996) 78, hep-th/9509146. }  
\REFSCON\rKLT{V. Kaplunovsky, J. Louis and S. Theisen,\\ Phys. Lett. {\bf 357B}
(1995) 71, hep-th/9506110.}
\REFSCON\rKKLMV{S. Kachru, A. Klemm, W. Lerche, P. Mayr and C. Vafa,\\
Nucl. Phys. {\bf B459} (1996) 537, hep-th/9508155.}
\refsend
.

A large class of \cys\ can be realised as hypersurfaces in toric varieties and,
in virtue of a construction of Batyrev~
\REFS\rBat{V.~Batyrev, Duke Math. Journ. {\bf 69} (1993) 349.}
\REFSCON\rCOK{P.~Candelas, X.~de la Ossa and S.~Katz,
Nucl. Phys. {\bf B450} (1995) 267,\\ hep-th/9412117.}
\refsend , these have a nice description in terms of a dual pair $(\D, \nabla)$
of reflexive polyhedra. This being so we may regard the \cym\ \ca{M} as being
specified by a polyhedron $\ca{M}=\ca{M}_{\D}$. Thus, it is natural to suppose
that the polyhedron determines the gauge group $G$ that appears on the
heterotic side. This was the point of view adopted in~   
\REFS\rCF{P.~Candelas and A.~Font, hep-th/9603170.}
\REFSCON\rBS{P.~Candelas, E.~Perevalov and G.~Rajesh, hep-th/9606133.}
\REFSCON\rBC{P.~Candelas, E.~Perevalov and G.~Rajesh, hep-th/9703148.}
\refsend
.
In these articles a first dictionary between perturbative symmetry restoration
on the heterotic side and toric data was established. The
duality~\eqref{duality} has far-reaching consequences. It is believed to apply
most directly to \cys\ that are both elliptic and $K3$ fibrations. Many of
these can be described by reflexive polyhedra and so it is natural to suppose
that there is a correspondence $G=G_{\nabla}$. This was shown to be the case
for certain simple cases in~\cite{\rCF-\rBC}, the correspondence being
stated most simply in terms of the dual polyhedron $\nabla$. 

The point we make here is that if we take~\eqref{duality} seriously, then
there should be a Heterotic theory for a great many polyhedra $\nabla$ that
correspond to elliptic and $K3$ fibrations. To appreciate the consequence of
this, consider the \cym\ $$\ca{M}_{E_8}~=~\IP_{4}^{(1,1,12,28,42)}[84]$$ which
is
an elliptic $K3$ fibration and corresponds to a gauge group $E_8$. Now 
$\ca{M}_{E_8}$
has Hodge numbers $h_{11}=11$ and $h_{21}=491$. The Hodge number $h_{11}$ is
related to the rank of the group and the number of tensor multiplets, $n_T$, by
a general relation~
$$h_{11}= \hbox{rank}(G) + n_T + 2 \eqlabel{kclasses}$$
which is satisfied in this case with $G=E_8$ and $n_T=1$. Now it turns out, by
virtue of the work of Avram \etal~
\REFS\rAS{A. Avram, M. Kreuzer, M. Mandelberg and H. Skarke,
hep-th/9610154.}
\refsend\ 
that the mirror of $\ca{M}_{E_8}$ is also an elliptic-$K3$ fibration. The
mirror, however, has $h_{11}=491$ which suggests, in virtue
of~\eqref{kclasses} that $G$ will have large rank. In fact this is so. This
is the manifold studied by Aspinwall and Gross~
\Ref\rAGun{P.~S.~Aspinwall and M.~Gross, unpublished.}\
who find that $G$ is a group of
rank 296~
$$G =  E_8^{17}{\times}F_4^{16}{\times}G_2^{32}{\times}SU(2)^{32}\ \hbox{ and }
\ n_T=193 .$$
{}From the toric perspective the group is large because the dual
polyhedron which was `small' for the cases considered in
\cite{\rCF-\rBC} has now many points. The purpose of this paper is to
present an algorithm that allows the group to be read off from the dual
polyhedron. Typically one obtains in this way groups of large rank
corresponding to the fact that~\eqref{duality} obtains for many manifolds
\ca{M}, and the typical \ca{M} has large $h_{11}$. The mirrors of the
manifolds discussed in
\cite{\rCF} are a case in point and provide many examples for
which large gauge groups arise. 

The organisation of this paper is as follows. In \SS{2} we review relevant
aspects of toric geometry, principally the construction of Refs.~       
\REFS\rMVI{D.~R.~Morrison and C.~Vafa, Nucl. Phys. {\bf B473} (1996) 74,
hep-th/9602114.}
\REFSCON\rMVII{D.~R.~Morrison and C.~Vafa, Nucl. Phys. {\bf B476} (1996) 437, 
hep-th/9603161.}
\REFSCON\rBer{M.~Bershadsky \etal, hep-th/9605200.}
\refsend\
of F-Theory duals of Heterotic vacua in six dimensions and the observation of
\cite{\rAS} that the mirror of an elliptic \cym\ is frequently also an elliptic
\cym. In \SS{3}, we describe the algorithm which allows us to read off
the vector and tensor multiplet content of the effective theory from the
toric data. In~\SS{4}, we illustrate this approach with some examples. The
procedure can also be applied to elliptic fourfolds, and we present the gauge
content for two of the fourfolds studied by Klemm \etal~
\Ref\rKL{A.~Klemm, B.~Lian, S.-S.~Roan and S.-T.~Yau, hep-th/9701023.}.
\SS{5}~discusses subtleties which arise in the toric picture.
\SS{6} summarizes our results. 
Tables giving the groups and tensor multiplet content of the models that we
study are given in the Appendix.
\newpage
\section{poly}{F-Theory Compactified on Elliptic \cy\ Threefolds}
In this section, we review $N=1$ vacua in six dimensions that result from
compactification of F-theory on elliptic \cy\ threefolds, largely
following~\cite{\rMVI,\rMVII}.

Recall that an elliptic \cy\ threefold can be described by the Weierstrass
equation
$$y^2=x^3+f(z,z^{\prime})x+g(z,z^{\prime}),
\eqlabel{Wei}$$
where $z$ and $z^{\prime }$ are affine coordinates on the base.
At the divisors on the base given by the zero loci of the discriminant,
$$ D =4f^3+27g^2,
\eqlabel{dis}$$ 
the torus degenerates. In many cases when this happens the effective four or
six dimensional effective theory develops a nonabelian gauge symmetry. The
singularities determine the gauge 
group and matter content of the F-theory compactification. The type of the
singularity, and hence the resulting gauge group, depends on the form of the
polynomials $f$ and $g$. A dictionary
relating the singularities of such elliptic fibrations and gauge
symmetry enhancement was given in \cite\rBer. 

One can arrive at a singular locus by adjusting the coefficients in the 
polynomials $f$ and $g$, or, in other words, by varying the complex structure
parameters of the threefold. It is then possible to resolve the singularities
by sequences of blow-ups, \ie by varying the \K\ class parameters. The smooth
\cym\ that
results still contains all the information about the enhanced gauge symmetry.
Moreover, in many relevant cases, it can be conveniently represented in terms
of the toric data as was shown in \cite{\rCF-\rBC}.

Suppose we are interested in F-theory compactifications dual to 
$E_8\times E_8$ heterotic models on a $K3$ with instanton numbers 
$(12+n, 12-n)$ in the two $E_8$'s and blowups thereof. Then the starting
point is the hypersurface in the toric variety defined by the data displayed
in Table~\tabref{scaling}~\cite{\rMVI}. Namely, start with homogeneous 
coordinates $s,t,u,v,x,y,w$, remove the loci $\{s=t=0\}$, $\{u=v=0\}$,
$\{x=y=w=0\}$, take the quotient by three scalings $(\l,\m,\n)$ with the
exponents shown in Table~\tabref{scaling} and restrict
to the solution set of (homogeneous version of) Eq.~\eqref{Wei}.

$$\vbox{\offinterlineskip\halign{
&\strut\vrule height 12pt depth 6pt #&\hfil\quad$#$\quad\vrule
&\hfil \quad$#$\quad&\hfil\qquad$#$\quad&\hfil\qquad$#$\quad
&\hfil\qquad$#$\quad&\hfil\qquad$#$\quad&\hfil\qquad$#$\quad
&\hfil\qquad$#$\quad\vrule&\quad$#$\quad\hfil\vrule\cr
\noalign{\hrule}
&&s&t&u&v&x&y&w&\hfil\hbox{degrees}\cr
\noalign{\hrule\vskip3pt\hrule}
&\l&1&1&\hidewidth{n}&0&\hidewidth{2n{+}4}&\hidewidth{3n{+}6}&0&6n+12\cr
&\m&0&0&1&1&4&6&0&12\cr
&\n&0&0&0&0&2&3&1&6\cr
\noalign{\hrule}
}}
$$
\nobreak\tablecaption{scaling}{The scaling weights of the elliptic
fibration over $\IF_{n}$.}
\bigskip
These data define a \cyt . F-theory compactified on this manifold is dual 
to the $E_8\times E_8$ heterotic model with instanton numbers $(12+n, 12-n)$
and maximally Higgsed gauge group. We can now construct the Newton polyhedron
describing this hypersurface.
Concretely, first we find all possible nonnegative powers of our homogeneous 
variables compatible with the constraints. We get in this way a number of
points in $\IR^{7}$. Since there is a 
$(\IC ^{\ast})^3$ action these points actually lie in a four-dimensional
subspace. Having chosen a basis we take the convex hull of the points and
obtain our Newton polyhedron $\D$. 

We can vary the complex structure parameters to introduce 
singularities into our \cyt\ and nonabelian enhanced gauge symmetry into the 
corresponding effective theory. For example, it was shown in \cite\rMVII\ that 
if we introduce a curve of singularities at $z=0$\Footnote{The base
of the elliptic fibration \IF$_n$ is a fiber bundle
over \IP $_1$ with fiber \IP $_1$. We denote, following \cite\rMVII, the 
affine
coordinate on the base by $z^{\prime}$, and the affine coordinate on the fiber
by $z$.},
then, on the heterotic side, this will have the effect of unhiggsing a certain
subgroup of the first $E_8$. Suppose now that we want to construct the 
corresponding polyhedron. To do it in the most direct way, it is easiest
to go back to our seven-dimensional points which simply give us various
powers of the homogeneous coordinates and use the results of \cite\rBer\
which relate the types of singularities to the degrees of vanishing 
of certain polynomials\Footnote{The authors of \cite\rBer\ rewrite the
Weierstrass equation in a more general form 
$$y^2+a_1xy+a_3y=x^3+a_2x^2+a_4x+a_6,$$
and the polynomials are just the $a_i = a_i(z,z')$.} on the base of the
elliptic fibration.
Thus, in toric language, the introduction of a curve of singularities at $z=0$
means simply eliminating a certain number of seven-dimensional points, which
in turn results in the disappearance of the corresponding four-dimensional
points from the Newton polyhedron $\D$. Since the Newton polyhedron is
diminished, the dual polyhedron $\nabla$ acquires
some additional points in the process. It is useful to emphasize here
that the 
manifolds corresponding to the polyhedra which result are {\sl smooth\/}
and correspond, upon compactification to
four dimensions on a $T^2$, to a theory in the Coulomb phase.

It was observed in \cite\rCF\ that it is the dual polyhedron which
exhibits a regular structure which makes possible, in particular, to 
determine the enhanced gauge symmetry given~$\nabla$. It was noticed there 
that 
in all the examples of heterotic/type II dual pairs the $K3$ and elliptic
fibration structure shows itself in the existence of three- and two-dimensional
reflexive subpolyhedra, respectively, inside the dual polyhedron of the \cym .
Moreover, the three-dimensional reflexive subpolyhedron which was conjectured
to represent the generic $K3$ fiber was shown to contain the information
about the part of the total gauge group (the only part in the examples 
considered in \cite{\rCF}) which has perturbative interpretation on the 
heterotic side.  
\newpage
\section{algo}{Identifying the Groups}
\subsection {Generalities}
In this section, we describe the algorithm that allows us to read off the gauge
content from the toric data. First we introduce a generalization of the
conjecture of Ref.~\cite{\rCF}, using the results of~\cite{\rAS} to state this
for \cys\
that are described by reflexive polyhedra, the integral points of the polyhedra
being points in a lattice $\L$.
It has been shown there that in order for a \cy\ $n$-fold to be
a fibration with generic fiber a \cy\ $(n-k)$-fold it is necessary and 
sufficient that\Footnote{We denote, as is standard, the lattice dual to $\L$
(where $\D$ lives) by $V$, and its real extension by $V_{\IR}$.} 
\item\ {(i) There is a projection operator $\P$: $\L\rightarrow \L_{n-k}$,
where 
$\L_{n-k}$ is an $n-k$ dimensional sublattice, such that $\P(\D)$ is a
reflexive polyhedron in $\L_{n-k}$, or}
\item\ {(ii) There is a lattice plane in $V_{\IR}$ through the origin 
whose intersection with $\nabla$ is an $n-k$ dimensional reflexive polyhedron,
{\it i.e.\/} it is a slice of the polyhedron.}

\noindent (i) and (ii) are equivalent conditions. In case (i) the
polyhedron of the fiber appears
as a {\sl projection\/} while in case (ii) it appears as an
{\sl injection\/}, the projection and the injection being related by mirror
symmetry (see Figure~\figref{fibrations}).
In particular, if the polyhedron of the $(n-k)$-dimensional \cym\ exists as
both a projection and an injection (\eg the image of the projection coincides
with a slice of the $\D$ by a plane as sketched in the second row of
Figure~\figref{fibrations}),
then the intersection in $\nabla$
is also a certain projection implying that the mirror manifold is a fibration
with an $n-k$ dimensional \cym\ as the typical fiber.
If (i) or (ii) hold there is also a way to to see the base of the fibration
torically~\REFS\rSp{M.~Kreuzer and H.~Skarke, hep-th/9701175.}\refsend
.
The hyperplane $H$ generates a $n-k$ dimensional sublattice of $V$. Denote this
lattice $V_{\rm fiber}$. Then the quotient lattice 
$V_{\rm base}=V/V_{\rm fiber}$ is the lattice in which the fan of the base 
lives. The fan itself can be constructed as follows. Let $\P_B$ be a projection
operator acting in $V$, of rank dim$(V)-2$, such that it projects $H$ onto
a point. Then
$\P_B(V)=V_{\rm base}$. When $\P_B$ acts on $\nabla$ the result is a $k$ 
dimensional
set of points in $V_{\rm base}$ which gives us the fan of the base if we draw
rays through each point in the set.

Recall now that the Hodge numbers for three-dimensional \cy\ hypersurfaces
are given by
$$\eqalign{&h_{21}={\rm pts}(\D) \ -\hskip-10pt
\sum_{{\rm codim}(\theta)=1}\hskip-10pt{\rm int}(\theta) \ +\hskip-10pt
\sum_{{\rm codim}(\theta)=2}\hskip-10pt{\rm int}(\theta)
{\rm int}(\tilde{\theta}) \ - \ 5,\cropen{8pt}
&h_{11}={\rm pts}(\nabla) \ -
\hskip-10pt\sum_{{\rm codim}(\tilde{\theta})=1}\hskip-10pt{\rm int}
(\tilde{\theta}) \ +\hskip-10pt
\sum_{{\rm codim}(\tilde{\theta})=2}\hskip-10pt
{\rm int}(\tilde{\theta}){\rm int}(\theta) \ - \ 5}
\eqlabel{h11}$$
where pts$(\D)$ denotes the number of integral points of $\D$, int$(\theta)$
stands for the number of integral points interior to a face $\theta$ and 
similar quantities pts($\nabla$) and int$(\tilde{\theta})$ are defined for
$\nabla$.
Equation \eqref{h11} expresses the number of deformations
of complex structure and \K\ classes in terms of the number of
points of the polyhedra. The terms in these expressions that involve
codimension-1 faces account, in the case of $h_{21}$, for the freedom to make
redefinitions of the homogeneous variables, and in the case of $h_{11}$, for
the singularities of the toric variety which do not intersect the
hypersurface. We will call these points `irrelevant'. The third terms in both
equations are `correction' terms, the numbers of deformations of the
corresponding hypersurface which are not visible torically. (Note that in
many cases it turns out to be possible to add a certain number of
points to the polyhedron under consideration so that the correction vanishes.)
\midinsert
\def\fibrations{
\vbox{\vskip10pt\hskip0.7in\hbox{\epsfxsize=5truein\epsfbox{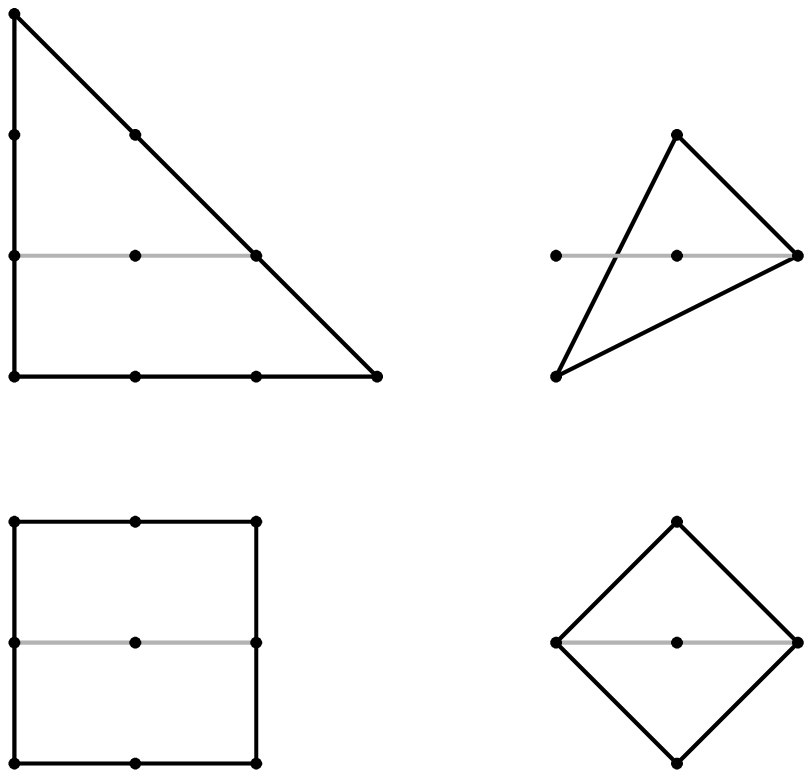}}}}
\figbox{\fibrations\vskip0pt}{\figlabel{fibrations}}{Two examples of
fibrations visible as projections or slices of the polyhedra. $\D$ and $\nabla$
are a dual pair of reflexive polyhedra corresponding to 1-dimensional \cys.
$\D$ encodes a zero-dimensional \cym\ $(=\IZ_{2})$ 
as a slice (but not a projection).
The dual polyhedron $\nabla$ encodes the $\IZ_{2}$ as a projection
(but not an injection). According to the criterion of Ref.~\cite{\rAS}, 
$\nabla$ is the Newton polyhedron of a $\IZ_{2}$ fibration.
$\D^{\prime}$ encodes the $\IZ_{2}$ fiber as both an injection as well as a
projection, hence the mirror $\nabla^{\prime}$ does too, so that both the
manifold and its mirror are $\IZ_{2}$ fibrations.} 
\place{1}{5.5}{$\D:$}
\place{3.7}{5.5}{$\nabla:$}
\place{1}{3.5}{$\D^{\prime}:$}
\place{3.7}{3.5}{$\nabla^{\prime}:$}
\endinsert
It is worth emphasizing that the statement about the possibility
of seeing the base of the \cy\ fibration torically holds provided the 
correction term in \eqref{h11} vanishes. Otherwise we may miss some of the
one-dimensional cones in the fan describing the base as a toric variety.

Suppose now that we are given an elliptic \cyt. The theorem of~\cite\rAS\
tells us that in this case it is possible to find a two-dimensional hyperplane 
$H$ in $V_{\IR}$ through the origin such that its intersection with 
$\nabla$ is a two-dimensional reflexive polyhedron representing the typical 
fiber. Let us denote it by $\nabla^{\ca{E}} =\nabla\cap H$. 
Projecting $\nabla$ with $\P_B$ such that $\P_B(\nabla^{\ca{E}})=(0,0)$ yields
a set of 
points living in a two-dimensional lattice which is what we call 
$V_{\rm base}$. We will denote this set of points $\S_B$. 
Drawing a ray through from the origin $(0,0)$ through every 
other point gives us the fan of the base. Note that a ray may pass through
more than one point and hence the number of rays, or one-dimensional cones,
is generically less than the number of non-zero points in 
$V_{\rm base}$. 
In many examples of \cite{\rCF-\rBC},
$\S_B$ coincides with a
two-dimensional face of the dual polyhedron $\nabla$ 
orthogonal to the hyperplane $H$.

The methods of toric geometry allow us to read off some topological invariants
of $B$. In particular it is known 
\ \REFS\rFul{W. Fulton, Introduction to Toric Varieties, Princeton University
Press, 1993.}
\refsend\
that for a nonsingular $n$-dimensional toric variety the Betti numbers are 
given by
$$b_{2k}=\sum_{i=k}^{n}(-1)^{i-k}\left( \matrix{i \cr k} \right)d_{n-i},
\eqlabel{betti}$$  
where $d_k$ is the number of $k$-dimensional cones in the fan.
In our case \eqref{betti} yields \hbox{$b_2=d_1-2d_0$} or, since
$h_{20}(B)=0$ 
for a toric variety, $$h_{11}(B)=d_1-2.\eqlabel{h11b} $$

Compactifying Type IIA strings on our \cym\ yields a four-dimensional $N=2$
vacuum in its Coulomb phase. This statement is intrinsically six-dimensional,
meaning that the compactification of F-theory on (a blown-down version of) the
same \cym\ yields an $N=1$ vacuum in 6D which contains essentially the same 
information. We want to determine the spectrum of this six-dimensional theory.
The first observation is that the number of massless tensor multiplets is
already determined by \eqref{h11b} since it is simply \cite\rMVII
$$n_T=h_{11}(B)-1.\eqlabel{nt}$$ 
Our next task is to find the massless vector multiplet content. Comparison
of~\eqref{kclasses} with~\eqref{h11} shows
us that essentially each `relevant' point in $\nabla$ corresponds either 
to a massless tensor multiplet or to a massless vector in the Cartan
subalgebra of $G$
provided the correction vanishes.
In view of \eqref{h11b} it is reasonable to conjecture that the points 
corresponding to tensor multiplets are those in $\S_B$, moreover, 
since each 
one-dimensional cone in that projection may contain more than one point, 
we will claim that the `tensor multiplet' points are those closest to the 
origin of $\S_B$ (apart from three of them since 
$n_T=d_1-3$).
Thus, in order to find the vector multiplet content, we should try to sort 
out the rest of the `relevant' points in $\nabla$. It has been conjectured 
and illustrated by many examples in \cite\rCF\ that in cases dual to 
$E_8\times E_8$ heterotic on a $K3$  the points corresponding to vectors
in the Cartan subalgebra of $G$ had the following properties.
\item\ {1. Under the projection $\P_B$: $\nabla\rightarrow \S_B$  the 
points corresponding to the subgroup of the first $E_8$ project onto points
in $\S_B$ of the form $(0,-b)$ (in a certain basis), $b>0$ and the 
points corresponding
to the unbroken subgroup of the second $E_8$ project onto points of which can
be written as $(0,c)$, $c>0$.}
\item\ {2. Under the projection ${\P_{\ca{E}}}$: $\nabla\rightarrow H$ 
these two sets
of `relevant' points projected onto certain points of $\nabla^{\ca{E}}$, and
the information about the precise nature of the unbroken group (or, in 
geometric terms, about the type of singularity along the corresponding divisor
of $B$) was contained in the numbers
$n_{i}=\vert{\P_{\ca{E}}}^{-1}(pt_i)\vert$, 
where $pt_i\in\nabla^{\ca{E}}$ (see Fig.~\figref{torus}) and, as usual,
$\vert A\vert$ denotes the number of elements in set $A$.}
\midinsert
\def\torus{
\vbox{\vskip10pt\hbox{\epsfxsize=2.0truein\epsfbox{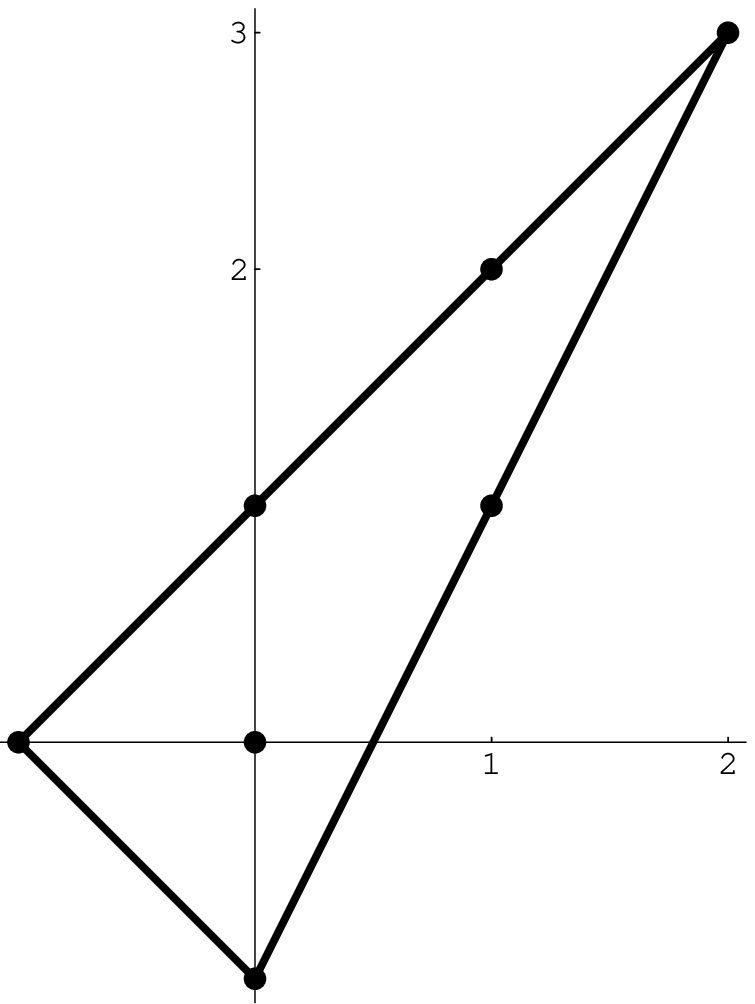}}}}
\figbox{\torus\vskip0pt}{\figlabel{torus}}
{The polyhedron, $^2\nabla$, of $\cp2^{(1,2,3)}[6]$. The
$pt_r^{\prime}$ are the points directly below the indicated
points of the plot, with the number of primes denoting the depth.}
\place{4.05}{4.0}{$pt_1$}
\place{3.4}{3.4}{$pt_2$}
\place{3.7}{2.65}{$pt_3$}
\place{2.7}{2.7}{$pt_4$}
\place{2.75}{1.8}{$pt_5$}
\place{2.05}{2.05}{$pt_6$}
\place{3.1}{1.4}{$pt_7$}
\endinsert
\subsection {The Algorithm}
We are in a position now to propose our algorithm. Each
one-dimensional cone
in the fan of a toric variety corresponds to a divisor \cite\rFul. In our
case the complex dimension of the base is two. So the divisors are complex
curves. In the examples of \cite\rCF\ the one-dimensional cones described
above (those containing points of the form $(0,-c)$ and $(0,b)$)
corresponded to the divisors described by the equations $z=0$ and
$z=\infty$, respectively. Let us call these cones (rays) $R_1$ and $R_2$. 
The points describing the unbroken subgroups of the first and second
$E_8$ were those of the form $\P_B^{-1}(R_1)$ and $\P_B^{-1}(R_2)$. In
geometrical
language, they are the points describing the types of singularities along
the two divisors, the corresponding set $\P_{\ca{E}}^{-1}$ encoding the type
of the
singularity. In general, the base may have many more curves of singularities 
than those two. Torically they will show themselves as one-dimensional cones.
There is actually a singularity along a given divisor represented by the ray
$R_k$ provided the set $\P_B^{-1}(R_k)$ contains more than one point. To
determine the actual type of singularity we have to act in exactly the same
way as we would have acted if the divisor in question had been given by
$z=0$. Namely we take $\P_B^{-1}(R_k)$ and project it onto $\nabla^{\ca{E}}$.
The 
singularity type is then encoded in the corresponding numbers 
$$n_{ik}=\vert {\P_{\ca{E}}}^{-1}(pt_i)\cap\P_B^{-1}(R_k)\vert
\eqlabel{nik}$$
in exactly the same way as it would have been encoded had the singularity
occurred along $\{z=0\}$\Footnote{In all previous examples
$pt_i\in\nabla^{\ca{E}}$. But
in general there are cases in which we should regard some of $pt_i$'s as 
belonging to $H$, not necessarily to $\nabla^{\ca{E}}$.}.
The dictionary between the numbers $\{n_i\}$ and
singularity types was described in \cite\rCF\ (see Table~\tabref{bottoms}) and
carries over to this more general case.
\pageinsert
$${
\def\skip{\hskip4pt}
\vbox{\offinterlineskip\halign{
\strut # height 12pt depth 5pt&\quad $#$ \skip \hfil\vrule
&\hskip10pt  $#$ \skip \hfil\vrule\cr
\noalign{\hrule}
\vrule&\hfil H&\hfil\hbox{Bottom}\cr
\noalign{\hrule\vskip3pt\hrule}
\vrule&SU(1) &\{ pt_1' \}\cr
\vrule&SU(2) &\{ pt_1', pt_2' \}\cr
\vrule&SU(3) &\{ pt_1', pt_2', pt_3' \}\cr
\vrule&G_2   &\{ pt_1'', pt_2', pt_3' \}\cr
\vrule&SO(5) &\{ pt_1', pt_2', pt_4' \}\cr
\vrule&SU(4) &\{ pt_1', pt_2', pt_3', pt_4' \}\cr
\vrule&SO(7) &\{ pt_1'', pt_2', pt_3', pt_4' \}\cr
\vrule&Sp_3 &\{ pt_1', pt_2', pt_4', pt_6' \}\cr
\vrule&SU(5) &\{ pt_1', pt_2', pt_3', pt_4', pt_5' \}\cr
\vrule&SO(9) &\{ pt_1'', pt_2'', pt_3', pt_4' \}\cr
\vrule&F_4   &\{ pt_1''', pt_2'', pt_3', pt_4' \}\cr
\vrule&SU(6)&\{ pt_1', pt_2', pt_3', pt_4', pt_5', pt_6' \}\cr
\vrule&SU(6)_b&\{ pt_1', pt_2', pt_3', pt_4', pt_5', pt_7' \}\cr
\vrule&SO(10)&\{ pt_1'', pt_2'', pt_3', pt_4', pt_5' \}\cr
\vrule&SO(11)&\{ pt_1'', pt_2'', pt_3', pt_4'', pt_5' \}\cr
\vrule&SO(12) &\{ pt_1'', pt_2'', pt_3', pt_4'', pt_5', pt_6'  \}\cr
\vrule&E_6   &\{ pt_1''', pt_2'', pt_3'', pt_4', pt_5' \}\cr
\vrule&E_7   & \{ pt_1'''', pt_2''', pt_3'', pt_4'', pt_5' \}\cr
\noalign{\hrule\vskip3pt\hrule}
\vrule&SU(6)_c&\{ pt_1', pt_2', pt_3', pt_4', pt_5', pt_6', pt_7'\}\cr
\vrule&SO(13) &\{ pt_1'', pt_2'', pt_3', pt_4'', pt_5', pt_6'' \}\cr
\vrule&E_{6B} &\{ pt_1''', pt_2'', pt_3'', pt_4', pt_5', pt_7' \}\cr
\vrule&E_{7B}&\{ pt_1'''', pt_2''', pt_3'', pt_4'', pt_5', pt_6' \}\cr
\vrule&E_8 &\{ pt_1^{(6)}, pt_2^{(4)}, pt_3''', pt_4'', pt_5'\}\cr
\noalign{\hrule}
}}
\hskip5pt
\def\skip{\hskip4pt}
\vbox{\offinterlineskip\halign{
\strut # height 12pt depth 5pt&\quad $#$ \skip \hfil\vrule
&\hskip10pt  $#$ \skip \hfil\vrule\cr
\noalign{\hrule}
\vrule&\hfil H&\hfil\hbox{Bottom}\cr
\noalign{\hrule\vskip3pt\hrule}
\vrule&SU(2)_b &\{ pt_2' \}\cr
\vrule&SU(2)_c &\{ pt_4' \}\cr
\vrule&SU(2)_d &\{ pt_6' \}\cr
\vrule&SU(2){\times}SU(2) &\{ pt_2', pt_4' \}\cr
\vrule&(SU(2){\times}SU(2))_b &\{ pt_4', pt_6' \}\cr
\vrule&SU(3){\times}SU(2) &\{ pt_2', pt_4', pt_5' \}\cr
\vrule&(SU(3){\times}SU(2))_b &\{ pt_3', pt_5' \}\cr
\vrule&(SU(3){\times}SU(2))_c &\{ pt_5' \}\cr
\vrule&SO(5){\times}SU(2) &\{ pt_2', pt_4', pt_6' \}\cr
\vrule&G_2{\times}SU(2) &\{ pt_2', pt_4'', pt_5' \}\cr
\vrule&SU(4){\times}SU(2) &\{ pt_2', pt_4', pt_5', pt_6' \}\cr
\vrule&SO(7){\times}SU(2) &\{ pt_2', pt_4'', pt_5', pt_6' \}\cr
\vrule&SO(9){\times}SU(2) &\{ pt_2', pt_4'', pt_5', pt_6'' \}\cr
\vrule&&\cr
\vrule&SU(3)_b &\{ pt_3' \}\cr
\vrule&SU(3)_c &\{ pt_7' \}\cr
\vrule&{SU(3)\times}SU(3) &\{ pt_3', pt_5', pt_7' \}\cr
\vrule&&\cr
\noalign{\hrule\vskip3pt\hrule}
\vrule&G_2{\times}SU(3)   &\{ pt_3', pt_5', pt_7'' \}\cr
\vrule&F_4{\times}SU(2) &\{ pt_2', pt_4'', pt_5', pt_6''' \}\cr
\vrule&&\cr
\vrule&&\cr
\vrule&&\cr
\noalign{\hrule}
}}
}$$
\tablecaption{bottoms}{\baselineskip=13pt The table on the left gives the
bottoms containing $pt'_1$ formed
by adding the points $pt_r^{(j)}$ to $\nab2$.
In each case the points of $\nab2$ are understood and the points that are
written are the lowest members of columns. Thus $pt'''_2$ for example implies
the
presence of $pt''_2$ and $pt'_2$. The bottoms that appear in the lower block
correspond to nonperturbatively realised groups.
The table on the right gives the bottoms that do not contain $pt'_1$. The
bottoms that are given in the lower
block again correspond to groups that are realised nonperturbatively.}
\endinsert
\subsection {Subtleties}
There are subtleties in the picture described above which are worth
mentioning. As was pointed out in
\ \REFS\rAG{P. S. Aspinwall and M. Gross, Phys. Lett. {\bf B387} (1996) 735.}
\refsend\
and \cite\rBer, the gauge symmetry appearing in uncompactified dimensions
turns out to be a subgroup of the singularity type $ADE$ series group.
These cases are due to the monodromy action on vanishing cycles if the 
monodromy happens to be an outer automorphism as opposed to a Weyl group 
element. These two cases were called `non-split' and `split', respectively, in
\cite\rBer. 
Also different kinds of singularities may collide as discussed for example
in~
\REFS\rAsp{P.~S.~Aspinwall, hep-th/9612108.}
\REFSCON\rBJ{M. Bershadsky and A. Johansen, hep-th/9610111.}
\refsend
, and when they do a new phase of the theory appears characterized by new
tensor multiplets and enhanced gauge symmetry. We believe that these subtle
points are already taken care of by the dual polyhedron and the points
we observe provide us with information about the spectrum of the corresponding
six-dimensional theory (and not quite, strictly speaking, about the types
of singularities along divisors).

There are also certain additional subtleties that arise in the toric picture.
This is because toric geometry
does not always encode all the deformations as points in the polyhedra. There
are often non-toric deformations coming from points interior to codimension-2
faces and points in the dual 1-faces (codimension-3 in the dual polyhedron).
Thus some of the deformations are not explicitly seen as points in the
polyhedron, but hidden away in the dual polyhedron. Thus our identification of
the gauge theory is incomplete until we can interpret all these non-toric data.
Furthermore, we also encounter situations where the polyhedron contains points
which are interior to codimension-2 faces (and hence relevant), but which are
interior to codimension-1 faces of the $K3$ fibers. For example, the point
$pt_{5}^{\prime}$ in the $E_8$ top becomes relevant in some polyhedra that
contain this top. This situation is tricky because it does not happen all the
time --- there are polyhedra containing the $E_8$ top where $pt_{5}^{\prime}$
is irrelevant, and there are others containing the $E_8$ top in which
$pt_{5}^{\prime}$ is relevant. There are even polyhedra where both kinds of
$E_8$ tops occur. Thus our algorithm is incomplete until we specify how to
handle such cases. While we do not have a general theorem that achieves this,
we have been able to study many such cases and have found consistent patterns
that allow us to treat this and many similar situations which arise in these
polyhedra. The prescription for handling such situations is best described by
giving examples, which we postpone until \SS{5}.      
\newpage
\section{mir}{Examples}
We turn now to the application of our algorithm to some
examples. Our first example is the well-known case of `simple' (in the
terminology of \cite\rAsp) point-like instantons in the Spin(32)/\IZ$_2$
heterotic string compactified on a $K3$.
\subsection {A simple example in detail}
As is well known, the $SO(32)$ Heterotic theory compactified on a $K3$
requires 24 instantons to cancel the anomaly. Consider the situation when two
of the instantons shrink to zero at the same point.
As was shown in~
\Ref\rWsi{E. Witten, Nucl. Phys. {\bf B460} (1996) 541, hep-th/9511030.},
the effective theory develops a nonperturbative $Sp(2)$ gauge symmetry. In
addition, since there are only 22 finite size instantons left, an $SO(10)$
subgroup of the primordial $SO(32)$ (or, more precisely, Spin(32)/\IZ$_2$)
that was previously broken to $SO(8)$ is now restored. One can easily
calculate that there are 231 neutral hypermultiplets and a single
massless tensor multiplet in the six-dimensional spectrum. If we
compactify
further to four dimensions on a torus and go to the Coulomb phase of the   
resulting $N=2$ theory, we obtain rank($Sp(2)$)+rank($SO(10)$)+3$=$10 massless
vector multiplets and 231 massless hypermultiplets. So, if we find the Type IIA
dual then the corresponding \cym\ will have $h_{11}=10$ and $h_{21}=230$. If we
can represent this manifold as a hypersurface in a toric variety, then we
should be able to read the $Sp(2)\times SO(10)$ gauge group off the dual
polyhedron. 

There is indeed a very simple procedure for generating the \cym\ in question
torically. As was conjectured in \cite\rMVII, the $SO(32)$ heterotic string
on a $K3$ is T-dual to the $E_{8}\times E_{8}$ heterotic string on another 
$K3$ with instanton numbers $(16,8)$ which in turn can be described by 
F-theory compactified on an \cy\ threefold elliptically fibered over \IF $_4$.
So, what we have to do is to take the \cym\ defined by Table~\tabref{scaling}
for $n=4$ and introduce an $Sp(2)$ singularity by the methods described
in~\SS{2}. The resulting dual polyhedron then consists of the points displayed
in Table~\tabref{2si}.
\midinsert
$$\vbox{\offinterlineskip\halign{\strut # height 9pt depth 4pt
&\hfil\quad \eightrm #\quad\hfil\vrule
&\hfil\quad \eightrm #\quad\hfil\vrule
&\hfil\quad \eightrm #\quad\hfil\vrule 
&\hfil\quad \eightrm #\quad\hfil\vrule
&\hfil\quad \eightrm #\quad\hfil\vrule \cr
\noalign{\hrule}
\omit{\vrule height3pt}&&&&&\cr
\vrule&$\P_B^{-1}(0,0)$ 
       &$\P_B^{-1}(R_1)$
        &$\P_B^{-1}(R_2)$ 
        &$\P_B^{-1}(R_3)$
        &$\P_B^{-1}(R_4)$\cr
\omit{\vrule height3pt}&&&&&\cr
\noalign{\hrule\vskip3pt\hrule}
\omit{\vrule height2pt}&&&&&\cr
\vrule&(0,\- 0,\- 2,\- 3)&(1,\- 4,\- 2,\- 3)&(0,\- 2,\- 2,\- 3)&(-1,\- 0,\-
2,\- 3)&(0, -1,\- 2,\- 3)\cr
\vrule&(0,\- 0,\- 1,\- 2)&(1,\- 4,\- 1,\- 2)&(0,\- 1,\- 2,\- 3)&&\cr
\vrule&(0,\- 0,\- 0,\- 1)&(1,\- 4,\- 0,\- 1)&(0,\- 2,\- 1,\- 2)&&\cr
\vrule&(0,\- 0,\- 1,\- 1)&&(0,\- 1,\- 1,\- 2)&&\cr
\vrule&(0,\- 0,\- 0,\- 0)&&(0,\- 1,\- 1,\- 1)&&\cr
\vrule&(0,\- 0,\- 0, -1)&&(0,\- 1,\- 0,\- 1)&&\cr
\vrule&(0,\- 0, -1,\- 0)&&(0,\- 1,\- 0,\- 0)&&\cr
\omit{\vrule height2pt}&&&&&\cr
\noalign{\hrule}
}}
$$
\nobreak\tablecaption{2si}{Points of the dual polyhedron describing the 
$SO(32)$ heterotic vacuum with two point-like instantons, sorted according to
how they project onto the fan of the base of the fibration.}
\bigskip
\endinsert

Points in the first column of the Table lie in the slice of the 
polyhedron by the plane $x_1=0$, $x_2=0$, and this set of
points is a copy of the torus of Figure~\figref{torus}. Notice also that if 
we project
$\nabla$ onto this plane we obtain the same points. The slice itself forms
a two-dimensional reflexive polyhedron describing a torus. We learn from our
rules that the threefold is
an elliptic fibration which in this case is true by construction.
Next, project onto the first two coordinates. This is our projection $\P_B$.
As we see from the Table, we obtain six points
$$\P_B(\nabla)=\{ (0,0),~(-1,0),~(0,-1),~(0,1),~(0,2),~(1,4)\},$$ 
that form the fan of \IF$_4$. We conclude that 
\IF$_4$ is the base of the elliptic fibration, which we knew to be true by
consruction in this case. Using \eqref{h11b} we find that $h_{11}=2$ 
for this base, and from \eqref{nt} that there is indeed only one massless
tensor multiplet present in the six-dimensional spectrum. 

Let us now turn to the enhanced gauge symmetry content of $\nabla$.
As was claimed in~\SS{2}, above each one-dimensional cone/ray
of the fan of the base there is a simple factor of the total gauge
group. Some of these may be the trivial group, which, for want of better 
notation, we denote by $SU(1)$. More precisely, the points $\P_B^{-1}(R)$, 
where $R$ is one of the rays of Figure~\figref{fan},
give us a simple factor in the enhanced gauge symmetry of the effective 
theory. In this case there are four such rays. 
\midinsert
\def\fan{
\vbox{\vskip10pt\hbox{\epsfxsize=1.5truein\epsfbox{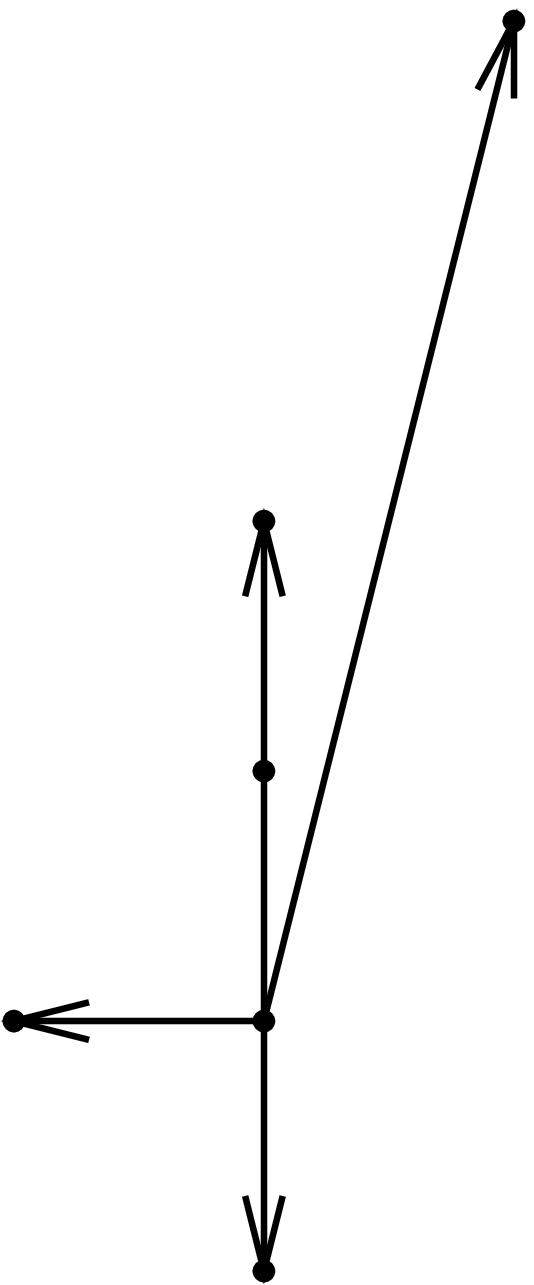}}}}
\figbox{\fan\vskip0pt}{\figlabel{fan}}
{The fan of \IF$_4$.}
\place{3.2}{0.7}{(0,-1)}
\place{3.4}{1.1}{$R_4$}
\place{3.4}{1.6}{(0,0)}
\place{3.75}{3.1}{$R_1$}
\place{4.1}{4.5}{(1,4)}
\place{3}{2.4}{$R_2$}
\place{3.2}{3.2}{(0,2)}
\place{2.8}{1.4}{$R_3$}
\place{2.1}{1.6}{(-1,0)}
\endinsert

The points of $\P_B^{-1}(R)$, for some ray $R$, are of the form $(ka,kb,c,d)$
where $(c,d)$ is a point of $\nabla^{\ca{E}}$, $(a,b)$ is the first integral
point along $R$ and $k$ is a positive integer. The point $(c,d)$ is one of
the points of $\nabla^{\ca{E}}$ that are denoted $pt_i$ and we associate
with $(ka,kb,c,d)$ the point $pt_i^{(k)}$.
Consider first $\P_B^{-1}(R_2)$. We see from Table~\tabref{2si} that
$$\P_B^{-1}(R_2)=\{pt_1'',pt_1',pt_2'',pt_2',pt_3',pt_4',pt_5'\}\simeq
SO(10),$$
the second equivalence following from a comparison with Table~\tabref{bottoms}.
For the ray $R_1$ we have
$$\P_B^{-1}(R_1)=\{pt_1',pt_2',pt_4'\}\simeq Sp(2)$$
while for $\P_B^{-1}(R_3)$ and $\P_B^{-1}(R_4)$ we find $\{pt_1'\}$
corresponding
to the trivial group $SU(1)$. Thus, combining the groups, we find the gauge
group $SO(10)\times Sp(2)$, in precise agreement with our expectations. 

Now we can study the gauge content of F-Theory
compactified on the mirrors of the \cys\ of Ref.~\cite{\rCF}. We are going to
make use of the fact that these manifolds are elliptic fibrations, moreover,
the polyhedron describing the generic fiber is visible in the dual polyhedron
as an injection. This property is shared by the direct polyhedron, hence
the mirror \cym\ is also an elliptic fibration. Therefore,
we can take the corresponding direct polyhedra regarding them
as dual polyhedra of the mirrors, and read off the tensor and vector multiplet 
spectra. Since the $h_{21}$'s of the original \cym\ are rather large, so are
$h_{11}$'s of their mirrors, suggesting that the enhanced gauge symmetry in
four dimensions we are about to uncover is bigger than what we are accustomed
to. In particular, if heterotic duals of these models exist, most of the gauge
symmetry in question is bound to have a nonperturbative origin. We describe a
few examples below. The complete results for this class of models are listed
in the~Appendix.
\subsection {The mirror of the manifold with Hodge numbers (3,243)}
The \cyt\ defined by the data in Table~\tabref{scaling} for $n=0$ has
$(h_{11},h_{21})=(3,243)$ and provides a dual to the compactification
of heterotic strings with instanton numbers $(12,12)$ in the two $E_8$'s. There
is no enhanced gauge symmetry in six dimensions. The mirror \cym\ has
$(h_{11},h_{21})=(243,3)$ of course which tells us that
${\rm rank}(G) + n_T =241$. Using our methods we find that $\S_B$
has 96 one-dimensional cones which seems to suggest that we have 93 massless
tensors in the spectrum. Then when we begin to ``sort out'' the points 
corresponding to vector multiplets we find that there are 8 $E_8$ factors but 
in 4 of them the point which projects on the $pt_5$ in Fig.~\figref{torus}
becomes `relevant' as opposed to the usual $E_8$ `top' described in \cite\rCF\
where it was interior to a facet. This is the case which is encountered
in almost all in the examples considered in this section. We will interpret
the extra `relevant' points as additional tensor multiplets (see~\SS{5}). So
the six-dimensional theory corresponding to the mirror has gauge group
$$G = E_8^8{\times}F_4^8{\times}G_2^{16}{\times}SU(2)^{16} \ \hbox{ and } \
n_T = 97.$$

Now unhiggs an $SU(2)$ subgroup of the first $E_8$ on the
heterotic side. The corresponding \cym\ has $(h_{11},h_{21})=(4,214)$ and the 
six-dimensional theory being a perturbative heterotic vacuum still has just
one massless tensor. The  $h_{11}$ of the dual \cym\ is now 214 suggesting that
${\rm rank}(G) + n_T =212$. We have 82 one-dimensional cones in 
$\S_B$ and two sets of $E_8$ points with one extra `relevant' point.
That brings the number of tensors to $82-3+2=81$ and in addition there is 
a gauge group of rank 131:
$E_8^5{\times}E_7^3{\times}F_4^6{\times}G_2^{12}{\times}SO(7)^2{\times}
SU(2)^{16}$.
\subsection {The mirror of the manifold with Hodge numbers (11,491)}
Consider the model dual to the heterotic vacuum obtained by 
compactification with instanton numbers $(24,0)$. There is an unbroken $E_8$
and the Hodge numbers are $(h_{11},h_{21})=(11,491)$. This manifold
is interesting because it has the
maximal $h_{21}$ encountered in those examples and, moreover, the maximal 
$h_{21}$ found in~
\REFS\rKS{A. Klemm and R. Schimmrigk, Nucl. Phys. {\bf B411} (1994) 559,
hep-th/9204060. \\
M. Kreuzer and H. Skarke, Nucl. Phys. {\bf B388} (1992) 113, hep-th/9205004.}
\REFSCON\rS{H. Skarke, alg-geom/9603007.}
\refsend\
for hypersurfaces in toric varieties. 
The mirror thus yields the biggest $h_{11}$ suggesting that this may be the
case
in which the maximal possible gauge symmetry is found in type II (and, perhaps,
heterotic) compactifications to six or four dimensions, and was studied
in~\cite{\rAGun}. On applying the algorithm we find the gauge
group of rank 296 mentioned in the introduction
$$G_{296} =  E_8^{17}{\times}F_4^{16}{\times}G_2^{32}{\times}SU(2)^{32} \ 
\hbox{ and } \ n_T = 193.$$
\subsection {The self-mirror manifold with Hodge numbers (251,251)}
The previous example is related in an interesting way to the
self-mirror manifold with Hodge numbers $(251,251)$. This is the largest
value of the Hodge numbers among the self-mirror examples listed
in~\cite{\rKS,\rS} (see also the interesting comments on this space and its
relation to the previous one in~\cite{\rKL}).
Furthermore, the sum $h_{11} + h_{21}$ is the
same as for
the previous example. We find a gauge group of rank 152 which is
$$G_{152} = E_8^{9}{\times}F_4^{8}{\times}G_2^{16}{\times}SU(2)^{16} \
\hbox{and} \ n_T = 97.$$
Note, as a curiosity, that this gauge group is the `square root' of the
product of the gauge groups in the preceding example,
\ie $G_{152}\times G_{152} = E_8\times G_{296}$, and the numbers of tensors
are also related : $2\times 97 = 1 + 193$.

The algorithm for finding the gauge content applies equally well to elliptic
fourfolds, the only difference from the threefold case being that the base is
now a three dimensional toric variety. Since F-theory compactified on a
fourfold yeilds a four-dimensional theory, there are no tensor multiplets, and
the rank of the gauge group is simply~
\REFS\rMoh{K.~Mohri, hep-th/9701147.}
\REFSCON\rCL{G.~Curio and D.~L\"{u}st, hep-th/9703007.}
\refsend\
$$\hbox{rank}(G) = h_{11} - h_{11}(B) - 1 + h_{21}(B)~.$$
For the examples we study, the base is a toric variety, so that $h_{21}(B)=0$.
Also, $h_{11}(B)$ is related to the number of rays in the fan of the base in a
simple way. The last two examples that we give here are the fourfold analogues
of the
last two examples~\cite{\rKL}. The relevant Hodge numbers are now $h_{11}$ and
$h_{31}$
and these two fourfolds are distinguished among the Fermat hypersurfaces in
weighted projective spaces by having $h_{11}+h_{31}$ as large as possible.
In one case $h_{31}-h_{11}$ is also a maximum while in the other the
manifold is self-mirror (and hence has $h_{11}=h_{31}$). We include these
examples here to show that for fourfolds the groups can become very large
indeed. Also, it was pointed out in~
\Ref\rBJPS{M.~Bershadsky, A.~Johansen, T.~Pantev and V.~Sadov,
hep-th/9701165.}\ that there is an additional contribution to the total
gauge group coming from three-branes which are necessary for anomaly
cancellation. Generically, this additional factor is
$U(1)^{\vert\chi\vert/24}$, provided there are no instantons in the
seven-branes. 
\subsection{The elliptic fourfold $\IP_5^{(1,1,84,516,1204,1806)}[3612]$}
This manifold has $h_{11}=252$ and $h_{31}=303148$. If we apply the
algorithm to $\D$ we obtain the group $G_{152}$ of the previous example.
Applying the algorithm to $\nabla$, we obtain a group of rank 121328
$$
G_{121328}~=~E_8^{2561}\times F_4^{7576}\times G_2^{20168}\times
SU(2)^{30200}~.$$
\subsection{The elliptic fourfold
$\IP_5^{(1,1806, 75894, 466206, 1087814, 1631721)}[3263442]$}
This manifold is self mirror with Hodge numbers $h_{11}=h_{31}=151700$. Both
the
manifold and its mirror correspond to the same group, $G_{60740}$, of the rank
indicated. Rather than write this group out explicitly, we simply note that
this fourfold and the previous one manifest the same curious group property
as their threefold analogues
 $$
G_{60740}\times G_{60740}~=~G_{152}\times G_{121328}~.$$
\newpage
\section{subtle} {Subtleties Revisited}
While presenting the algorithm, we mentioned certain 
subtleties that arise in the toric picture. Here we give prescriptions for
handling some of the common difficulties that we encounter. This is best done
by giving examples.
Consider the $n=3$ model, which has gauge group $SU(3)$, and Hodge
numbers $(5, 251)$. The dual
polyhedron for this threefold contains the top that we identify
with $SU(2)$. So
naively, one might expect the effective theory to be an $SU(2)$ gauge theory.
This is wrong, since the correct answer is $SU(3)$. However, the number of
non-toric deformations can be readily
counted, and is seen to be~1. Thus the ``missing'' rank of the gauge group
appears non-torically. If we were to add a point to the top to make the $SU(3)$
structure explicit, then we find that the Hodge numbers are unchanged, but 
the number of non-toric deformations is now zero. We interpret this to mean
that the same manifold can be described by different polyhedra, but for our
purposes, the most useful description is the one with zero non-toric
deformations. The cases we discuss below are all ultimately dealt with in the
same way. In all the cases that we have analysed in detail, we have been able
to reduce the correction to zero by adding points to the polyhedron while
holding the Hodge numbers fixed. We have made the assumption that this process
does not change the manifold and we have found that in all the cases we have
analysed in detail, this enables the algorithm to produce the correct answer.

Another interesting situation occurs when we unhiggs an $E_8$ gauge group. For
concreteness, take $n=0$. We find Hodge numbers $(23, 143)$. In this case
$\delta$ is 11. We know that unhiggsing $E_8$ causes
all the instantons in it to become tensor multiplets, so we expect 12 extra
tensors. Our first instinct
is to say that the extra tensors are encoded in the non-toric data, and that
it should be possible to add points to the polyhedron to make $\delta =0$.
This is, in fact, true. We can add 11 points to the dual polyhedron in such a
way as to leave the Hodge numbers invariant and reduce the number of non-toric
deformations to zero, as well as make 11 extra tensors explicit in the
polyhedron. However, we are still short one tensor multiplet. The extra tensor
multiplet is manifested as $pt_5'$ which is interior to a codimension one face
of the $K3$ polyhedron but {\sl not\/} interior to a codimension one face of
the threefold polyhedron. Points for which this is true can, by an enumeration
of cases, be interpreted as tensor multiplets or group factors. 
Similar situations result when
we unhiggs $E_{6b}$, $E_{7b}$, $SO(12)$, $SO(9){\times}SU(2)$,
$G_2{\times}SU(3)$ and $SU(6)_c$, when points $pt_{7}^{\prime}$,
$pt_{6}^{\prime}$, $pt_{6}^{\prime \prime}$, $pt_{6}^{\prime \prime \prime}$,
$pt_{5}^{\prime}$ and $pt_{7}^{\prime}$ respectively are relevant (\ie they
lie in codimension-2 faces), and they are also accompanied by non-zero values
of the non-toric correction $\delta$. These are the ``blue points'' of the
figures of~\cite{\rCF}. In all these cases, they correspond to extra tensors.
Curiously, we can add extra points to the polyhedra for each of these groups
in such a way that the Hodge numbers do not change, but $\delta$ vanishes and
the ``blue point''
becomes irrelevant since it now lies in the interior of a codimension one face
of the threefold polyhedron. Other more complicated situations occasionally
arise. However, in all cases that we have examined we are able to apply the
algorithm unambiguously after adding points appropriately to the polyhedron. 

\newpage
\section{con}{Discussion}
In this paper, we have presented an extension of the dictionary between toric
geometry data and and the spectrum of type IIA strings (F-theory) compactified
to four (six) dimensions on a \cyt\ described as a hypersurface in a toric 
variety. Specifically, we were able to find the enhanced gauge symmetry 
as well as the number of massless tensor multiplets (in six dimensions)
observed in the resulting vacuum. Apart from the models dual to perturbative
heterotic vacua and those resulting from ``simple'' point-like instantons
which was analyzed previously, our methods permit the analysis of cases with
a large gauge group as well as a large
number of massless tensors. We present many such 
examples. The algorithm generalises readily to the case of elliptic fourfolds.
We are currently studying the gauge content of a large class of such fourfolds,
and hope to report on our progress in future.
Our method can only be applied fully if the correction term,
$\delta$, vanishes. Otherwise there remains a contribution, $\delta$, to be
apportioned between the rank of the group and the number of tensor multiplets.
In all the cases that we have analysed in detail, it has proved possible to
add points to the polyhedron so as to reduce the correction term to zero while
holding the Hodge numbers fixed. It seems improbable that this will always be
possible. One obvious omission from our toric geometry---physics
dictionary is the information about charged matter content. We hope
to report on the progress in this direction in future.

Another interesting question to answer, we believe, would be, given such
a compactification of the type II theory, what is its heterotic dual.
It is fairly obvious that in order to provide such a gauge symmetry  by means
of a heterotic compactification on a $K3$, it would be necessary to combine
singularities of the gauge bundle with singularities of the internal manifold.
It would be very interesting to be able to specify the exact singularity 
structure which yields the duals to our models. 
\vskip5pt
\noindent {\bf Acknowledgements}
\vskip5pt
\noindent We wish to thank A.~Avram and H.~Skarke for useful discussions.
This work was supported in part by the Robert Welch Foundation and NSF grant
PHY-9511632.
\newpage
\chapno=-1
\section{appendix} {Appendix: Tables of Gauge Groups}
We list tables of groups that we have identified using the methods discussed
above. These groups were calculated for the manifolds of Ref.~\cite\rCF\ and
their mirrors. The tables list the groups, the rank of the groups, the number
of tensor multiplets and the relevant Hodge number. To conserve space, we list
the groups in subscripted form, thus, $SU_{3}^{2}$ represents~$SU(3)^2$. For
the original manifolds, the gauge and matter content is completely known from
heterotic/Type II duality, hence it is easy to identify the non-toric
deformations as either contributions to the gauge group or the number of
tensors. For the mirror manifolds, however, we cannot always identify the
non-toric corrections with certainty, since the gauge content of the
heterotic dual (if any) is not known. Thus, our knowledge of the gauge content
of the
theories encoded by the mirror manifolds is incomplete in those cases where
$\tilde \d$ is non-zero. We
therefore also list the value of $\tilde \d$ for each of the mirrors.  
\newpage

\vsize=9.5truein
\voffset-0.4truein

\def\tablebodya{%
&{SU}_{1}&0&1&3&E_{8}^{8}F_{4}^{8}G_{2}^{16}{SU}_{2}^{16}&144&97&0&243\cr
&{SU}_{2}&1&1&4&E_{7}^{3}E_{8}^{5}F_{4}^{6}G_{2}^{12}{SO}_{7}^{2}{SU}_{2}^{16}&131&81&0&214\cr
&{SU}_{2b}&1&1&4&E_{7}^{3}E_{8}^{5}F_{4}^{6}G_{2}^{12}{SO}_{7}^{2}{SU}_{2}^{16}&131&81&0&214\cr
&{SU}_{2c}&1&2&5&E_{8}^{5}F_{4}^{4}G_{2}^{10}{SO}_{5}^{2}{SO}_{9}^{2}{SO}_{11}^{2}{SO}_{13}{SU}_{2}^{12}&116&67&0&185\cr
&{SU}_{2d}&1&3&6&E_{8}^{5}F_{4}^{5}G_{2}^{10}{SU}_{2}^{11}&91&63&0&156\cr
&{SU}_{3}&2&1&5&E_{6}^{3}E_{8}^{5}F_{4}^{6}G_{2}^{10}{SU}_{2}^{10}{SU}_{3}^{4}&120&75&0&197\cr
&{SU}_{3b}&2&1&5&E_{6}^{3}E_{8}^{5}F_{4}^{6}G_{2}^{10}{SU}_{2}^{10}{SU}_{3}^{4}&120&75&0&197\cr
&{SU}_{3c}&2&3&7&E_{8}^{5}F_{4}^{4}G_{2}^{9}{SU}_{2}^{12}{SU}_{3}&88&61&0&151\cr
&{SO}_{5}&2&1&5&E_{8}^{5}F_{4}^{4}G_{2}^{10}{SO}_{5}^{2}{SO}_{9}^{2}{SO}_{11}^{3}{SU}_{2}^{12}&115&67&1&185\cr
&G_{2}&2&1&5&E_{8}^{5}F_{4}^{9}G_{2}^{10}{SU}_{2}^{14}&110&75&10&197\cr
&{SU}_{2}^{2}&2&1&5&E_{8}^{5}F_{4}^{4}G_{2}^{10}{SO}_{5}^{2}{SO}_{9}^{2}{SO}_{11}^{2}{SO}_{12}{SU}_{2}^{12}&116&67&0&185\cr
&{SU}_{2}{SU}_{2b}&2&2&6&E_{8}^{5}F_{4}^{4}G_{2}^{10}{SO}_{9}{SU}_{2}^{11}&91&63&0&156\cr
&{SU}_{4}&3&1&6&E_{8}^{5}F_{4}^{4}G_{2}^{10}{SO}_{9}^{2}{SO}_{10}^{3}{SU}_{2}^{14}&113&67&0&182\cr
&{SO}_{7}&3&1&6&E_{8}^{5}F_{4}^{4}G_{2}^{10}{SO}_{9}^{5}{SU}_{2}^{14}&110&67&3&182\cr
&{Sp}_{3}&3&1&6&E_{8}^{5}F_{4}^{4}G_{2}^{11}{SU}_{2}^{11}&89&63&2&156\cr
&{SU}_{2}{SU}_{3}&3&1&6&E_{8}^{5}F_{4}^{4}G_{2}^{8}{SU}_{2}^{10}{SU}_{3}^{2}{SU}_{4}^{2}{SU}_{5}^{2}{SU}_{6}&105&61&0&168\cr
&{SU}_{3}{SU}_{2b}&3&1&6&E_{8}^{5}F_{4}^{4}G_{2}^{8}{SU}_{2}^{10}{SU}_{3}^{2}{SU}_{4}^{2}{SU}_{5}^{2}{SU}_{6}&105&61&0&168\cr
&{SU}_{3}{SU}_{2c}&3&1&6&E_{8}^{5}F_{4}^{4}G_{2}^{8}{SU}_{2}^{10}{SU}_{3}^{2}{SU}_{4}^{2}{SU}_{5}^{2}{SU}_{6}&105&61&0&168\cr
&{SO}_{5}{SU}_{2}&3&1&6&E_{8}^{5}F_{4}^{4}G_{2}^{10}{SO}_{7}{SU}_{2}^{11}&90&63&1&156\cr
&G_{2}{SU}_{2}&3&1&6&E_{8}^{5}F_{4}^{4}G_{2}^{8}{SO}_{5}^{4}{Sp}_{3}{SU}_{2}^{12}&95&61&10&168\cr
&{SU}_{5}&4&1&7&E_{8}^{5}F_{4}^{4}G_{2}^{8}{SU}_{2}^{10}{SU}_{3}^{2}{SU}_{4}^{2}{SU}_{5}^{3}&104&61&0&167\cr
&{SO}_{9}&4&1&7&E_{8}^{5}F_{4}^{4}G_{2}^{10}{SO}_{7}^{5}{SU}_{2}^{10}&101&67&5&175\cr
&F_{4}&4&1&7&E_{8}^{5}F_{4}^{4}G_{2}^{15}{SU}_{2}^{10}&96&67&10&175\cr
&{SU}_{3}^{2}&4&1&7&E_{8}^{5}F_{4}^{4}G_{2}^{8}{SU}_{2}^{12}{SU}_{3}^{2}&88&61&0&151\cr
&G_{2}{SU}_{3}&4&1&7&E_{8}^{5}F_{4}^{4}G_{2}^{8}{SU}_{2}^{10}{SU}_{3}&84&55&10&151\cr
&{SU}_{2}{SU}_{4}&4&1&7&E_{8}^{5}F_{4}^{4}G_{2}^{8}{SU}_{2}^{11}{SU}_{3}^{2}{SU}_{4}&90&61&0&153\cr
&{SO}_{7}{SU}_{2}&4&1&7&E_{8}^{5}F_{4}^{4}G_{2}^{8}{SO}_{5}{SU}_{2}^{13}&87&61&3&153\cr
&{SU}_{6}&5&1&8&E_{8}^{5}F_{4}^{4}G_{2}^{8}{SU}_{2}^{11}{SU}_{3}^{3}&89&61&0&152\cr
&{SU}_{6b}&5&1&8&E_{8}^{5}F_{4}^{4}G_{2}^{8}{SU}_{2}^{13}{SU}_{3}&87&61&0&150\cr
&{SU}_{6c}&5&3&10&E_{8}^{5}F_{4}^{4}G_{2}^{8}{SU}_{2}^{11}{SU}_{3}&85&61&0&148\cr
&{SO}_{10}&5&1&8&E_{8}^{5}F_{4}^{4}G_{2}^{8}{SU}_{2}^{10}{SU}_{3}^{2}{SU}_{4}^{5}&101&61&0&164\cr
&{SO}_{11}&5&1&8&E_{8}^{5}F_{4}^{4}G_{2}^{8}{SO}_{5}^{5}{SU}_{2}^{12}&94&61&7&164\cr
&{SO}_{9}{SU}_{2}&5&1&8&E_{8}^{5}F_{4}^{4}G_{2}^{8}{SU}_{2}^{10}&82&57&5&146\cr
&F_{4}{SU}_{2}&5&1&8&E_{8}^{5}F_{4}^{4}G_{2}^{8}{SU}_{2}^{9}&81&53&10&146\cr
&{SO}_{12}&6&1&9&E_{8}^{5}F_{4}^{4}G_{2}^{8}{SU}_{2}^{14}&86&61&0&149\cr
&{SO}_{13}&6&5&13&E_{8}^{5}F_{4}^{4}G_{2}^{8}{SU}_{2}^{9}&81&57&5&145\cr
&E_{6}&6&1&9&E_{8}^{5}F_{4}^{4}G_{2}^{8}{SU}_{2}^{10}{SU}_{3}^{7}&96&61&0&159\cr
&E_{6b}&6&7&15&E_{8}^{5}F_{4}^{4}G_{2}^{8}{SU}_{2}^{10}{SU}_{3}&84&61&0&147\cr
&E_{7}&7&1&10&E_{8}^{5}F_{4}^{4}G_{2}^{8}{SU}_{2}^{17}&89&61&0&152\cr
&E_{7b}&7&9&18&E_{8}^{5}F_{4}^{4}G_{2}^{8}{SU}_{2}^{9}&81&61&0&144\cr
&E_{8}&8&13&23&E_{8}^{5}F_{4}^{4}G_{2}^{8}{SU}_{2}^{8}&80&61&0&143\cr
}
\def\tablebodyb{%
&{SU}_{1}&0&1&3&E_{8}^{8}F_{4}^{8}G_{2}^{16}{SU}_{2}^{16}&144&97&0&243\cr
&{SU}_{2}&1&1&4&E_{7}^{4}E_{8}^{4}F_{4}^{5}G_{2}^{10}{SO}_{7}^{3}{SU}_{2}^{16}&125&75&0&202\cr
&{SU}_{2b}&1&1&4&E_{7}^{4}E_{8}^{4}F_{4}^{5}G_{2}^{10}{SO}_{7}^{3}{SU}_{2}^{16}&125&75&0&202\cr
&{SU}_{2c}&1&1&4&E_{8}^{4}F_{4}^{3}G_{2}^{8}{SO}_{5}^{2}{SO}_{9}^{2}{SO}_{11}^{2}{SO}_{13}^{2}{Sp}_{3}{SU}_{2}^{10}&107&57&0&166\cr
&{SU}_{2d}&1&1&4&E_{8}^{4}F_{4}^{5}G_{2}^{8}{SU}_{2}^{10}&78&55&1&136\cr
&{SU}_{3}&2&1&5&E_{6}^{4}E_{8}^{4}F_{4}^{5}G_{2}^{8}{SU}_{2}^{8}{SU}_{3}^{5}&110&67&0&179\cr
&{SU}_{3b}&2&1&5&E_{6}^{4}E_{8}^{4}F_{4}^{5}G_{2}^{8}{SU}_{2}^{8}{SU}_{3}^{5}&110&67&0&179\cr
&{SU}_{3c}&2&1&5&E_{8}^{4}F_{4}^{3}G_{2}^{8}{SU}_{2}^{10}{SU}_{3}&72&51&0&125\cr
&{SO}_{5}&2&1&5&E_{8}^{4}F_{4}^{3}G_{2}^{8}{SO}_{5}^{3}{SO}_{9}^{2}{SO}_{11}^{4}{SU}_{2}^{10}&104&57&2&165\cr
&G_{2}&2&1&5&E_{8}^{4}F_{4}^{9}G_{2}^{8}{SU}_{2}^{13}&97&67&13&179\cr
&{SU}_{2}^{2}&2&1&5&E_{8}^{4}F_{4}^{3}G_{2}^{8}{SO}_{5}^{3}{SO}_{9}^{2}{SO}_{11}^{2}{SO}_{12}^{2}{SU}_{2}^{10}&106&57&0&165\cr
&{SU}_{2}{SU}_{2b}&2&1&5&E_{8}^{4}F_{4}^{3}G_{2}^{8}{SO}_{9}^{2}{SU}_{2}^{10}&78&53&0&133\cr
&{SU}_{4}&3&1&6&E_{8}^{4}F_{4}^{3}G_{2}^{8}{SO}_{9}^{2}{SO}_{10}^{4}{SU}_{2}^{13}&101&57&0&160\cr
&{SO}_{7}&3&1&6&E_{8}^{4}F_{4}^{3}G_{2}^{8}{SO}_{9}^{6}{SU}_{2}^{13}&97&57&4&160\cr
&{Sp}_{3}&3&1&6&E_{8}^{4}F_{4}^{3}G_{2}^{10}{SU}_{2}^{9}&73&53&4&132\cr
&{SU}_{2}{SU}_{3}&3&1&6&E_{8}^{4}F_{4}^{3}G_{2}^{6}{SU}_{2}^{8}{SU}_{3}^{2}{SU}_{4}^{2}{SU}_{5}^{2}{SU}_{6}^{2}&92&50&0&144\cr
&{SU}_{3}{SU}_{2b}&3&1&6&E_{8}^{4}F_{4}^{3}G_{2}^{6}{SU}_{2}^{8}{SU}_{3}^{2}{SU}_{4}^{2}{SU}_{5}^{2}{SU}_{6}^{2}&92&50&0&144\cr
&{SU}_{3}{SU}_{2c}&3&1&6&E_{8}^{4}F_{4}^{3}G_{2}^{6}{SU}_{2}^{8}{SU}_{3}^{2}{SU}_{4}^{2}{SU}_{5}^{2}{SU}_{6}^{2}&92&50&0&144\cr
&{SO}_{5}{SU}_{2}&3&1&6&E_{8}^{4}F_{4}^{3}G_{2}^{8}{SO}_{7}^{2}{SU}_{2}^{9}&75&53&2&132\cr
&G_{2}{SU}_{2}&3&1&6&E_{8}^{4}F_{4}^{3}G_{2}^{6}{SO}_{5}^{4}{Sp}_{3}^{2}{SU}_{2}^{10}&80&50&12&144\cr
&{SU}_{5}&4&1&7&E_{8}^{4}F_{4}^{3}G_{2}^{6}{SU}_{2}^{8}{SU}_{3}^{2}{SU}_{4}^{2}{SU}_{5}^{4}&90&50&0&142\cr
&{SO}_{9}&4&1&7&E_{8}^{4}F_{4}^{3}G_{2}^{8}{SO}_{7}^{6}{SU}_{2}^{8}&86&57&6&151\cr
&F_{4}&4&1&7&E_{8}^{4}F_{4}^{3}G_{2}^{14}{SU}_{2}^{8}&80&57&12&151\cr
&{SU}_{3}^{2}&4&1&7&E_{8}^{4}F_{4}^{3}G_{2}^{6}{SU}_{2}^{10}{SU}_{3}^{3}&72&50&0&124\cr
&G_{2}{SU}_{3}&4&2&8&E_{8}^{4}F_{4}^{3}G_{2}^{6}{SU}_{2}^{8}{SU}_{3}&66&44&10&122\cr
&{SU}_{2}{SU}_{4}&4&1&7&E_{8}^{4}F_{4}^{3}G_{2}^{6}{SU}_{2}^{9}{SU}_{3}^{2}{SU}_{4}^{2}&75&50&0&127\cr
&{SO}_{7}{SU}_{2}&4&1&7&E_{8}^{4}F_{4}^{3}G_{2}^{6}{SO}_{5}^{2}{SU}_{2}^{11}&71&50&4&127\cr
&{SU}_{6}&5&1&8&E_{8}^{4}F_{4}^{3}G_{2}^{6}{SU}_{2}^{9}{SU}_{3}^{4}&73&50&0&125\cr
&{SU}_{6b}&5&1&8&E_{8}^{4}F_{4}^{3}G_{2}^{6}{SU}_{2}^{12}{SU}_{3}&70&50&0&122\cr
&{SU}_{6c}&5&4&11&E_{8}^{4}F_{4}^{3}G_{2}^{6}{SU}_{2}^{9}{SU}_{3}&67&50&0&119\cr
&{SO}_{10}&5&1&8&E_{8}^{4}F_{4}^{3}G_{2}^{6}{SU}_{2}^{8}{SU}_{3}^{2}{SU}_{4}^{6}&86&50&0&138\cr
&{SO}_{11}&5&1&8&E_{8}^{4}F_{4}^{3}G_{2}^{6}{SO}_{5}^{6}{SU}_{2}^{10}&78&50&8&138\cr
&{SO}_{9}{SU}_{2}&5&1&8&E_{8}^{4}F_{4}^{3}G_{2}^{6}{SU}_{2}^{9}&65&46&5&118\cr
&F_{4}{SU}_{2}&5&2&9&E_{8}^{4}F_{4}^{3}G_{2}^{6}{SU}_{2}^{7}&63&42&10&117\cr
&{SO}_{12}&6&1&9&E_{8}^{4}F_{4}^{3}G_{2}^{6}{SU}_{2}^{13}&69&50&0&121\cr
&{SO}_{13}&6&6&14&E_{8}^{4}F_{4}^{3}G_{2}^{6}{SU}_{2}^{7}&63&46&5&116\cr
&E_{6}&6&1&9&E_{8}^{4}F_{4}^{3}G_{2}^{6}{SU}_{2}^{8}{SU}_{3}^{8}&80&50&0&132\cr
&E_{6b}&6&8&16&E_{8}^{4}F_{4}^{3}G_{2}^{6}{SU}_{2}^{8}{SU}_{3}&66&50&0&118\cr
&E_{7}&7&1&10&E_{8}^{4}F_{4}^{3}G_{2}^{6}{SU}_{2}^{16}&72&50&0&124\cr
&E_{7b}&7&10&19&E_{8}^{4}F_{4}^{3}G_{2}^{6}{SU}_{2}^{7}&63&50&0&115\cr
&E_{8}&8&14&24&E_{8}^{4}F_{4}^{3}G_{2}^{6}{SU}_{2}^{6}&62&50&0&114\cr
}
\def\tablebodyc{%
&{SU}_{1}&0&1&3&E_{8}^{8}F_{4}^{8}G_{2}^{16}{SU}_{2}^{16}&144&96&1&243\cr
&{SU}_{2}&1&1&4&E_{7}^{5}E_{8}^{3}F_{4}^{4}G_{2}^{8}{SO}_{7}^{4}{SU}_{2}^{16}&119&68&1&190\cr
&{SU}_{2b}&1&1&4&E_{7}^{5}E_{8}^{3}F_{4}^{4}G_{2}^{8}{SO}_{7}^{4}{SU}_{2}^{16}&119&68&1&190\cr
&{SU}_{2c}&1&1&4&E_{8}^{3}F_{4}^{2}G_{2}^{6}{SO}_{5}^{2}{SO}_{9}^{2}{SO}_{11}^{2}{SO}_{13}^{3}{Sp}_{3}^{2}{SU}_{2}^{8}&98&46&2&148\cr
&{SU}_{2d}&1&1&4&E_{8}^{3}F_{4}^{5}G_{2}^{6}{SU}_{2}^{9}&65&46&5&118\cr
&{SU}_{3}&2&1&5&E_{6}^{5}E_{8}^{3}F_{4}^{4}G_{2}^{6}{SU}_{2}^{6}{SU}_{3}^{6}&100&58&1&161\cr
&{SU}_{3b}&2&1&5&E_{6}^{5}E_{8}^{3}F_{4}^{4}G_{2}^{6}{SU}_{2}^{6}{SU}_{3}^{6}&100&58&1&161\cr
&{SU}_{3c}&2&1&5&E_{8}^{3}F_{4}^{2}G_{2}^{7}{SU}_{2}^{8}{SU}_{3}&56&40&3&101\cr
&{SO}_{5}&2&1&5&E_{8}^{3}F_{4}^{2}G_{2}^{6}{SO}_{5}^{4}{SO}_{9}^{2}{SO}_{11}^{5}{SU}_{2}^{8}&93&46&4&145\cr
&G_{2}&2&1&5&E_{8}^{3}F_{4}^{9}G_{2}^{6}{SU}_{2}^{12}&84&58&17&161\cr
&{SU}_{2}^{2}&2&1&5&E_{8}^{3}F_{4}^{2}G_{2}^{6}{SO}_{5}^{4}{SO}_{9}^{2}{SO}_{11}^{2}{SO}_{12}^{3}{SU}_{2}^{8}&96&46&1&145\cr
&{SU}_{2}{SU}_{2b}&2&1&5&E_{8}^{3}F_{4}^{2}G_{2}^{6}{SO}_{9}^{3}{SU}_{2}^{9}&65&42&2&111\cr
&{SU}_{4}&3&1&6&E_{8}^{3}F_{4}^{2}G_{2}^{6}{SO}_{9}^{2}{SO}_{10}^{5}{SU}_{2}^{12}&89&46&1&138\cr
&{SO}_{7}&3&1&6&E_{8}^{3}F_{4}^{2}G_{2}^{6}{SO}_{9}^{7}{SU}_{2}^{12}&84&46&6&138\cr
&{Sp}_{3}&3&1&6&E_{8}^{3}F_{4}^{2}G_{2}^{9}{SU}_{2}^{7}&57&42&7&108\cr
&{SU}_{2}{SU}_{3}&3&1&6&E_{8}^{3}F_{4}^{2}G_{2}^{4}{SU}_{2}^{6}{SU}_{3}^{2}{SU}_{4}^{2}{SU}_{5}^{2}{SU}_{6}^{3}&79&38&1&120\cr
&{SU}_{3}{SU}_{2b}&3&1&6&E_{8}^{3}F_{4}^{2}G_{2}^{4}{SU}_{2}^{6}{SU}_{3}^{2}{SU}_{4}^{2}{SU}_{5}^{2}{SU}_{6}^{3}&79&38&1&120\cr
&{SU}_{3}{SU}_{2c}&3&1&6&E_{8}^{3}F_{4}^{2}G_{2}^{4}{SU}_{2}^{6}{SU}_{3}^{2}{SU}_{4}^{2}{SU}_{5}^{2}{SU}_{6}^{3}&79&38&1&120\cr
&{SO}_{5}{SU}_{2}&3&1&6&E_{8}^{3}F_{4}^{2}G_{2}^{6}{SO}_{7}^{3}{SU}_{2}^{7}&60&42&4&108\cr
&G_{2}{SU}_{2}&3&1&6&E_{8}^{3}F_{4}^{2}G_{2}^{4}{SO}_{5}^{4}{Sp}_{3}^{3}{SU}_{2}^{8}&65&38&15&120\cr
&{SU}_{5}&4&1&7&E_{8}^{3}F_{4}^{2}G_{2}^{4}{SU}_{2}^{6}{SU}_{3}^{2}{SU}_{4}^{2}{SU}_{5}^{5}&76&38&1&117\cr
&{SO}_{9}&4&1&7&E_{8}^{3}F_{4}^{2}G_{2}^{6}{SO}_{7}^{7}{SU}_{2}^{6}&71&46&8&127\cr
&F_{4}&4&1&7&E_{8}^{3}F_{4}^{2}G_{2}^{13}{SU}_{2}^{6}&64&46&15&127\cr
&{SU}_{3}^{2}&4&1&7&E_{8}^{3}F_{4}^{2}G_{2}^{4}{SU}_{2}^{8}{SU}_{3}^{4}&56&38&1&97\cr
&G_{2}{SU}_{3}&4&3&9&E_{8}^{3}F_{4}^{2}G_{2}^{4}{SU}_{2}^{6}{SU}_{3}&48&32&11&93\cr
&{SU}_{2}{SU}_{4}&4&1&7&E_{8}^{3}F_{4}^{2}G_{2}^{4}{SU}_{2}^{7}{SU}_{3}^{2}{SU}_{4}^{3}&60&38&1&101\cr
&{SO}_{7}{SU}_{2}&4&1&7&E_{8}^{3}F_{4}^{2}G_{2}^{4}{SO}_{5}^{3}{SU}_{2}^{9}&55&38&6&101\cr
&{SU}_{6}&5&1&8&E_{8}^{3}F_{4}^{2}G_{2}^{4}{SU}_{2}^{7}{SU}_{3}^{5}&57&38&1&98\cr
&{SU}_{6b}&5&1&8&E_{8}^{3}F_{4}^{2}G_{2}^{4}{SU}_{2}^{11}{SU}_{3}&53&38&1&94\cr
&{SU}_{6c}&5&5&12&E_{8}^{3}F_{4}^{2}G_{2}^{4}{SU}_{2}^{7}{SU}_{3}&49&38&1&90\cr
&{SO}_{10}&5&1&8&E_{8}^{3}F_{4}^{2}G_{2}^{4}{SU}_{2}^{6}{SU}_{3}^{2}{SU}_{4}^{7}&71&38&1&112\cr
&{SO}_{11}&5&1&8&E_{8}^{3}F_{4}^{2}G_{2}^{4}{SO}_{5}^{7}{SU}_{2}^{8}&62&38&10&112\cr
&{SO}_{9}{SU}_{2}&5&1&8&E_{8}^{3}F_{4}^{2}G_{2}^{4}{SU}_{2}^{8}&48&34&6&90\cr
&F_{4}{SU}_{2}&5&3&10&E_{8}^{3}F_{4}^{2}G_{2}^{4}{SU}_{2}^{5}&45&30&11&88\cr
&{SO}_{12}&6&1&9&E_{8}^{3}F_{4}^{2}G_{2}^{4}{SU}_{2}^{12}&52&38&1&93\cr
&{SO}_{13}&6&7&15&E_{8}^{3}F_{4}^{2}G_{2}^{4}{SU}_{2}^{5}&45&34&6&87\cr
&E_{6}&6&1&9&E_{8}^{3}F_{4}^{2}G_{2}^{4}{SU}_{2}^{6}{SU}_{3}^{9}&64&38&1&105\cr
&E_{6b}&6&9&17&E_{8}^{3}F_{4}^{2}G_{2}^{4}{SU}_{2}^{6}{SU}_{3}&48&38&1&89\cr
&E_{7}&7&1&10&E_{8}^{3}F_{4}^{2}G_{2}^{4}{SU}_{2}^{15}&55&38&1&96\cr
&E_{7b}&7&11&20&E_{8}^{3}F_{4}^{2}G_{2}^{4}{SU}_{2}^{5}&45&38&1&86\cr
&E_{8}&8&15&25&E_{8}^{3}F_{4}^{2}G_{2}^{4}{SU}_{2}^{4}&44&38&1&85\cr
}
\def\tablebodyd{%
&{SU}_{1}&2&1&5&E_{8}^{8}F_{4}^{9}G_{2}^{17}{SU}_{2}^{16}&150&99&0&251\cr
&{SU}_{2}&3&1&6&E_{7}^{6}E_{8}^{2}F_{4}^{4}G_{2}^{7}{SO}_{7}^{5}{SU}_{2}^{16}&119&65&0&186\cr
&{SU}_{2b}&3&1&6&E_{7}^{6}E_{8}^{2}F_{4}^{4}G_{2}^{7}{SO}_{7}^{5}{SU}_{2}^{16}&119&65&0&186\cr
&{SU}_{2c}&3&1&6&E_{8}^{2}F_{4}^{2}G_{2}^{5}{SO}_{5}^{2}{SO}_{9}^{2}{SO}_{11}^{2}{SO}_{13}^{4}{Sp}_{3}^{3}{SU}_{2}^{6}&95&39&2&138\cr
&{SU}_{2d}&3&1&6&E_{8}^{2}F_{4}^{6}G_{2}^{5}{SU}_{2}^{8}&58&41&7&108\cr
&{SU}_{3}&4&1&7&E_{6}^{6}E_{8}^{2}F_{4}^{4}G_{2}^{5}{SU}_{2}^{4}{SU}_{3}^{7}&96&53&0&151\cr
&{SU}_{3b}&4&1&7&E_{6}^{6}E_{8}^{2}F_{4}^{4}G_{2}^{5}{SU}_{2}^{4}{SU}_{3}^{7}&96&53&0&151\cr
&{SU}_{3c}&4&1&7&E_{8}^{2}F_{4}^{2}G_{2}^{7}{SU}_{2}^{6}{SU}_{3}&46&33&4&85\cr
&{SO}_{5}&4&1&7&E_{8}^{2}F_{4}^{2}G_{2}^{5}{SO}_{5}^{5}{SO}_{9}^{2}{SO}_{11}^{6}{SU}_{2}^{6}&88&39&4&133\cr
&G_{2}&4&1&7&E_{8}^{2}F_{4}^{10}G_{2}^{5}{SU}_{2}^{11}&77&53&19&151\cr
&{SU}_{2}^{2}&4&1&7&E_{8}^{2}F_{4}^{2}G_{2}^{5}{SO}_{5}^{5}{SO}_{9}^{2}{SO}_{11}^{2}{SO}_{12}^{4}{SU}_{2}^{6}&92&39&0&133\cr
&{SU}_{2}{SU}_{2b}&4&1&7&E_{8}^{2}F_{4}^{2}G_{2}^{5}{SO}_{9}^{4}{SU}_{2}^{8}&58&35&2&97\cr
&{SU}_{4}&5&1&8&E_{8}^{2}F_{4}^{2}G_{2}^{5}{SO}_{9}^{2}{SO}_{10}^{6}{SU}_{2}^{11}&83&39&0&124\cr
&{SO}_{7}&5&1&8&E_{8}^{2}F_{4}^{2}G_{2}^{5}{SO}_{9}^{8}{SU}_{2}^{11}&77&39&6&124\cr
&{Sp}_{3}&5&1&8&E_{8}^{2}F_{4}^{2}G_{2}^{9}{SU}_{2}^{5}&47&35&8&92\cr
&{SU}_{2}{SU}_{3}&5&1&8&E_{8}^{2}F_{4}^{2}G_{2}^{3}{SU}_{2}^{4}{SU}_{3}^{2}{SU}_{4}^{2}{SU}_{5}^{2}{SU}_{6}^{4}&72&30&0&104\cr
&{SU}_{3}{SU}_{2b}&5&1&8&E_{8}^{2}F_{4}^{2}G_{2}^{3}{SU}_{2}^{4}{SU}_{3}^{2}{SU}_{4}^{2}{SU}_{5}^{2}{SU}_{6}^{4}&72&30&0&104\cr
&{SU}_{3}{SU}_{2c}&5&1&8&E_{8}^{2}F_{4}^{2}G_{2}^{3}{SU}_{2}^{4}{SU}_{3}^{2}{SU}_{4}^{2}{SU}_{5}^{2}{SU}_{6}^{4}&72&30&0&104\cr
&{SO}_{5}{SU}_{2}&5&1&8&E_{8}^{2}F_{4}^{2}G_{2}^{5}{SO}_{7}^{4}{SU}_{2}^{5}&51&35&4&92\cr
&G_{2}{SU}_{2}&5&1&8&E_{8}^{2}F_{4}^{2}G_{2}^{3}{SO}_{5}^{4}{Sp}_{3}^{4}{SU}_{2}^{6}&56&30&16&104\cr
&{SU}_{5}&6&1&9&E_{8}^{2}F_{4}^{2}G_{2}^{3}{SU}_{2}^{4}{SU}_{3}^{2}{SU}_{4}^{2}{SU}_{5}^{6}&68&30&0&100\cr
&{SO}_{9}&6&1&9&E_{8}^{2}F_{4}^{2}G_{2}^{5}{SO}_{7}^{8}{SU}_{2}^{4}&62&39&8&111\cr
&F_{4}&6&1&9&E_{8}^{2}F_{4}^{2}G_{2}^{13}{SU}_{2}^{4}&54&39&16&111\cr
&{SU}_{3}^{2}&6&1&9&E_{8}^{2}F_{4}^{2}G_{2}^{3}{SU}_{2}^{6}{SU}_{3}^{5}&46&30&0&78\cr
&G_{2}{SU}_{3}&6&4&12&E_{8}^{2}F_{4}^{2}G_{2}^{3}{SU}_{2}^{4}{SU}_{3}&36&24&10&72\cr
&{SU}_{2}{SU}_{4}&6&1&9&E_{8}^{2}F_{4}^{2}G_{2}^{3}{SU}_{2}^{5}{SU}_{3}^{2}{SU}_{4}^{4}&51&30&0&83\cr
&{SO}_{7}{SU}_{2}&6&1&9&E_{8}^{2}F_{4}^{2}G_{2}^{3}{SO}_{5}^{4}{SU}_{2}^{7}&45&30&6&83\cr
&{SU}_{6}&7&1&10&E_{8}^{2}F_{4}^{2}G_{2}^{3}{SU}_{2}^{5}{SU}_{3}^{6}&47&30&0&79\cr
&{SU}_{6b}&7&1&10&E_{8}^{2}F_{4}^{2}G_{2}^{3}{SU}_{2}^{10}{SU}_{3}&42&30&0&74\cr
&{SU}_{6c}&7&6&15&E_{8}^{2}F_{4}^{2}G_{2}^{3}{SU}_{2}^{5}{SU}_{3}&37&30&0&69\cr
&{SO}_{10}&7&1&10&E_{8}^{2}F_{4}^{2}G_{2}^{3}{SU}_{2}^{4}{SU}_{3}^{2}{SU}_{4}^{8}&62&30&0&94\cr
&{SO}_{11}&7&1&10&E_{8}^{2}F_{4}^{2}G_{2}^{3}{SO}_{5}^{8}{SU}_{2}^{6}&52&30&10&94\cr
&{SO}_{9}{SU}_{2}&7&1&10&E_{8}^{2}F_{4}^{2}G_{2}^{3}{SU}_{2}^{7}&37&26&5&70\cr
&F_{4}{SU}_{2}&7&4&13&E_{8}^{2}F_{4}^{2}G_{2}^{3}{SU}_{2}^{3}&33&22&10&67\cr
&{SO}_{12}&8&1&11&E_{8}^{2}F_{4}^{2}G_{2}^{3}{SU}_{2}^{11}&41&30&0&73\cr
&{SO}_{13}&8&8&18&E_{8}^{2}F_{4}^{2}G_{2}^{3}{SU}_{2}^{3}&33&26&5&66\cr
&E_{6}&8&1&11&E_{8}^{2}F_{4}^{2}G_{2}^{3}{SU}_{2}^{4}{SU}_{3}^{10}&54&30&0&86\cr
&E_{6b}&8&10&20&E_{8}^{2}F_{4}^{2}G_{2}^{3}{SU}_{2}^{4}{SU}_{3}&36&30&0&68\cr
&E_{7}&9&1&12&E_{8}^{2}F_{4}^{2}G_{2}^{3}{SU}_{2}^{14}&44&30&0&76\cr
&E_{7b}&9&12&23&E_{8}^{2}F_{4}^{2}G_{2}^{3}{SU}_{2}^{3}&33&30&0&65\cr
&E_{8}&10&16&28&E_{8}^{2}F_{4}^{2}G_{2}^{3}{SU}_{2}^{2}&32&30&0&64\cr
}
\def\tablebodye{%
&{SU}_{1}&4&1&7&E_{8}^{9}F_{4}^{9}G_{2}^{18}{SU}_{2}^{18}&162&107&0&271\cr
&{SU}_{2}&5&1&8&E_{7}^{7}E_{8}^{2}F_{4}^{3}G_{2}^{6}{SO}_{7}^{6}{SU}_{2}^{18}&125&67&0&194\cr
&{SU}_{2b}&5&1&8&E_{7}^{7}E_{8}^{2}F_{4}^{3}G_{2}^{6}{SO}_{7}^{6}{SU}_{2}^{18}&125&67&0&194\cr
&{SU}_{2c}&5&1&8&E_{8}^{2}F_{4}G_{2}^{4}{SO}_{5}^{2}{SO}_{9}^{2}{SO}_{11}^{2}{SO}_{13}^{5}{Sp}_{3}^{4}{SU}_{2}^{6}&98&37&3&140\cr
&{SU}_{2d}&5&1&8&E_{8}^{2}F_{4}^{6}G_{2}^{4}{SU}_{2}^{9}&57&41&10&110\cr
&{SU}_{3}&6&1&9&E_{6}^{7}E_{8}^{2}F_{4}^{3}G_{2}^{4}{SU}_{2}^{4}{SU}_{3}^{8}&98&53&0&153\cr
&{SU}_{3b}&6&1&9&E_{6}^{7}E_{8}^{2}F_{4}^{3}G_{2}^{4}{SU}_{2}^{4}{SU}_{3}^{8}&98&53&0&153\cr
&{SU}_{3c}&6&1&9&E_{8}^{2}F_{4}G_{2}^{7}{SU}_{2}^{6}{SU}_{3}&42&31&6&81\cr
&{SO}_{5}&6&1&9&E_{8}^{2}F_{4}G_{2}^{4}{SO}_{5}^{6}{SO}_{9}^{2}{SO}_{11}^{7}{SU}_{2}^{6}&89&37&5&133\cr
&G_{2}&6&1&9&E_{8}^{2}F_{4}^{10}G_{2}^{4}{SU}_{2}^{12}&76&53&22&153\cr
&{SU}_{2}^{2}&6&1&9&E_{8}^{2}F_{4}G_{2}^{4}{SO}_{5}^{6}{SO}_{9}^{2}{SO}_{11}^{2}{SO}_{12}^{5}{SU}_{2}^{6}&94&37&0&133\cr
&{SU}_{2}{SU}_{2b}&6&1&9&E_{8}^{2}F_{4}G_{2}^{4}{SO}_{9}^{5}{SU}_{2}^{9}&57&33&3&95\cr
&{SU}_{4}&7&1&10&E_{8}^{2}F_{4}G_{2}^{4}{SO}_{9}^{2}{SO}_{10}^{7}{SU}_{2}^{12}&83&37&0&122\cr
&{SO}_{7}&7&1&10&E_{8}^{2}F_{4}G_{2}^{4}{SO}_{9}^{9}{SU}_{2}^{12}&76&37&7&122\cr
&{Sp}_{3}&7&1&10&E_{8}^{2}F_{4}G_{2}^{9}{SU}_{2}^{5}&43&33&10&88\cr
&{SU}_{2}{SU}_{3}&7&1&10&E_{8}^{2}F_{4}G_{2}^{2}{SU}_{2}^{4}{SU}_{3}^{2}{SU}_{4}^{2}{SU}_{5}^{2}{SU}_{6}^{5}&71&27&0&100\cr
&{SU}_{3}{SU}_{2b}&7&1&10&E_{8}^{2}F_{4}G_{2}^{2}{SU}_{2}^{4}{SU}_{3}^{2}{SU}_{4}^{2}{SU}_{5}^{2}{SU}_{6}^{5}&71&27&0&100\cr
&{SU}_{3}{SU}_{2c}&7&1&10&E_{8}^{2}F_{4}G_{2}^{2}{SU}_{2}^{4}{SU}_{3}^{2}{SU}_{4}^{2}{SU}_{5}^{2}{SU}_{6}^{5}&71&27&0&100\cr
&{SO}_{5}{SU}_{2}&7&1&10&E_{8}^{2}F_{4}G_{2}^{4}{SO}_{7}^{5}{SU}_{2}^{5}&48&33&5&88\cr
&G_{2}{SU}_{2}&7&1&10&E_{8}^{2}F_{4}G_{2}^{2}{SO}_{5}^{4}{Sp}_{3}^{5}{SU}_{2}^{6}&53&27&18&100\cr
&{SU}_{5}&8&1&11&E_{8}^{2}F_{4}G_{2}^{2}{SU}_{2}^{4}{SU}_{3}^{2}{SU}_{4}^{2}{SU}_{5}^{7}&66&27&0&95\cr
&{SO}_{9}&8&1&11&E_{8}^{2}F_{4}G_{2}^{4}{SO}_{7}^{9}{SU}_{2}^{4}&59&37&9&107\cr
&F_{4}&8&1&11&E_{8}^{2}F_{4}G_{2}^{13}{SU}_{2}^{4}&50&37&18&107\cr
&{SU}_{3}^{2}&8&1&11&E_{8}^{2}F_{4}G_{2}^{2}{SU}_{2}^{6}{SU}_{3}^{6}&42&27&0&71\cr
&G_{2}{SU}_{3}&8&5&15&E_{8}^{2}F_{4}G_{2}^{2}{SU}_{2}^{4}{SU}_{3}&30&21&10&63\cr
&{SU}_{2}{SU}_{4}&8&1&11&E_{8}^{2}F_{4}G_{2}^{2}{SU}_{2}^{5}{SU}_{3}^{2}{SU}_{4}^{5}&48&27&0&77\cr
&{SO}_{7}{SU}_{2}&8&1&11&E_{8}^{2}F_{4}G_{2}^{2}{SO}_{5}^{5}{SU}_{2}^{7}&41&27&7&77\cr
&{SU}_{6}&9&1&12&E_{8}^{2}F_{4}G_{2}^{2}{SU}_{2}^{5}{SU}_{3}^{7}&43&27&0&72\cr
&{SU}_{6b}&9&1&12&E_{8}^{2}F_{4}G_{2}^{2}{SU}_{2}^{11}{SU}_{3}&37&27&0&66\cr
&{SU}_{6c}&9&7&18&E_{8}^{2}F_{4}G_{2}^{2}{SU}_{2}^{5}{SU}_{3}&31&27&0&60\cr
&{SO}_{10}&9&1&12&E_{8}^{2}F_{4}G_{2}^{2}{SU}_{2}^{4}{SU}_{3}^{2}{SU}_{4}^{9}&59&27&0&88\cr
&{SO}_{11}&9&1&12&E_{8}^{2}F_{4}G_{2}^{2}{SO}_{5}^{9}{SU}_{2}^{6}&48&27&11&88\cr
&{SO}_{9}{SU}_{2}&9&1&12&E_{8}^{2}F_{4}G_{2}^{2}{SU}_{2}^{8}&32&23&5&62\cr
&F_{4}{SU}_{2}&9&5&16&E_{8}^{2}F_{4}G_{2}^{2}{SU}_{2}^{3}&27&19&10&58\cr
&{SO}_{12}&10&1&13&E_{8}^{2}F_{4}G_{2}^{2}{SU}_{2}^{12}&36&27&0&65\cr
&{SO}_{13}&10&9&21&E_{8}^{2}F_{4}G_{2}^{2}{SU}_{2}^{3}&27&23&5&57\cr
&E_{6}&10&1&13&E_{8}^{2}F_{4}G_{2}^{2}{SU}_{2}^{4}{SU}_{3}^{11}&50&27&0&79\cr
&E_{6b}&10&11&23&E_{8}^{2}F_{4}G_{2}^{2}{SU}_{2}^{4}{SU}_{3}&30&27&0&59\cr
&E_{7}&11&1&14&E_{8}^{2}F_{4}G_{2}^{2}{SU}_{2}^{15}&39&27&0&68\cr
&E_{7b}&11&13&26&E_{8}^{2}F_{4}G_{2}^{2}{SU}_{2}^{3}&27&27&0&56\cr
&E_{8}&12&17&31&E_{8}^{2}F_{4}G_{2}^{2}{SU}_{2}^{2}&26&27&0&55\cr
}
\def\tablebodyf{%
&{SU}_{1}&4&1&7&E_{8}^{10}F_{4}^{9}G_{2}^{20}{SU}_{2}^{20}&176&117&0&295\cr
&{SU}_{2}&5&1&8&E_{7}^{8}E_{8}^{2}F_{4}^{2}G_{2}^{6}{SO}_{7}^{7}{SU}_{2}^{20}&133&71&0&206\cr
&{SU}_{2b}&5&1&8&E_{7}^{8}E_{8}^{2}F_{4}^{2}G_{2}^{6}{SO}_{7}^{7}{SU}_{2}^{20}&133&71&0&206\cr
&{SU}_{2c}&5&1&8&E_{8}^{2}G_{2}^{4}{SO}_{5}^{2}{SO}_{9}^{2}{SO}_{11}^{2}{SO}_{13}^{6}{Sp}_{3}^{5}{SU}_{2}^{6}&103&37&4&146\cr
&{SU}_{2d}&5&1&8&E_{8}^{2}F_{4}^{6}G_{2}^{4}{SU}_{2}^{10}&58&43&13&116\cr
&{SU}_{3}&6&1&9&E_{6}^{8}E_{8}^{2}F_{4}^{2}G_{2}^{4}{SU}_{2}^{4}{SU}_{3}^{9}&102&55&0&159\cr
&{SU}_{3b}&6&1&9&E_{6}^{8}E_{8}^{2}F_{4}^{2}G_{2}^{4}{SU}_{2}^{4}{SU}_{3}^{9}&102&55&0&159\cr
&{SU}_{3c}&6&1&9&E_{8}^{2}G_{2}^{8}{SU}_{2}^{6}{SU}_{3}&40&31&8&81\cr
&{SO}_{5}&6&1&9&E_{8}^{2}G_{2}^{4}{SO}_{5}^{7}{SO}_{9}^{2}{SO}_{11}^{8}{SU}_{2}^{6}&92&37&6&137\cr
&G_{2}&6&1&9&E_{8}^{2}F_{4}^{10}G_{2}^{4}{SU}_{2}^{13}&77&55&25&159\cr
&{SU}_{2}^{2}&6&1&9&E_{8}^{2}G_{2}^{4}{SO}_{5}^{7}{SO}_{9}^{2}{SO}_{11}^{2}{SO}_{12}^{6}{SU}_{2}^{6}&98&37&0&137\cr
&{SU}_{2}{SU}_{2b}&6&1&9&E_{8}^{2}G_{2}^{4}{SO}_{9}^{6}{SU}_{2}^{10}&58&33&4&97\cr
&{SU}_{4}&7&1&10&E_{8}^{2}G_{2}^{4}{SO}_{9}^{2}{SO}_{10}^{8}{SU}_{2}^{13}&85&37&0&124\cr
&{SO}_{7}&7&1&10&E_{8}^{2}G_{2}^{4}{SO}_{9}^{10}{SU}_{2}^{13}&77&37&8&124\cr
&{Sp}_{3}&7&1&10&E_{8}^{2}G_{2}^{10}{SU}_{2}^{5}&41&33&12&88\cr
&{SU}_{2}{SU}_{3}&7&1&10&E_{8}^{2}G_{2}^{2}{SU}_{2}^{4}{SU}_{3}^{2}{SU}_{4}^{2}{SU}_{5}^{2}{SU}_{6}^{6}&72&26&0&100\cr
&{SU}_{3}{SU}_{2b}&7&1&10&E_{8}^{2}G_{2}^{2}{SU}_{2}^{4}{SU}_{3}^{2}{SU}_{4}^{2}{SU}_{5}^{2}{SU}_{6}^{6}&72&26&0&100\cr
&{SU}_{3}{SU}_{2c}&7&1&10&E_{8}^{2}G_{2}^{2}{SU}_{2}^{4}{SU}_{3}^{2}{SU}_{4}^{2}{SU}_{5}^{2}{SU}_{6}^{6}&72&26&0&100\cr
&{SO}_{5}{SU}_{2}&7&1&10&E_{8}^{2}G_{2}^{4}{SO}_{7}^{6}{SU}_{2}^{5}&47&33&6&88\cr
&G_{2}{SU}_{2}&7&1&10&E_{8}^{2}G_{2}^{2}{SO}_{5}^{4}{Sp}_{3}^{6}{SU}_{2}^{6}&52&26&20&100\cr
&{SU}_{5}&8&1&11&E_{8}^{2}G_{2}^{2}{SU}_{2}^{4}{SU}_{3}^{2}{SU}_{4}^{2}{SU}_{5}^{8}&66&26&0&94\cr
&{SO}_{9}&8&1&11&E_{8}^{2}G_{2}^{4}{SO}_{7}^{10}{SU}_{2}^{4}&58&37&10&107\cr
&F_{4}&8&1&11&E_{8}^{2}G_{2}^{14}{SU}_{2}^{4}&48&37&20&107\cr
&{SU}_{3}^{2}&8&1&11&E_{8}^{2}G_{2}^{2}{SU}_{2}^{6}{SU}_{3}^{7}&40&26&0&68\cr
&G_{2}{SU}_{3}&8&6&16&E_{8}^{2}G_{2}^{2}{SU}_{2}^{4}{SU}_{3}&26&20&10&58\cr
&{SU}_{2}{SU}_{4}&8&1&11&E_{8}^{2}G_{2}^{2}{SU}_{2}^{5}{SU}_{3}^{2}{SU}_{4}^{6}&47&26&0&75\cr
&{SO}_{7}{SU}_{2}&8&1&11&E_{8}^{2}G_{2}^{2}{SO}_{5}^{6}{SU}_{2}^{7}&39&26&8&75\cr
&{SU}_{6}&9&1&12&E_{8}^{2}G_{2}^{2}{SU}_{2}^{5}{SU}_{3}^{8}&41&26&0&69\cr
&{SU}_{6b}&9&1&12&E_{8}^{2}G_{2}^{2}{SU}_{2}^{12}{SU}_{3}&34&26&0&62\cr
&{SU}_{6c}&9&8&19&E_{8}^{2}G_{2}^{2}{SU}_{2}^{5}{SU}_{3}&27&26&0&55\cr
&{SO}_{10}&9&1&12&E_{8}^{2}G_{2}^{2}{SU}_{2}^{4}{SU}_{3}^{2}{SU}_{4}^{10}&58&26&0&86\cr
&{SO}_{11}&9&1&12&E_{8}^{2}G_{2}^{2}{SO}_{5}^{10}{SU}_{2}^{6}&46&26&12&86\cr
&{SO}_{9}{SU}_{2}&9&1&12&E_{8}^{2}G_{2}^{2}{SU}_{2}^{9}&29&22&5&58\cr
&F_{4}{SU}_{2}&9&6&17&E_{8}^{2}G_{2}^{2}{SU}_{2}^{3}&23&18&10&53\cr
&{SO}_{12}&10&1&13&E_{8}^{2}G_{2}^{2}{SU}_{2}^{13}&33&26&0&61\cr
&{SO}_{13}&10&10&22&E_{8}^{2}G_{2}^{2}{SU}_{2}^{3}&23&22&5&52\cr
&E_{6}&10&1&13&E_{8}^{2}G_{2}^{2}{SU}_{2}^{4}{SU}_{3}^{12}&48&26&0&76\cr
&E_{6b}&10&12&24&E_{8}^{2}G_{2}^{2}{SU}_{2}^{4}{SU}_{3}&26&26&0&54\cr
&E_{7}&11&1&14&E_{8}^{2}G_{2}^{2}{SU}_{2}^{16}&36&26&0&64\cr
&E_{7b}&11&14&27&E_{8}^{2}G_{2}^{2}{SU}_{2}^{3}&23&26&0&51\cr
&E_{8}&12&18&32&E_{8}^{2}G_{2}^{2}{SU}_{2}^{2}&22&26&0&50\cr
}
\def\tablebodyg{%
&{SU}_{1}&6&1&9&E_{8}^{11}F_{4}^{10}G_{2}^{21}{SU}_{2}^{22}&192&127&0&321\cr
&{SU}_{2}&7&1&10&E_{7}^{9}E_{8}^{2}F_{4}^{2}G_{2}^{5}{SO}_{7}^{8}{SU}_{2}^{22}&143&75&0&220\cr
&{SU}_{2b}&7&1&10&E_{7}^{9}E_{8}^{2}F_{4}^{2}G_{2}^{5}{SO}_{7}^{8}{SU}_{2}^{22}&143&75&0&220\cr
&{SU}_{2c}&7&1&10&E_{8}^{2}G_{2}^{3}{SO}_{5}^{2}{SO}_{9}^{2}{SO}_{11}^{2}{SO}_{13}^{7}{Sp}_{3}^{6}{SU}_{2}^{6}&110&37&5&154\cr
&{SU}_{2d}&7&1&10&E_{8}^{2}F_{4}^{7}G_{2}^{3}{SU}_{2}^{11}&61&45&16&124\cr
&{SU}_{3}&8&1&11&E_{6}^{9}E_{8}^{2}F_{4}^{2}G_{2}^{3}{SU}_{2}^{4}{SU}_{3}^{10}&108&57&0&167\cr
&{SU}_{3b}&8&1&11&E_{6}^{9}E_{8}^{2}F_{4}^{2}G_{2}^{3}{SU}_{2}^{4}{SU}_{3}^{10}&108&57&0&167\cr
&{SU}_{3c}&8&1&11&E_{8}^{2}G_{2}^{8}{SU}_{2}^{6}{SU}_{3}&40&31&10&83\cr
&{SO}_{5}&8&1&11&E_{8}^{2}G_{2}^{3}{SO}_{5}^{8}{SO}_{9}^{2}{SO}_{11}^{9}{SU}_{2}^{6}&97&37&7&143\cr
&G_{2}&8&1&11&E_{8}^{2}F_{4}^{11}G_{2}^{3}{SU}_{2}^{14}&80&57&28&167\cr
&{SU}_{2}^{2}&8&1&11&E_{8}^{2}G_{2}^{3}{SO}_{5}^{8}{SO}_{9}^{2}{SO}_{11}^{2}{SO}_{12}^{7}{SU}_{2}^{6}&104&37&0&143\cr
&{SU}_{2}{SU}_{2b}&8&1&11&E_{8}^{2}G_{2}^{3}{SO}_{9}^{7}{SU}_{2}^{11}&61&33&5&101\cr
&{SU}_{4}&9&1&12&E_{8}^{2}G_{2}^{3}{SO}_{9}^{2}{SO}_{10}^{9}{SU}_{2}^{14}&89&37&0&128\cr
&{SO}_{7}&9&1&12&E_{8}^{2}G_{2}^{3}{SO}_{9}^{11}{SU}_{2}^{14}&80&37&9&128\cr
&{Sp}_{3}&9&1&12&E_{8}^{2}G_{2}^{10}{SU}_{2}^{5}&41&33&14&90\cr
&{SU}_{2}{SU}_{3}&9&1&12&E_{8}^{2}G_{2}{SU}_{2}^{4}{SU}_{3}^{2}{SU}_{4}^{2}{SU}_{5}^{2}{SU}_{6}^{7}&75&25&0&102\cr
&{SU}_{3}{SU}_{2b}&9&1&12&E_{8}^{2}G_{2}{SU}_{2}^{4}{SU}_{3}^{2}{SU}_{4}^{2}{SU}_{5}^{2}{SU}_{6}^{7}&75&25&0&102\cr
&{SU}_{3}{SU}_{2c}&9&1&12&E_{8}^{2}G_{2}{SU}_{2}^{4}{SU}_{3}^{2}{SU}_{4}^{2}{SU}_{5}^{2}{SU}_{6}^{7}&75&25&0&102\cr
&{SO}_{5}{SU}_{2}&9&1&12&E_{8}^{2}G_{2}^{3}{SO}_{7}^{7}{SU}_{2}^{5}&48&33&7&90\cr
&G_{2}{SU}_{2}&9&1&12&E_{8}^{2}G_{2}{SO}_{5}^{4}{Sp}_{3}^{7}{SU}_{2}^{6}&53&25&22&102\cr
&{SU}_{5}&10&1&13&E_{8}^{2}G_{2}{SU}_{2}^{4}{SU}_{3}^{2}{SU}_{4}^{2}{SU}_{5}^{9}&68&25&0&95\cr
&{SO}_{9}&10&1&13&E_{8}^{2}G_{2}^{3}{SO}_{7}^{11}{SU}_{2}^{4}&59&37&11&109\cr
&F_{4}&10&1&13&E_{8}^{2}G_{2}^{14}{SU}_{2}^{4}&48&37&22&109\cr
&{SU}_{3}^{2}&10&1&13&E_{8}^{2}G_{2}{SU}_{2}^{6}{SU}_{3}^{8}&40&25&0&67\cr
&G_{2}{SU}_{3}&10&7&19&E_{8}^{2}G_{2}{SU}_{2}^{4}{SU}_{3}&24&19&10&55\cr
&{SU}_{2}{SU}_{4}&10&1&13&E_{8}^{2}G_{2}{SU}_{2}^{5}{SU}_{3}^{2}{SU}_{4}^{7}&48&25&0&75\cr
&{SO}_{7}{SU}_{2}&10&1&13&E_{8}^{2}G_{2}{SO}_{5}^{7}{SU}_{2}^{7}&39&25&9&75\cr
&{SU}_{6}&11&1&14&E_{8}^{2}G_{2}{SU}_{2}^{5}{SU}_{3}^{9}&41&25&0&68\cr
&{SU}_{6b}&11&1&14&E_{8}^{2}G_{2}{SU}_{2}^{13}{SU}_{3}&33&25&0&60\cr
&{SU}_{6c}&11&9&22&E_{8}^{2}G_{2}{SU}_{2}^{5}{SU}_{3}&25&25&0&52\cr
&{SO}_{10}&11&1&14&E_{8}^{2}G_{2}{SU}_{2}^{4}{SU}_{3}^{2}{SU}_{4}^{11}&59&25&0&86\cr
&{SO}_{11}&11&1&14&E_{8}^{2}G_{2}{SO}_{5}^{11}{SU}_{2}^{6}&46&25&13&86\cr
&{SO}_{9}{SU}_{2}&11&1&14&E_{8}^{2}G_{2}{SU}_{2}^{10}&28&21&5&56\cr
&F_{4}{SU}_{2}&11&7&20&E_{8}^{2}G_{2}{SU}_{2}^{3}&21&17&10&50\cr
&{SO}_{12}&12&1&15&E_{8}^{2}G_{2}{SU}_{2}^{14}&32&25&0&59\cr
&{SO}_{13}&12&11&25&E_{8}^{2}G_{2}{SU}_{2}^{3}&21&21&5&49\cr
&E_{6}&12&1&15&E_{8}^{2}G_{2}{SU}_{2}^{4}{SU}_{3}^{13}&48&25&0&75\cr
&E_{6b}&12&13&27&E_{8}^{2}G_{2}{SU}_{2}^{4}{SU}_{3}&24&25&0&51\cr
&E_{7}&13&1&16&E_{8}^{2}G_{2}{SU}_{2}^{17}&35&25&0&62\cr
&E_{7b}&13&15&30&E_{8}^{2}G_{2}{SU}_{2}^{3}&21&25&0&48\cr
&E_{8}&14&19&35&E_{8}^{2}G_{2}{SU}_{2}^{2}&20&25&0&47\cr
}
\def\tablebodyh{%
&{SU}_{1}&7&1&10&E_{8}^{12}F_{4}^{11}G_{2}^{22}{SU}_{2}^{24}&208&138&0&348\cr
&{SU}_{2}&8&1&11&E_{7}^{10}E_{8}^{2}F_{4}^{2}G_{2}^{4}{SO}_{7}^{9}{SU}_{2}^{24}&153&80&0&235\cr
&{SU}_{2b}&8&1&11&E_{7}^{10}E_{8}^{2}F_{4}^{2}G_{2}^{4}{SO}_{7}^{9}{SU}_{2}^{24}&153&80&0&235\cr
&{SU}_{2c}&8&1&11&E_{8}^{2}G_{2}^{2}{SO}_{5}^{2}{SO}_{9}^{2}{SO}_{11}^{2}{SO}_{13}^{8}{Sp}_{3}^{7}{SU}_{2}^{6}&117&38&6&163\cr
&{SU}_{2d}&8&1&11&E_{8}^{2}F_{4}^{8}G_{2}^{2}{SU}_{2}^{12}&64&48&19&133\cr
&{SU}_{3}&9&1&12&E_{6}^{10}E_{8}^{2}F_{4}^{2}G_{2}^{2}{SU}_{2}^{4}{SU}_{3}^{11}&114&60&0&176\cr
&{SU}_{3b}&9&1&12&E_{6}^{10}E_{8}^{2}F_{4}^{2}G_{2}^{2}{SU}_{2}^{4}{SU}_{3}^{11}&114&60&0&176\cr
&{SU}_{3c}&9&1&12&E_{8}^{2}G_{2}^{8}{SU}_{2}^{6}{SU}_{3}&40&32&12&86\cr
&{SO}_{5}&9&1&12&E_{8}^{2}G_{2}^{2}{SO}_{5}^{9}{SO}_{9}^{2}{SO}_{11}^{10}{SU}_{2}^{6}&102&38&8&150\cr
&G_{2}&9&1&12&E_{8}^{2}F_{4}^{12}G_{2}^{2}{SU}_{2}^{15}&83&60&31&176\cr
&{SU}_{2}^{2}&9&1&12&E_{8}^{2}G_{2}^{2}{SO}_{5}^{9}{SO}_{9}^{2}{SO}_{11}^{2}{SO}_{12}^{8}{SU}_{2}^{6}&110&38&0&150\cr
&{SU}_{2}{SU}_{2b}&9&1&12&E_{8}^{2}G_{2}^{2}{SO}_{9}^{8}{SU}_{2}^{12}&64&34&6&106\cr
&{SU}_{4}&10&1&13&E_{8}^{2}G_{2}^{2}{SO}_{9}^{2}{SO}_{10}^{10}{SU}_{2}^{15}&93&38&0&133\cr
&{SO}_{7}&10&1&13&E_{8}^{2}G_{2}^{2}{SO}_{9}^{12}{SU}_{2}^{15}&83&38&10&133\cr
&{Sp}_{3}&10&1&13&E_{8}^{2}G_{2}^{10}{SU}_{2}^{5}&41&34&16&93\cr
&{SU}_{2}{SU}_{3}&10&1&13&E_{8}^{2}{SU}_{2}^{4}{SU}_{3}^{2}{SU}_{4}^{2}{SU}_{5}^{2}{SU}_{6}^{8}&78&25&0&105\cr
&{SU}_{3}{SU}_{2b}&10&1&13&E_{8}^{2}{SU}_{2}^{4}{SU}_{3}^{2}{SU}_{4}^{2}{SU}_{5}^{2}{SU}_{6}^{8}&78&25&0&105\cr
&{SU}_{3}{SU}_{2c}&10&1&13&E_{8}^{2}{SU}_{2}^{4}{SU}_{3}^{2}{SU}_{4}^{2}{SU}_{5}^{2}{SU}_{6}^{8}&78&25&0&105\cr
&{SO}_{5}{SU}_{2}&10&1&13&E_{8}^{2}G_{2}^{2}{SO}_{7}^{8}{SU}_{2}^{5}&49&34&8&93\cr
&G_{2}{SU}_{2}&10&1&13&E_{8}^{2}{SO}_{5}^{4}{Sp}_{3}^{8}{SU}_{2}^{6}&54&25&24&105\cr
&{SU}_{5}&11&1&14&E_{8}^{2}{SU}_{2}^{4}{SU}_{3}^{2}{SU}_{4}^{2}{SU}_{5}^{10}&70&25&0&97\cr
&{SO}_{9}&11&1&14&E_{8}^{2}G_{2}^{2}{SO}_{7}^{12}{SU}_{2}^{4}&60&38&12&112\cr
&F_{4}&11&1&14&E_{8}^{2}G_{2}^{14}{SU}_{2}^{4}&48&38&24&112\cr
&{SU}_{3}^{2}&11&1&14&E_{8}^{2}{SU}_{2}^{6}{SU}_{3}^{9}&40&25&0&67\cr
&G_{2}{SU}_{3}&11&8&21&E_{8}^{2}{SU}_{2}^{4}{SU}_{3}&22&19&10&53\cr
&{SU}_{2}{SU}_{4}&11&1&14&E_{8}^{2}{SU}_{2}^{5}{SU}_{3}^{2}{SU}_{4}^{8}&49&25&0&76\cr
&{SO}_{7}{SU}_{2}&11&1&14&E_{8}^{2}{SO}_{5}^{8}{SU}_{2}^{7}&39&25&10&76\cr
&{SU}_{6}&12&1&15&E_{8}^{2}{SU}_{2}^{5}{SU}_{3}^{10}&41&25&0&68\cr
&{SU}_{6b}&12&1&15&E_{8}^{2}{SU}_{2}^{14}{SU}_{3}&32&25&0&59\cr
&{SU}_{6c}&12&10&24&E_{8}^{2}{SU}_{2}^{5}{SU}_{3}&23&25&0&50\cr
&{SO}_{10}&12&1&15&E_{8}^{2}{SU}_{2}^{4}{SU}_{3}^{2}{SU}_{4}^{12}&60&25&0&87\cr
&{SO}_{11}&12&1&15&E_{8}^{2}{SO}_{5}^{12}{SU}_{2}^{6}&46&25&14&87\cr
&{SO}_{9}{SU}_{2}&12&1&15&E_{8}^{2}{SU}_{2}^{11}&27&21&5&55\cr
&F_{4}{SU}_{2}&12&8&22&E_{8}^{2}{SU}_{2}^{3}&19&17&10&48\cr
&{SO}_{12}&13&1&16&E_{8}^{2}{SU}_{2}^{15}&31&25&0&58\cr
&{SO}_{13}&13&12&27&E_{8}^{2}{SU}_{2}^{3}&19&21&5&47\cr
&E_{6}&13&1&16&E_{8}^{2}{SU}_{2}^{4}{SU}_{3}^{14}&48&25&0&75\cr
&E_{6b}&13&14&29&E_{8}^{2}{SU}_{2}^{4}{SU}_{3}&22&25&0&49\cr
&E_{7}&14&1&17&E_{8}^{2}{SU}_{2}^{18}&34&25&0&61\cr
&E_{7b}&14&16&32&E_{8}^{2}{SU}_{2}^{3}&19&25&0&46\cr
&E_{8}&15&20&37&E_{8}^{2}{SU}_{2}^{2}&18&25&0&45\cr
}
\def\tablebodyi{%
&{SU}_{1}&7&1&10&E_{8}^{13}F_{4}^{12}G_{2}^{24}{SU}_{2}^{25}&225&149&0&376\cr
&{SU}_{2}&8&1&11&E_{7}^{11}E_{8}^{2}F_{4}^{2}G_{2}^{4}{SO}_{7}^{10}{SU}_{2}^{25}&164&85&0&251\cr
&{SU}_{2b}&8&1&11&E_{7}^{11}E_{8}^{2}F_{4}^{2}G_{2}^{4}{SO}_{7}^{10}{SU}_{2}^{25}&164&85&0&251\cr
&{SU}_{2c}&8&1&11&E_{8}^{2}G_{2}^{2}{SO}_{5}^{2}{SO}_{9}^{2}{SO}_{11}^{2}{SO}_{13}^{9}{Sp}_{3}^{8}{SU}_{2}^{5}&125&39&7&173\cr
&{SU}_{2d}&8&1&11&E_{8}^{2}F_{4}^{9}G_{2}^{2}{SU}_{2}^{12}&68&51&22&143\cr
&{SU}_{3}&9&1&12&E_{6}^{11}E_{8}^{2}F_{4}^{2}G_{2}^{2}{SU}_{2}^{3}{SU}_{3}^{12}&121&63&0&186\cr
&{SU}_{3b}&9&1&12&E_{6}^{11}E_{8}^{2}F_{4}^{2}G_{2}^{2}{SU}_{2}^{3}{SU}_{3}^{12}&121&63&0&186\cr
&{SU}_{3c}&9&1&12&E_{8}^{2}G_{2}^{9}{SU}_{2}^{5}{SU}_{3}&41&33&14&90\cr
&{SO}_{5}&9&1&12&E_{8}^{2}G_{2}^{2}{SO}_{5}^{10}{SO}_{9}^{2}{SO}_{11}^{11}{SU}_{2}^{5}&108&39&9&158\cr
&G_{2}&9&1&12&E_{8}^{2}F_{4}^{13}G_{2}^{2}{SU}_{2}^{15}&87&63&34&186\cr
&{SU}_{2}^{2}&9&1&12&E_{8}^{2}G_{2}^{2}{SO}_{5}^{10}{SO}_{9}^{2}{SO}_{11}^{2}{SO}_{12}^{9}{SU}_{2}^{5}&117&39&0&158\cr
&{SU}_{2}{SU}_{2b}&9&1&12&E_{8}^{2}G_{2}^{2}{SO}_{9}^{9}{SU}_{2}^{12}&68&35&7&112\cr
&{SU}_{4}&10&1&13&E_{8}^{2}G_{2}^{2}{SO}_{9}^{2}{SO}_{10}^{11}{SU}_{2}^{15}&98&39&0&139\cr
&{SO}_{7}&10&1&13&E_{8}^{2}G_{2}^{2}{SO}_{9}^{13}{SU}_{2}^{15}&87&39&11&139\cr
&{Sp}_{3}&10&1&13&E_{8}^{2}G_{2}^{11}{SU}_{2}^{4}&42&35&18&97\cr
&{SU}_{2}{SU}_{3}&10&1&13&E_{8}^{2}{SU}_{2}^{3}{SU}_{3}^{2}{SU}_{4}^{2}{SU}_{5}^{2}{SU}_{6}^{9}&82&25&0&109\cr
&{SU}_{3}{SU}_{2b}&10&1&13&E_{8}^{2}{SU}_{2}^{3}{SU}_{3}^{2}{SU}_{4}^{2}{SU}_{5}^{2}{SU}_{6}^{9}&82&25&0&109\cr
&{SU}_{3}{SU}_{2c}&10&1&13&E_{8}^{2}{SU}_{2}^{3}{SU}_{3}^{2}{SU}_{4}^{2}{SU}_{5}^{2}{SU}_{6}^{9}&82&25&0&109\cr
&{SO}_{5}{SU}_{2}&10&1&13&E_{8}^{2}G_{2}^{2}{SO}_{7}^{9}{SU}_{2}^{4}&51&35&9&97\cr
&G_{2}{SU}_{2}&10&1&13&E_{8}^{2}{SO}_{5}^{4}{Sp}_{3}^{9}{SU}_{2}^{5}&56&25&26&109\cr
&{SU}_{5}&11&1&14&E_{8}^{2}{SU}_{2}^{3}{SU}_{3}^{2}{SU}_{4}^{2}{SU}_{5}^{11}&73&25&0&100\cr
&{SO}_{9}&11&1&14&E_{8}^{2}G_{2}^{2}{SO}_{7}^{13}{SU}_{2}^{3}&62&39&13&116\cr
&F_{4}&11&1&14&E_{8}^{2}G_{2}^{15}{SU}_{2}^{3}&49&39&26&116\cr
&{SU}_{3}^{2}&11&1&14&E_{8}^{2}{SU}_{2}^{5}{SU}_{3}^{10}&41&25&0&68\cr
&G_{2}{SU}_{3}&11&9&22&E_{8}^{2}{SU}_{2}^{3}{SU}_{3}&21&19&10&52\cr
&{SU}_{2}{SU}_{4}&11&1&14&E_{8}^{2}{SU}_{2}^{4}{SU}_{3}^{2}{SU}_{4}^{9}&51&25&0&78\cr
&{SO}_{7}{SU}_{2}&11&1&14&E_{8}^{2}{SO}_{5}^{9}{SU}_{2}^{6}&40&25&11&78\cr
&{SU}_{6}&12&1&15&E_{8}^{2}{SU}_{2}^{4}{SU}_{3}^{11}&42&25&0&69\cr
&{SU}_{6b}&12&1&15&E_{8}^{2}{SU}_{2}^{14}{SU}_{3}&32&25&0&59\cr
&{SU}_{6c}&12&11&25&E_{8}^{2}{SU}_{2}^{4}{SU}_{3}&22&25&0&49\cr
&{SO}_{10}&12&1&15&E_{8}^{2}{SU}_{2}^{3}{SU}_{3}^{2}{SU}_{4}^{13}&62&25&0&89\cr
&{SO}_{11}&12&1&15&E_{8}^{2}{SO}_{5}^{13}{SU}_{2}^{5}&47&25&15&89\cr
&{SO}_{9}{SU}_{2}&12&1&15&E_{8}^{2}{SU}_{2}^{11}&27&21&5&55\cr
&F_{4}{SU}_{2}&12&9&23&E_{8}^{2}{SU}_{2}^{2}&18&17&10&47\cr
&{SO}_{12}&13&1&16&E_{8}^{2}{SU}_{2}^{15}&31&25&0&58\cr
&{SO}_{13}&13&13&28&E_{8}^{2}{SU}_{2}^{2}&18&21&5&46\cr
&E_{6}&13&1&16&E_{8}^{2}{SU}_{2}^{3}{SU}_{3}^{15}&49&25&0&76\cr
&E_{6b}&13&15&30&E_{8}^{2}{SU}_{2}^{3}{SU}_{3}&21&25&0&48\cr
&E_{7}&14&1&17&E_{8}^{2}{SU}_{2}^{18}&34&25&0&61\cr
&E_{7b}&14&17&33&E_{8}^{2}{SU}_{2}^{2}&18&25&0&45\cr
&E_{8}&15&21&38&E_{8}^{2}{SU}_{2}&17&25&0&44\cr
}
\def\tablebodyj{%
&{SU}_{1}&8&4&14&E_{8}^{14}F_{4}^{13}G_{2}^{26}{SU}_{2}^{26}&242&160&0&404\cr
&{SU}_{2}&9&4&15&E_{7}^{12}E_{8}^{2}F_{4}^{2}G_{2}^{4}{SO}_{7}^{11}{SU}_{2}^{26}&175&90&0&267\cr
&{SU}_{2b}&9&4&15&E_{7}^{12}E_{8}^{2}F_{4}^{2}G_{2}^{4}{SO}_{7}^{11}{SU}_{2}^{26}&175&90&0&267\cr
&{SU}_{2c}&9&4&15&E_{8}^{2}G_{2}^{2}{SO}_{5}^{2}{SO}_{9}^{2}{SO}_{11}^{2}{SO}_{13}^{10}{Sp}_{3}^{9}{SU}_{2}^{4}&133&40&8&183\cr
&{SU}_{2d}&9&4&15&E_{8}^{2}F_{4}^{10}G_{2}^{2}{SU}_{2}^{12}&72&54&25&153\cr
&{SU}_{3}&10&4&16&E_{6}^{12}E_{8}^{2}F_{4}^{2}G_{2}^{2}{SU}_{2}^{2}{SU}_{3}^{13}&128&66&0&196\cr
&{SU}_{3b}&10&4&16&E_{6}^{12}E_{8}^{2}F_{4}^{2}G_{2}^{2}{SU}_{2}^{2}{SU}_{3}^{13}&128&66&0&196\cr
&{SU}_{3c}&10&4&16&E_{8}^{2}G_{2}^{10}{SU}_{2}^{4}{SU}_{3}&42&34&16&94\cr
&{SO}_{5}&10&4&16&E_{8}^{2}G_{2}^{2}{SO}_{5}^{11}{SO}_{9}^{2}{SO}_{11}^{12}{SU}_{2}^{4}&114&40&10&166\cr
&G_{2}&10&4&16&E_{8}^{2}F_{4}^{14}G_{2}^{2}{SU}_{2}^{15}&91&66&37&196\cr
&{SU}_{2}^{2}&10&4&16&E_{8}^{2}G_{2}^{2}{SO}_{5}^{11}{SO}_{9}^{2}{SO}_{11}^{2}{SO}_{12}^{10}{SU}_{2}^{4}&124&40&0&166\cr
&{SU}_{2}{SU}_{2b}&10&4&16&E_{8}^{2}G_{2}^{2}{SO}_{9}^{10}{SU}_{2}^{12}&72&36&8&118\cr
&{SU}_{4}&11&4&17&E_{8}^{2}G_{2}^{2}{SO}_{9}^{2}{SO}_{10}^{12}{SU}_{2}^{15}&103&40&0&145\cr
&{SO}_{7}&11&4&17&E_{8}^{2}G_{2}^{2}{SO}_{9}^{14}{SU}_{2}^{15}&91&40&12&145\cr
&{Sp}_{3}&11&4&17&E_{8}^{2}G_{2}^{12}{SU}_{2}^{3}&43&36&20&101\cr
&{SU}_{2}{SU}_{3}&11&4&17&E_{8}^{2}{SU}_{2}^{2}{SU}_{3}^{2}{SU}_{4}^{2}{SU}_{5}^{2}{SU}_{6}^{10}&86&25&0&113\cr
&{SU}_{3}{SU}_{2b}&11&4&17&E_{8}^{2}{SU}_{2}^{2}{SU}_{3}^{2}{SU}_{4}^{2}{SU}_{5}^{2}{SU}_{6}^{10}&86&25&0&113\cr
&{SU}_{3}{SU}_{2c}&11&4&17&E_{8}^{2}{SU}_{2}^{2}{SU}_{3}^{2}{SU}_{4}^{2}{SU}_{5}^{2}{SU}_{6}^{10}&86&25&0&113\cr
&{SO}_{5}{SU}_{2}&11&4&17&E_{8}^{2}G_{2}^{2}{SO}_{7}^{10}{SU}_{2}^{3}&53&36&10&101\cr
&G_{2}{SU}_{2}&11&4&17&E_{8}^{2}{SO}_{5}^{4}{Sp}_{3}^{10}{SU}_{2}^{4}&58&25&28&113\cr
&{SU}_{5}&12&4&18&E_{8}^{2}{SU}_{2}^{2}{SU}_{3}^{2}{SU}_{4}^{2}{SU}_{5}^{12}&76&25&0&103\cr
&{SO}_{9}&12&4&18&E_{8}^{2}G_{2}^{2}{SO}_{7}^{14}{SU}_{2}^{2}&64&40&14&120\cr
&F_{4}&12&4&18&E_{8}^{2}G_{2}^{16}{SU}_{2}^{2}&50&40&28&120\cr
&{SU}_{3}^{2}&12&4&18&E_{8}^{2}{SU}_{2}^{4}{SU}_{3}^{11}&42&25&0&69\cr
&G_{2}{SU}_{3}&12&13&27&E_{8}^{2}{SU}_{2}^{2}{SU}_{3}&20&19&10&51\cr
&{SU}_{2}{SU}_{4}&12&4&18&E_{8}^{2}{SU}_{2}^{3}{SU}_{3}^{2}{SU}_{4}^{10}&53&25&0&80\cr
&{SO}_{7}{SU}_{2}&12&4&18&E_{8}^{2}{SO}_{5}^{10}{SU}_{2}^{5}&41&25&12&80\cr
&{SU}_{6}&13&4&19&E_{8}^{2}{SU}_{2}^{3}{SU}_{3}^{12}&43&25&0&70\cr
&{SU}_{6b}&13&4&19&E_{8}^{2}{SU}_{2}^{14}{SU}_{3}&32&25&0&59\cr
&{SU}_{6c}&13&15&30&E_{8}^{2}{SU}_{2}^{3}{SU}_{3}&21&25&0&48\cr
&{SO}_{10}&13&4&19&E_{8}^{2}{SU}_{2}^{2}{SU}_{3}^{2}{SU}_{4}^{14}&64&25&0&91\cr
&{SO}_{11}&13&4&19&E_{8}^{2}{SO}_{5}^{14}{SU}_{2}^{4}&48&25&16&91\cr
&{SO}_{9}{SU}_{2}&13&4&19&E_{8}^{2}{SU}_{2}^{11}&27&21&5&55\cr
&F_{4}{SU}_{2}&13&13&28&E_{8}^{2}{SU}_{2}&17&17&10&46\cr
&{SO}_{12}&14&4&20&E_{8}^{2}{SU}_{2}^{15}&31&25&0&58\cr
&{SO}_{13}&14&17&33&E_{8}^{2}{SU}_{2}&17&21&5&45\cr
&E_{6}&14&4&20&E_{8}^{2}{SU}_{2}^{2}{SU}_{3}^{16}&50&25&0&77\cr
&E_{6b}&14&19&35&E_{8}^{2}{SU}_{2}^{2}{SU}_{3}&20&25&0&47\cr
&E_{7}&15&4&21&E_{8}^{2}{SU}_{2}^{18}&34&25&0&61\cr
&E_{7b}&15&21&38&E_{8}^{2}{SU}_{2}&17&25&0&44\cr
&E_{8}&16&25&43&E_{8}^{2}&16&25&0&43\cr
}
\def\tablebodyk{%
&{SU}_{1}&8&3&13&E_{8}^{15}F_{4}^{14}G_{2}^{28}{SU}_{2}^{28}&260&171&0&433\cr
&{SU}_{2}&9&3&14&E_{7}^{13}E_{8}^{2}F_{4}^{2}G_{2}^{4}{SO}_{7}^{12}{SU}_{2}^{28}&187&95&0&284\cr
&{SU}_{2b}&9&3&14&E_{7}^{13}E_{8}^{2}F_{4}^{2}G_{2}^{4}{SO}_{7}^{12}{SU}_{2}^{28}&187&95&0&284\cr
&{SU}_{2c}&9&3&14&E_{8}^{2}G_{2}^{2}{SO}_{5}^{2}{SO}_{9}^{2}{SO}_{11}^{2}{SO}_{13}^{11}{Sp}_{3}^{10}{SU}_{2}^{4}&142&41&9&194\cr
&{SU}_{2d}&9&3&14&E_{8}^{2}F_{4}^{11}G_{2}^{2}{SU}_{2}^{13}&77&57&28&164\cr
&{SU}_{3}&10&3&15&E_{6}^{13}E_{8}^{2}F_{4}^{2}G_{2}^{2}{SU}_{2}^{2}{SU}_{3}^{14}&136&69&0&207\cr
&{SU}_{3b}&10&3&15&E_{6}^{13}E_{8}^{2}F_{4}^{2}G_{2}^{2}{SU}_{2}^{2}{SU}_{3}^{14}&136&69&0&207\cr
&{SU}_{3c}&10&3&15&E_{8}^{2}G_{2}^{11}{SU}_{2}^{4}{SU}_{3}&44&35&18&99\cr
&{SO}_{5}&10&3&15&E_{8}^{2}G_{2}^{2}{SO}_{5}^{12}{SO}_{9}^{2}{SO}_{11}^{13}{SU}_{2}^{4}&121&41&11&175\cr
&G_{2}&10&3&15&E_{8}^{2}F_{4}^{15}G_{2}^{2}{SU}_{2}^{16}&96&69&40&207\cr
&{SU}_{2}^{2}&10&3&15&E_{8}^{2}G_{2}^{2}{SO}_{5}^{12}{SO}_{9}^{2}{SO}_{11}^{2}{SO}_{12}^{11}{SU}_{2}^{4}&132&41&0&175\cr
&{SU}_{2}{SU}_{2b}&10&3&15&E_{8}^{2}G_{2}^{2}{SO}_{9}^{11}{SU}_{2}^{13}&77&37&9&125\cr
&{SU}_{4}&11&3&16&E_{8}^{2}G_{2}^{2}{SO}_{9}^{2}{SO}_{10}^{13}{SU}_{2}^{16}&109&41&0&152\cr
&{SO}_{7}&11&3&16&E_{8}^{2}G_{2}^{2}{SO}_{9}^{15}{SU}_{2}^{16}&96&41&13&152\cr
&{Sp}_{3}&11&3&16&E_{8}^{2}G_{2}^{13}{SU}_{2}^{3}&45&37&22&106\cr
&{SU}_{2}{SU}_{3}&11&3&16&E_{8}^{2}{SU}_{2}^{2}{SU}_{3}^{2}{SU}_{4}^{2}{SU}_{5}^{2}{SU}_{6}^{11}&91&25&0&118\cr
&{SU}_{3}{SU}_{2b}&11&3&16&E_{8}^{2}{SU}_{2}^{2}{SU}_{3}^{2}{SU}_{4}^{2}{SU}_{5}^{2}{SU}_{6}^{11}&91&25&0&118\cr
&{SU}_{3}{SU}_{2c}&11&3&16&E_{8}^{2}{SU}_{2}^{2}{SU}_{3}^{2}{SU}_{4}^{2}{SU}_{5}^{2}{SU}_{6}^{11}&91&25&0&118\cr
&{SO}_{5}{SU}_{2}&11&3&16&E_{8}^{2}G_{2}^{2}{SO}_{7}^{11}{SU}_{2}^{3}&56&37&11&106\cr
&G_{2}{SU}_{2}&11&3&16&E_{8}^{2}{SO}_{5}^{4}{Sp}_{3}^{11}{SU}_{2}^{4}&61&25&30&118\cr
&{SU}_{5}&12&3&17&E_{8}^{2}{SU}_{2}^{2}{SU}_{3}^{2}{SU}_{4}^{2}{SU}_{5}^{13}&80&25&0&107\cr
&{SO}_{9}&12&3&17&E_{8}^{2}G_{2}^{2}{SO}_{7}^{15}{SU}_{2}^{2}&67&41&15&125\cr
&F_{4}&12&3&17&E_{8}^{2}G_{2}^{17}{SU}_{2}^{2}&52&41&30&125\cr
&{SU}_{3}^{2}&12&3&17&E_{8}^{2}{SU}_{2}^{4}{SU}_{3}^{12}&44&25&0&71\cr
&G_{2}{SU}_{3}&12&13&27&E_{8}^{2}{SU}_{2}^{2}{SU}_{3}&20&19&10&51\cr
&{SU}_{2}{SU}_{4}&12&3&17&E_{8}^{2}{SU}_{2}^{3}{SU}_{3}^{2}{SU}_{4}^{11}&56&25&0&83\cr
&{SO}_{7}{SU}_{2}&12&3&17&E_{8}^{2}{SO}_{5}^{11}{SU}_{2}^{5}&43&25&13&83\cr
&{SU}_{6}&13&3&18&E_{8}^{2}{SU}_{2}^{3}{SU}_{3}^{13}&45&25&0&72\cr
&{SU}_{6b}&13&3&18&E_{8}^{2}{SU}_{2}^{15}{SU}_{3}&33&25&0&60\cr
&{SU}_{6c}&13&15&30&E_{8}^{2}{SU}_{2}^{3}{SU}_{3}&21&25&0&48\cr
&{SO}_{10}&13&3&18&E_{8}^{2}{SU}_{2}^{2}{SU}_{3}^{2}{SU}_{4}^{15}&67&25&0&94\cr
&{SO}_{11}&13&3&18&E_{8}^{2}{SO}_{5}^{15}{SU}_{2}^{4}&50&25&17&94\cr
&{SO}_{9}{SU}_{2}&13&3&18&E_{8}^{2}{SU}_{2}^{12}&28&21&5&56\cr
&F_{4}{SU}_{2}&13&13&28&E_{8}^{2}{SU}_{2}&17&17&10&46\cr
&{SO}_{12}&14&3&19&E_{8}^{2}{SU}_{2}^{16}&32&25&0&59\cr
&{SO}_{13}&14&17&33&E_{8}^{2}{SU}_{2}&17&21&5&45\cr
&E_{6}&14&3&19&E_{8}^{2}{SU}_{2}^{2}{SU}_{3}^{17}&52&25&0&79\cr
&E_{6b}&14&19&35&E_{8}^{2}{SU}_{2}^{2}{SU}_{3}&20&25&0&47\cr
&E_{7}&15&3&20&E_{8}^{2}{SU}_{2}^{19}&35&25&0&62\cr
&E_{7b}&15&21&38&E_{8}^{2}{SU}_{2}&17&25&0&44\cr
&E_{8}&16&25&43&E_{8}^{2}&16&25&0&43\cr
}
\def\tablebodyl{%
&{SU}_{1}&8&2&12&E_{8}^{16}F_{4}^{15}G_{2}^{30}{SU}_{2}^{30}&278&182&0&462\cr
&{SU}_{2}&9&2&13&E_{7}^{14}E_{8}^{2}F_{4}^{2}G_{2}^{4}{SO}_{7}^{13}{SU}_{2}^{30}&199&100&0&301\cr
&{SU}_{2b}&9&2&13&E_{7}^{14}E_{8}^{2}F_{4}^{2}G_{2}^{4}{SO}_{7}^{13}{SU}_{2}^{30}&199&100&0&301\cr
&{SU}_{2c}&9&2&13&E_{8}^{2}G_{2}^{2}{SO}_{5}^{2}{SO}_{9}^{2}{SO}_{11}^{2}{SO}_{13}^{12}{Sp}_{3}^{11}{SU}_{2}^{4}&151&42&10&205\cr
&{SU}_{2d}&9&2&13&E_{8}^{2}F_{4}^{12}G_{2}^{2}{SU}_{2}^{14}&82&60&31&175\cr
&{SU}_{3}&10&2&14&E_{6}^{14}E_{8}^{2}F_{4}^{2}G_{2}^{2}{SU}_{2}^{2}{SU}_{3}^{15}&144&72&0&218\cr
&{SU}_{3b}&10&2&14&E_{6}^{14}E_{8}^{2}F_{4}^{2}G_{2}^{2}{SU}_{2}^{2}{SU}_{3}^{15}&144&72&0&218\cr
&{SU}_{3c}&10&2&14&E_{8}^{2}G_{2}^{12}{SU}_{2}^{4}{SU}_{3}&46&36&20&104\cr
&{SO}_{5}&10&2&14&E_{8}^{2}G_{2}^{2}{SO}_{5}^{13}{SO}_{9}^{2}{SO}_{11}^{14}{SU}_{2}^{4}&128&42&12&184\cr
&G_{2}&10&2&14&E_{8}^{2}F_{4}^{16}G_{2}^{2}{SU}_{2}^{17}&101&72&43&218\cr
&{SU}_{2}^{2}&10&2&14&E_{8}^{2}G_{2}^{2}{SO}_{5}^{13}{SO}_{9}^{2}{SO}_{11}^{2}{SO}_{12}^{12}{SU}_{2}^{4}&140&42&0&184\cr
&{SU}_{2}{SU}_{2b}&10&2&14&E_{8}^{2}G_{2}^{2}{SO}_{9}^{12}{SU}_{2}^{14}&82&38&10&132\cr
&{SU}_{4}&11&2&15&E_{8}^{2}G_{2}^{2}{SO}_{9}^{2}{SO}_{10}^{14}{SU}_{2}^{17}&115&42&0&159\cr
&{SO}_{7}&11&2&15&E_{8}^{2}G_{2}^{2}{SO}_{9}^{16}{SU}_{2}^{17}&101&42&14&159\cr
&{Sp}_{3}&11&2&15&E_{8}^{2}G_{2}^{14}{SU}_{2}^{3}&47&38&24&111\cr
&{SU}_{2}{SU}_{3}&11&2&15&E_{8}^{2}{SU}_{2}^{2}{SU}_{3}^{2}{SU}_{4}^{2}{SU}_{5}^{2}{SU}_{6}^{12}&96&25&0&123\cr
&{SU}_{3}{SU}_{2b}&11&2&15&E_{8}^{2}{SU}_{2}^{2}{SU}_{3}^{2}{SU}_{4}^{2}{SU}_{5}^{2}{SU}_{6}^{12}&96&25&0&123\cr
&{SU}_{3}{SU}_{2c}&11&2&15&E_{8}^{2}{SU}_{2}^{2}{SU}_{3}^{2}{SU}_{4}^{2}{SU}_{5}^{2}{SU}_{6}^{12}&96&25&0&123\cr
&{SO}_{5}{SU}_{2}&11&2&15&E_{8}^{2}G_{2}^{2}{SO}_{7}^{12}{SU}_{2}^{3}&59&38&12&111\cr
&G_{2}{SU}_{2}&11&2&15&E_{8}^{2}{SO}_{5}^{4}{Sp}_{3}^{12}{SU}_{2}^{4}&64&25&32&123\cr
&{SU}_{5}&12&2&16&E_{8}^{2}{SU}_{2}^{2}{SU}_{3}^{2}{SU}_{4}^{2}{SU}_{5}^{14}&84&25&0&111\cr
&{SO}_{9}&12&2&16&E_{8}^{2}G_{2}^{2}{SO}_{7}^{16}{SU}_{2}^{2}&70&42&16&130\cr
&F_{4}&12&2&16&E_{8}^{2}G_{2}^{18}{SU}_{2}^{2}&54&42&32&130\cr
&{SU}_{3}^{2}&12&2&16&E_{8}^{2}{SU}_{2}^{4}{SU}_{3}^{13}&46&25&0&73\cr
&G_{2}{SU}_{3}&12&13&27&E_{8}^{2}{SU}_{2}^{2}{SU}_{3}&20&19&10&51\cr
&{SU}_{2}{SU}_{4}&12&2&16&E_{8}^{2}{SU}_{2}^{3}{SU}_{3}^{2}{SU}_{4}^{12}&59&25&0&86\cr
&{SO}_{7}{SU}_{2}&12&2&16&E_{8}^{2}{SO}_{5}^{12}{SU}_{2}^{5}&45&25&14&86\cr
&{SU}_{6}&13&2&17&E_{8}^{2}{SU}_{2}^{3}{SU}_{3}^{14}&47&25&0&74\cr
&{SU}_{6b}&13&2&17&E_{8}^{2}{SU}_{2}^{16}{SU}_{3}&34&25&0&61\cr
&{SU}_{6c}&13&15&30&E_{8}^{2}{SU}_{2}^{3}{SU}_{3}&21&25&0&48\cr
&{SO}_{10}&13&2&17&E_{8}^{2}{SU}_{2}^{2}{SU}_{3}^{2}{SU}_{4}^{16}&70&25&0&97\cr
&{SO}_{11}&13&2&17&E_{8}^{2}{SO}_{5}^{16}{SU}_{2}^{4}&52&25&18&97\cr
&{SO}_{9}{SU}_{2}&13&2&17&E_{8}^{2}{SU}_{2}^{13}&29&21&5&57\cr
&F_{4}{SU}_{2}&13&13&28&E_{8}^{2}{SU}_{2}&17&17&10&46\cr
&{SO}_{12}&14&2&18&E_{8}^{2}{SU}_{2}^{17}&33&25&0&60\cr
&{SO}_{13}&14&17&33&E_{8}^{2}{SU}_{2}&17&21&5&45\cr
&E_{6}&14&2&18&E_{8}^{2}{SU}_{2}^{2}{SU}_{3}^{18}&54&25&0&81\cr
&E_{6b}&14&19&35&E_{8}^{2}{SU}_{2}^{2}{SU}_{3}&20&25&0&47\cr
&E_{7}&15&2&19&E_{8}^{2}{SU}_{2}^{20}&36&25&0&63\cr
&E_{7b}&15&21&38&E_{8}^{2}{SU}_{2}&17&25&0&44\cr
&E_{8}&16&25&43&E_{8}^{2}&16&25&0&43\cr
}
\def\tablebodym{%
&{SU}_{1}&8&1&11&E_{8}^{17}F_{4}^{16}G_{2}^{32}{SU}_{2}^{32}&296&193&0&491\cr
&{SU}_{2}&9&1&12&E_{7}^{15}E_{8}^{2}F_{4}^{2}G_{2}^{4}{SO}_{7}^{14}{SU}_{2}^{32}&211&105&0&318\cr
&{SU}_{2b}&9&1&12&E_{7}^{15}E_{8}^{2}F_{4}^{2}G_{2}^{4}{SO}_{7}^{14}{SU}_{2}^{32}&211&105&0&318\cr
&{SU}_{2c}&9&1&12&E_{8}^{2}G_{2}^{2}{SO}_{5}^{2}{SO}_{9}^{2}{SO}_{11}^{2}{SO}_{13}^{13}{Sp}_{3}^{12}{SU}_{2}^{4}&160&43&11&216\cr
&{SU}_{2d}&9&1&12&E_{8}^{2}F_{4}^{13}G_{2}^{2}{SU}_{2}^{15}&87&63&34&186\cr
&{SU}_{3}&10&1&13&E_{6}^{15}E_{8}^{2}F_{4}^{2}G_{2}^{2}{SU}_{2}^{2}{SU}_{3}^{16}&152&75&0&229\cr
&{SU}_{3b}&10&1&13&E_{6}^{15}E_{8}^{2}F_{4}^{2}G_{2}^{2}{SU}_{2}^{2}{SU}_{3}^{16}&152&75&0&229\cr
&{SU}_{3c}&10&1&13&E_{8}^{2}G_{2}^{13}{SU}_{2}^{4}{SU}_{3}&48&37&22&109\cr
&{SO}_{5}&10&1&13&E_{8}^{2}G_{2}^{2}{SO}_{5}^{14}{SO}_{9}^{2}{SO}_{11}^{15}{SU}_{2}^{4}&135&43&13&193\cr
&G_{2}&10&1&13&E_{8}^{2}F_{4}^{17}G_{2}^{2}{SU}_{2}^{18}&106&75&46&229\cr
&{SU}_{2}^{2}&10&1&13&E_{8}^{2}G_{2}^{2}{SO}_{5}^{14}{SO}_{9}^{2}{SO}_{11}^{2}{SO}_{12}^{13}{SU}_{2}^{4}&148&43&0&193\cr
&{SU}_{2}{SU}_{2b}&10&1&13&E_{8}^{2}G_{2}^{2}{SO}_{9}^{13}{SU}_{2}^{15}&87&39&11&139\cr
&{SU}_{4}&11&1&14&E_{8}^{2}G_{2}^{2}{SO}_{9}^{2}{SO}_{10}^{15}{SU}_{2}^{18}&121&43&0&166\cr
&{SO}_{7}&11&1&14&E_{8}^{2}G_{2}^{2}{SO}_{9}^{17}{SU}_{2}^{18}&106&43&15&166\cr
&{Sp}_{3}&11&1&14&E_{8}^{2}G_{2}^{15}{SU}_{2}^{3}&49&39&26&116\cr
&{SU}_{2}{SU}_{3}&11&1&14&E_{8}^{2}{SU}_{2}^{2}{SU}_{3}^{2}{SU}_{4}^{2}{SU}_{5}^{2}{SU}_{6}^{13}&101&25&0&128\cr
&{SU}_{3}{SU}_{2b}&11&1&14&E_{8}^{2}{SU}_{2}^{2}{SU}_{3}^{2}{SU}_{4}^{2}{SU}_{5}^{2}{SU}_{6}^{13}&101&25&0&128\cr
&{SU}_{3}{SU}_{2c}&11&1&14&E_{8}^{2}{SU}_{2}^{2}{SU}_{3}^{2}{SU}_{4}^{2}{SU}_{5}^{2}{SU}_{6}^{13}&101&25&0&128\cr
&{SO}_{5}{SU}_{2}&11&1&14&E_{8}^{2}G_{2}^{2}{SO}_{7}^{13}{SU}_{2}^{3}&62&39&13&116\cr
&G_{2}{SU}_{2}&11&1&14&E_{8}^{2}{SO}_{5}^{4}{Sp}_{3}^{13}{SU}_{2}^{4}&67&25&34&128\cr
&{SU}_{5}&12&1&15&E_{8}^{2}{SU}_{2}^{2}{SU}_{3}^{2}{SU}_{4}^{2}{SU}_{5}^{15}&88&25&0&115\cr
&{SO}_{9}&12&1&15&E_{8}^{2}G_{2}^{2}{SO}_{7}^{17}{SU}_{2}^{2}&73&43&17&135\cr
&F_{4}&12&1&15&E_{8}^{2}G_{2}^{19}{SU}_{2}^{2}&56&43&34&135\cr
&{SU}_{3}^{2}&12&1&15&E_{8}^{2}{SU}_{2}^{4}{SU}_{3}^{14}&48&25&0&75\cr
&G_{2}{SU}_{3}&12&13&27&E_{8}^{2}{SU}_{2}^{2}{SU}_{3}&20&19&10&51\cr
&{SU}_{2}{SU}_{4}&12&1&15&E_{8}^{2}{SU}_{2}^{3}{SU}_{3}^{2}{SU}_{4}^{13}&62&25&0&89\cr
&{SO}_{7}{SU}_{2}&12&1&15&E_{8}^{2}{SO}_{5}^{13}{SU}_{2}^{5}&47&25&15&89\cr
&{SU}_{6}&13&1&16&E_{8}^{2}{SU}_{2}^{3}{SU}_{3}^{15}&49&25&0&76\cr
&{SU}_{6b}&13&1&16&E_{8}^{2}{SU}_{2}^{17}{SU}_{3}&35&25&0&62\cr
&{SU}_{6c}&13&15&30&E_{8}^{2}{SU}_{2}^{3}{SU}_{3}&21&25&0&48\cr
&{SO}_{10}&13&1&16&E_{8}^{2}{SU}_{2}^{2}{SU}_{3}^{2}{SU}_{4}^{17}&73&25&0&100\cr
&{SO}_{11}&13&1&16&E_{8}^{2}{SO}_{5}^{17}{SU}_{2}^{4}&54&25&19&100\cr
&{SO}_{9}{SU}_{2}&13&1&16&E_{8}^{2}{SU}_{2}^{14}&30&21&5&58\cr
&F_{4}{SU}_{2}&13&13&28&E_{8}^{2}{SU}_{2}&17&17&10&46\cr
&{SO}_{12}&14&1&17&E_{8}^{2}{SU}_{2}^{18}&34&25&0&61\cr
&{SO}_{13}&14&17&33&E_{8}^{2}{SU}_{2}&17&21&5&45\cr
&E_{6}&14&1&17&E_{8}^{2}{SU}_{2}^{2}{SU}_{3}^{19}&56&25&0&83\cr
&E_{6b}&14&19&35&E_{8}^{2}{SU}_{2}^{2}{SU}_{3}&20&25&0&47\cr
&E_{7}&15&1&18&E_{8}^{2}{SU}_{2}^{21}&37&25&0&64\cr
&E_{7b}&15&21&38&E_{8}^{2}{SU}_{2}&17&25&0&44\cr
&E_{8}&16&25&43&E_{8}^{2}&16&25&0&43\cr
}

$$\vbox{
\def\skip{\hskip4pt}
\def\extraheight{\omit{\vrule height2pt}&&&&&&&&&\cr}
\offinterlineskip\halign{
\strut #\vrule height 9.1pt depth 3.1pt&\skip$#$\skip\hfil\vrule
&\skip$#$\skip\hfil\vrule&\skip$#$\skip\hfil\vrule
&\skip$#$\skip\hfil\vrule&\skip$#$\skip\hfil\vrule&\skip$#$\skip\hfil\vrule
&\skip$#$\skip\hfil\vrule&\skip$#$\skip\hfil\vrule&\skip$#$\skip\hfil\vrule \cr
\noalign{\hrule}
\omit{\vrule height11pt depth 5.1pt}&
\multispan{9}{\hfil\llap{$n~=~0$,~~ }\hbox{Groups} $SU_1\times H$,\skip
\hbox{Mirror Groups} $\widetilde H$\hfil\vrule}\cr
\noalign{\hrule}
\extraheight
&\hfil H&\hfil\hbox{rk}&\hfil n_T&\hfil h_{11}
&\hfil \widetilde H&\hfil \widetilde{\hbox{rk}} 
&\tilde n_T&\hfil \tilde\d&\hfil h_{21}\cr
\extraheight
\noalign{\hrule\vskip3pt\hrule}
\extraheight
\tablebodya
\extraheight
\noalign{\hrule}
}}
$$
\newpage
$$\vbox{
\def\skip{\hskip4pt}
\def\extraheight{\omit{\vrule height2pt}&&&&&&&&&\cr}
\offinterlineskip\halign{
\strut #\vrule height 9.1pt depth 3.1pt&\skip$#$\skip\hfil\vrule
&\skip$#$\skip\hfil\vrule&\skip$#$\skip\hfil\vrule&\skip$#$\skip\hfil\vrule
&\skip$#$\skip\hfil\vrule&\skip$#$\skip\hfil\vrule&\skip$#$\skip\hfil\vrule
&\skip$#$\skip\hfil\vrule&\skip$#$\skip\hfil\vrule\cr
\noalign{\hrule}
\omit{\vrule height11pt depth 5.1pt}&
\multispan{9}{\hfil\llap{$n~=~1$,~~ }\hbox{Groups} $SU_1\times H$,\skip
\hbox{Mirror Groups} $\widetilde H$\hfil\vrule}\cr
\noalign{\hrule}
\extraheight
&\hfil H&\hfil\hbox{rk}&\hfil n_T&\hfil h_{11}
&\hfil \widetilde H&\hfil \widetilde{\hbox{rk}} 
&\tilde n_T&\hfil \tilde\d&\hfil h_{21}\cr
\extraheight
\noalign{\hrule\vskip3pt\hrule}
\extraheight
\tablebodyb
\extraheight
\noalign{\hrule}
}}
$$
\newpage
$$\vbox{
\def\skip{\hskip4pt}
\def\extraheight{\omit{\vrule height2pt}&&&&&&&&&\cr}
\offinterlineskip\halign{
\strut #\vrule height 9.1pt depth 3.1pt&\skip$#$\skip\hfil\vrule
&\skip$#$\skip\hfil\vrule&\skip$#$\skip\hfil\vrule&\skip$#$\skip\hfil\vrule
&\skip$#$\skip\hfil\vrule&\skip$#$\skip\hfil\vrule&\skip$#$\skip\hfil\vrule
&\skip$#$\skip\hfil\vrule&\skip$#$\skip\hfil\vrule\cr
\noalign{\hrule}
\omit{\vrule height11pt depth 5.1pt}&
\multispan{9}{\hfil\llap{$n~=~2$,~~ }\hbox{Groups} $SU_1\times H$,\skip 
\hbox{Mirror Groups} $\widetilde H$\hfil\vrule}\cr
\noalign{\hrule}
\extraheight
&\hfil H&\hfil\hbox{rk}&\hfil n_T&\hfil h_{11}
&\hfil \widetilde H&\hfil \widetilde{\hbox{rk}} 
&\tilde n_T&\hfil \tilde\d&\hfil h_{21}\cr
\extraheight
\noalign{\hrule\vskip3pt\hrule}
\extraheight
\tablebodyc
\extraheight
\noalign{\hrule}
}}
$$
\newpage
$$\vbox{
\def\skip{\hskip4pt}
\def\extraheight{\omit{\vrule height2pt}&&&&&&&&&\cr}
\offinterlineskip\halign{
\strut #\vrule height 9.1pt depth 3.1pt&\skip$#$\skip\hfil\vrule
&\skip$#$\skip\hfil\vrule&\skip$#$\skip\hfil\vrule&\skip$#$\skip\hfil\vrule
&\skip$#$\skip\hfil\vrule&\skip$#$\skip\hfil\vrule&\skip$#$\skip\hfil\vrule
&\skip$#$\skip\hfil\vrule&\skip$#$\skip\hfil\vrule\cr
\noalign{\hrule}
\omit{\vrule height11pt depth 5.1pt}&
\multispan{9}{\hfil\llap{$n~=~3$,~~ }\hbox{Groups} $SU_3\times H$,\skip 
\hbox{Mirror Groups} $\widetilde H$\hfil\vrule}\cr
\noalign{\hrule}
\extraheight
&\hfil H&\hfil\hbox{rk}&\hfil n_T&\hfil h_{11}
&\hfil \widetilde H&\hfil \widetilde{\hbox{rk}} 
&\tilde n_T&\hfil \tilde\d&\hfil h_{21}\cr
\extraheight
\noalign{\hrule\vskip3pt\hrule}
\extraheight
\tablebodyd
\extraheight
\noalign{\hrule}
}}
$$
\newpage
$$\vbox{
\def\skip{\hskip4pt}
\def\extraheight{\omit{\vrule height2pt}&&&&&&&&&\cr}
\offinterlineskip\halign{
\strut #\vrule height 9.1pt depth 3.1pt&\skip$#$\skip\hfil\vrule
&\skip$#$\skip\hfil\vrule&\skip$#$\skip\hfil\vrule&\skip$#$\skip\hfil\vrule
&\skip$#$\skip\hfil\vrule&\skip$#$\skip\hfil\vrule&\skip$#$\skip\hfil\vrule
&\skip$#$\skip\hfil\vrule&\skip$#$\skip\hfil\vrule\cr
\noalign{\hrule}
\omit{\vrule height11pt depth 5.1pt}&
\multispan{9}{\hfil\llap{$n~=~4$,~~ }\hbox{Groups} $SO_8\times H$,\skip 
\hbox{Mirror Groups} $\widetilde H$\hfil\vrule}\cr
\noalign{\hrule}
\extraheight
&\hfil H&\hfil\hbox{rk}&\hfil n_T&\hfil h_{11}
&\hfil \widetilde H&\hfil \widetilde{\hbox{rk}} 
&\tilde n_T&\hfil \tilde\d&\hfil h_{21}\cr
\extraheight
\noalign{\hrule\vskip3pt\hrule}
\extraheight
\tablebodye
\extraheight
\noalign{\hrule}
}}
$$
\newpage
$$\vbox{
\def\skip{\hskip4pt}
\def\extraheight{\omit{\vrule height2pt}&&&&&&&&&\cr}
\offinterlineskip\halign{
\strut #\vrule height 9.1pt depth 3.1pt&\skip$#$\skip\hfil\vrule
&\skip$#$\skip\hfil\vrule&\skip$#$\skip\hfil\vrule&\skip$#$\skip\hfil\vrule
&\skip$#$\skip\hfil\vrule&\skip$#$\skip\hfil\vrule&\skip$#$\skip\hfil\vrule
&\skip$#$\skip\hfil\vrule&\skip$#$\skip\hfil\vrule\cr
\noalign{\hrule}
\omit{\vrule height11pt depth 5.1pt}&
\multispan{9}{\hfil\llap{$n~=~5$,~~ }\hbox{Groups} $F_4\times H$,\skip 
\hbox{Mirror Groups} $\widetilde H$\hfil\vrule}\cr
\noalign{\hrule}
\extraheight
&\hfil H&\hfil\hbox{rk}&\hfil n_T&\hfil h_{11}
&\hfil \widetilde H&\hfil \widetilde{\hbox{rk}} 
&\tilde n_T&\hfil \tilde\d&\hfil h_{21}\cr
\extraheight
\noalign{\hrule\vskip3pt\hrule}
\extraheight
\tablebodyf
\extraheight
\noalign{\hrule}
}}
$$
\newpage
$$\vbox{
\def\skip{\hskip4pt}
\def\extraheight{\omit{\vrule height2pt}&&&&&&&&&\cr}
\offinterlineskip\halign{
\strut #\vrule height 9.1pt depth 3.1pt&\skip$#$\skip\hfil\vrule
&\skip$#$\skip\hfil\vrule&\skip$#$\skip\hfil\vrule&\skip$#$\skip\hfil\vrule
&\skip$#$\skip\hfil\vrule&\skip$#$\skip\hfil\vrule&\skip$#$\skip\hfil\vrule
&\skip$#$\skip\hfil\vrule&\skip$#$\skip\hfil\vrule\cr
\noalign{\hrule}
\omit{\vrule height11pt depth 5.1pt}&
\multispan{9}{\hfil\llap{$n~=~6$,~~ }\hbox{Groups} $E_6\times H$,\skip
\hbox{Mirror Groups} $\widetilde H$\hfil\vrule}\cr
\noalign{\hrule}
\extraheight
&\hfil H&\hfil\hbox{rk}&\hfil n_T&\hfil h_{11}
&\hfil \widetilde H&\hfil \widetilde{\hbox{rk}} 
&\tilde n_T&\hfil \tilde\d&\hfil h_{21}\cr
\extraheight
\noalign{\hrule\vskip3pt\hrule}
\extraheight
\tablebodyg
\extraheight
\noalign{\hrule}
}}
$$
\newpage
$$\vbox{
\def\skip{\hskip4pt}
\def\extraheight{\omit{\vrule height2pt}&&&&&&&&&\cr}
\offinterlineskip\halign{
\strut #\vrule height 9.1pt depth 3.1pt&\skip$#$\skip\hfil\vrule
&\skip$#$\skip\hfil\vrule&\skip$#$\skip\hfil\vrule&\skip$#$\skip\hfil\vrule
&\skip$#$\skip\hfil\vrule&\skip$#$\skip\hfil\vrule&\skip$#$\skip\hfil\vrule
&\skip$#$\skip\hfil\vrule&\skip$#$\skip\hfil\vrule\cr
\noalign{\hrule}
\omit{\vrule height11pt depth 5.1pt}&
\multispan{9}{\hfil\llap{$n~=~7$,~~ }\hbox{Groups} $E_7\times H$,\skip
\hbox{Mirror Groups} $\widetilde H$\hfil\vrule}\cr
\noalign{\hrule}
\extraheight
&\hfil H&\hfil\hbox{rk}&\hfil n_T&\hfil h_{11}
&\hfil \widetilde H&\hfil \widetilde{\hbox{rk}} 
&\tilde n_T&\hfil \tilde\d&\hfil h_{21}\cr
\extraheight
\noalign{\hrule\vskip3pt\hrule}
\extraheight
\tablebodyh
\extraheight
\noalign{\hrule}
}}
$$
\newpage
$$\vbox{
\def\skip{\hskip4pt}
\def\extraheight{\omit{\vrule height2pt}&&&&&&&&&\cr}
\offinterlineskip\halign{
\strut #\vrule height 9.1pt depth 3.1pt&\skip$#$\skip\hfil\vrule
&\skip$#$\skip\hfil\vrule&\skip$#$\skip\hfil\vrule&\skip$#$\skip\hfil\vrule
&\skip$#$\skip\hfil\vrule&\skip$#$\skip\hfil\vrule&\skip$#$\skip\hfil\vrule
&\skip$#$\skip\hfil\vrule&\skip$#$\skip\hfil\vrule\cr
\noalign{\hrule}
\omit{\vrule height11pt depth 5.1pt}&
\multispan{9}{\hfil\llap{$n~=~8$,~~ }\hbox{Groups} $E_7\times H$,\skip
\hbox{Mirror Groups} $\widetilde H$\hfil\vrule}\cr
\noalign{\hrule}
\extraheight
&\hfil H&\hfil\hbox{rk}&\hfil n_T&\hfil h_{11}
&\hfil \widetilde H&\hfil \widetilde{\hbox{rk}} 
&\tilde n_T&\hfil \tilde\d&\hfil h_{21}\cr
\extraheight
\noalign{\hrule\vskip3pt\hrule}
\extraheight
\tablebodyi
\extraheight
\noalign{\hrule}
}}
$$
\newpage
$$\vbox{
\def\skip{\hskip4pt}
\def\extraheight{\omit{\vrule height2pt}&&&&&&&&&\cr}
\offinterlineskip\halign{
\strut #\vrule height 9.1pt depth 3.1pt&\skip$#$\skip\hfil\vrule
&\skip$#$\skip\hfil\vrule&\skip$#$\skip\hfil\vrule&\skip$#$\skip\hfil\vrule
&\skip$#$\skip\hfil\vrule&\skip$#$\skip\hfil\vrule&\skip$#$\skip\hfil\vrule
&\skip$#$\skip\hfil\vrule&\skip$#$\skip\hfil\vrule\cr
\noalign{\hrule}
\omit{\vrule height11pt depth 5.1pt}&
\multispan{9}{\hfil\llap{$n~=~9$,~~ }\hbox{Groups} $E_8\times H$,\skip
\hbox{Mirror Groups} $\widetilde H$\hfil\vrule}\cr
\noalign{\hrule}
\extraheight
&\hfil H&\hfil\hbox{rk}&\hfil n_T&\hfil h_{11}
&\hfil \widetilde H&\hfil \widetilde{\hbox{rk}} 
&\tilde n_T&\hfil \tilde\d&\hfil h_{21}\cr
\extraheight
\noalign{\hrule\vskip3pt\hrule}
\extraheight
\tablebodyj
\extraheight
\noalign{\hrule}
}}
$$
\newpage
$$\vbox{
\def\skip{\hskip4pt}
\def\extraheight{\omit{\vrule height2pt}&&&&&&&&&\cr}
\offinterlineskip\halign{
\strut #\vrule height 9.1pt depth 3.1pt&\skip$#$\skip\hfil\vrule
&\skip$#$\skip\hfil\vrule&\skip$#$\skip\hfil\vrule&\skip$#$\skip\hfil\vrule
&\skip$#$\skip\hfil\vrule&\skip$#$\skip\hfil\vrule&\skip$#$\skip\hfil\vrule
&\skip$#$\skip\hfil\vrule&\skip$#$\skip\hfil\vrule\cr
\noalign{\hrule}
\omit{\vrule height11pt depth 5.1pt}&
\multispan{9}{\hfil\llap{$n~=~10$,~~ }\hbox{Groups} $E_8\times H$,\skip
\hbox{Mirror Groups} $\widetilde H$\hfil\vrule}\cr
\noalign{\hrule}
\extraheight
&\hfil H&\hfil\hbox{rk}&\hfil n_T&\hfil h_{11}
&\hfil \widetilde H&\hfil \widetilde{\hbox{rk}} 
&\tilde n_T&\hfil \tilde\d&\hfil h_{21}\cr
\extraheight
\noalign{\hrule\vskip3pt\hrule}
\extraheight
\tablebodyk
\extraheight
\noalign{\hrule}
}}
$$
\newpage
$$\vbox{
\def\skip{\hskip4pt}
\def\extraheight{\omit{\vrule height2pt}&&&&&&&&&\cr}
\offinterlineskip\halign{
\strut #\vrule height 9.1pt depth 3.1pt&\skip$#$\skip\hfil\vrule
&\skip$#$\skip\hfil\vrule&\skip$#$\skip\hfil\vrule&\skip$#$\skip\hfil\vrule
&\skip$#$\skip\hfil\vrule&\skip$#$\skip\hfil\vrule&\skip$#$\skip\hfil\vrule
&\skip$#$\skip\hfil\vrule&\skip$#$\skip\hfil\vrule\cr
\noalign{\hrule}
\omit{\vrule height11pt depth 5.1pt}&
\multispan{9}{\hfil\llap{$n~=~11$,~~ }\hbox{Groups} $E_8\times H$,\skip
\hbox{Mirror Groups} $\widetilde H$\hfil\vrule}\cr
\noalign{\hrule}
\extraheight
&\hfil H&\hfil\hbox{rk}&\hfil n_T&\hfil h_{11}
&\hfil \widetilde H&\hfil \widetilde{\hbox{rk}} 
&\tilde n_T&\hfil \tilde\d&\hfil h_{21}\cr
\extraheight
\noalign{\hrule\vskip3pt\hrule}
\extraheight
\tablebodyl
\extraheight
\noalign{\hrule}
}}
$$
\newpage
$$\vbox{
\def\skip{\hskip4pt}
\def\extraheight{\omit{\vrule height2pt}&&&&&&&&&\cr}
\offinterlineskip\halign{
\strut #\vrule height 9.1pt depth 3.1pt&\skip$#$\skip\hfil\vrule
&\skip$#$\skip\hfil\vrule&\skip$#$\skip\hfil\vrule&\skip$#$\skip\hfil\vrule
&\skip$#$\skip\hfil\vrule&\skip$#$\skip\hfil\vrule&\skip$#$\skip\hfil\vrule
&\skip$#$\skip\hfil\vrule&\skip$#$\skip\hfil\vrule\cr
\noalign{\hrule}
\omit{\vrule height11pt depth 5.1pt}&
\multispan{9}{\hfil\llap{$n~=~12$,~~ }\hbox{Groups} $E_8\times H$,\skip
\hbox{Mirror Groups} $\widetilde H$\hfil\vrule}\cr
\noalign{\hrule}
\extraheight
&\hfil H&\hfil\hbox{rk}&\hfil n_T&\hfil h_{11}
&\hfil \widetilde H&\hfil \widetilde{\hbox{rk}} 
&\tilde n_T&\hfil \tilde\d&\hfil h_{21}\cr
\extraheight
\noalign{\hrule\vskip3pt\hrule}
\extraheight
\tablebodym
\extraheight
\noalign{\hrule}
}}
$$
\newpage
\immediate\closeout\referencewrite\referenceopenfalse
\line{\bf\hfil References\hfil}\bigskip\parindent=0pt\input referenc.texauxil

\bye